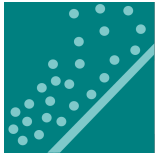

Review

# One-Dimensional Relativistic Self-Gravitating Systems

Robert B. Mann

## Special Issue
Statistical Mechanics of Self-Gravitating Systems

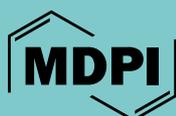

*Review*

# One-Dimensional Relativistic Self-Gravitating Systems [†]

**Robert B. Mann [1,2]**

[1] Department of Physics and Astronomy, University of Waterloo, Waterloo, ON N2L 3G1, Canada; rbmann@uwaterloo.ca
[2] Perimeter Institute for Theoretical Physics, 35 Caroline St., Waterloo, ON N2L 2Y5, Canada
[†] Dedicated to the memory of Tadayuki Ohta, who introduced me to this fascinating subject.

**Abstract:** One of the oldest problems in physics is that of calculating the motion of $N$ particles under a specified mutual force: the $N$-body problem. Much is known about this problem if the specified force is non-relativistic gravity, and considerable progress has been made by considering the problem in one spatial dimension. Here, I review what is known about the relativistic gravitational $N$-body problem. Reduction to one spatial dimension has the feature of the absence of gravitational radiation, thereby allowing for a clear comparison between the physics of one-dimensional relativistic and non-relativistic self-gravitating systems. After describing how to obtain a relativistic theory of gravity coupled to $N$ point particles, I discuss in turn the two-body, three-body, four-body, and $N$-body problems. Quite general exact solutions can be obtained for the two-body problem, unlike the situation in general relativity in three spatial dimensions for which only highly specified solutions exist. The three-body problem exhibits mild forms of chaos, and provides one of the first theoretical settings in which relativistic chaos can be studied. For $N \geq 4$, other interesting features emerge. Relativistic self-gravitating systems have a number of interesting problems awaiting further investigation, providing us with a new frontier for exploring relativistic many-body systems.

## 1. Introduction

One of the oldest problems in physics is the $N$-body problem: the determination of the motion of a system of $N$ particles mutually interacting through specified forces. This problem appears in a broad variety of subfields of physics, including cosmology, stellar dynamics, planetary motion, atomic physics, and nuclear physics. The $N$-body problem is a particular challenge if the interactions are purely gravitational. Although an exact solution is known for the two-body problem in pure Newtonian gravity in three spatial dimensions, there is no closed form solution for large $N$, even for $N = 3$ [1], although particular solutions exist in restricted cases [2]. No exact solution is known in the general-relativistic case, even for $N = 2$, since it experiences the dissipation of energy in the form of gravitational radiation.

One-dimensional self-gravitating systems (OGSs) have played an important role in advancing our understanding of the gravitational $N$-body problem [3]. Such systems have been of interest for over half a century, where they have played an important role in astrophysics and cosmology for more than 30 years [4]. Apart from being prototypes for studying the behaviour of gravity in higher dimensions, they also approximate the behaviour in three spatial dimensions of some physical systems. Examples include very long-lived core-halo configurations that model a dense massive core in near-equilibrium, surrounded by a halo of high-kinetic energy stars that feebly interact with the core [4–6]. Other examples include cosmological models [7,8], the dynamics of stars in a direction orthogonal to the plane of a highly flattened galaxy [9], shells of matter interacting with a spherical globular cluster [10], and the collisions of flat parallel domain walls moving in directions orthogonal to their surfaces. A recent review of the OGS [11] provides a description of its basic properties, its relaxation to equilibrium, and its application to dynamical structure formation in cosmology.

Although the connection between the idealized OGS and natural astrophysical systems can be tenuous, the accuracy and ease with which their dynamical evolution may be simulated has remained the principal motivation for continued study of the OGS. Unlike 3-dimensional self-gravitating systems, in which the motion of the (point) masses must be numerically integrated, the OGS admits direct computation of the particle (or sheet, or shell) crossings. This provides the accurate computation of the evolution of the system over many dynamical time scales. Furthermore, a number of interesting questions concerning the statistical properties of the OGS remain open, including whether it can attain a state of true equilibrium from arbitrary initial conditions [5], its ergodic behaviour, the circumstances (if any) under which equipartition of energy can be attained [12], and the appearance of fractal behaviour [3,13].

For three decades, studies of the OGS have been in a non-relativistic context, assuming Newtonian gravity with its standard causal structure [3,7,8,12,14–22]. Research into relativistic one-dimensional self-gravitating systems (ROGS) was generally ignored. In large part, this was because relativistic effects do not play a dominant role in stellar dynamics, but it was also due to the lack of a theoretical framework for relativistic gravity in one spatial dimension. The Einstein tensor is identically zero in this $(1 + 1)$-dimensional space–time context, and so Einstein's equation at face value would simply imply vanishing stress–energy. However, a reduction in the number of spatial dimensions in a relativistic context can be expected to be quite useful since gravitational radiation (at least due to spin-2 gravitons) cannot exist in less than three spatial dimensions. However, most (if not all) of the remaining conceptual features of relativistic gravity are retained in lower dimensions, and so one might hope to obtain insights into the nature of relativistic classical and quantum gravitation in a wide variety of physical situations by studying the ROGS.

It is straightforward to find a set of equations governing the motion of particles—these are furnished by the geodesic equations. In addition to this, what is needed to study ROGS is a set of equations governing the dynamics of the space–time metric in a self-consistent way. Early versions of $(1 + 1)$–dimensional gravity [23,24] set the Ricci curvature scalar equal to a constant, yielding trivial dynamics for the space–time metric (although containing sufficiently interesting features [25] such that this theory is still of interest today [26]). The intensive investigation of a wide variety of gravitational theories ensued a few years later, primarily motivated by a quest to understand quantum gravity in a simplified context [27]. The overwhelming majority of such investigations were concerned with the (quantum) dynamics of the space–time metric, and not with the dynamical motion of particles in such space–times.

At about the same time that interest in the $(1 + 1)$-dimensional quantum gravity began, investigations into the ROGS also began. The purpose of this article is to review the origins, results, and status of relativistic one-dimensional self-gravitating systems. After a brief review of the OGS, I begin by reviewing how the $D \to 2$ limit of $D$-dimensional general relativity [28] can be self-consistently coupled to point particles, thereby yielding the ROGS. The equations of motion for the particles are obtained using the canonical formalism, which I describe in some detail. I shall then consider in turn the 2-body, 3-body, 4-body, and $N$-body ROGS, discussing their distinctions from the OGS, their salient features, their chaotic behaviour, and their statistical properties, as relevant. I conclude by discussing a number of interesting open problems for relativistic one-dimensional self-gravitating systems. While other constants will retain their values throughout, the speed of light $c$ will generally be set to unity, and only explicitly written where relevant for instructive purposes.

## 2. Non-Relativistic Self-Gravitating Systems

For a system of particles, the Hamiltonian in Newtonian gravity in two dimensions is

$$H = \sum_a \frac{p_a^2}{2m_a} + \pi G \sum_{a,b=1}^N m_a m_b \mid z_a - z_b \mid \tag{1}$$

where $m_a$, $z_a$, and $p_a$ are the mass, the coordinate location, and the momentum of the $a$-th particle, respectively, and $G$ is the gravitational constant. The potential between any two particles is proportional to the product of their masses and the spatial separation between them, as expected from dimensionally continuing the well-known potential

$$V = -G_d \sum_{a \neq b}^{N} \frac{m_a m_b}{|z_a - z_b|^{d-2}} \tag{2}$$

of Newtonian gravity in $d$ spatial dimensions, where $G_1 = \pi G$. When $d = 1$ the potential in (1) vanishes, and so the restriction $a \neq b$ in (1) is not required.

The equations of motion of the OGS (1) are given by Hamilton's equations

$$\dot{z}_a = \frac{\partial H}{\partial p_a} = \frac{p_a}{m_a} \tag{3}$$

$$\dot{p}_a = -\frac{\partial H}{\partial z_a} = -2\pi G \sum_b m_a m_b \frac{\partial \mid z_a - z_b \mid}{\partial z_a}$$

$$= -2\pi G \sum_b m_a m_b \, \mathrm{sgn}(z_a - z_b) \tag{4}$$

yielding

$$\ddot{z}_a = -2\pi G \sum_b m_b \, \mathrm{sgn}(z_a - z_b) \tag{5}$$

for the acceleration of the $a$-th particle.

We see that each particle experiences a constant force from each of the other particles, where the sign ($-$ or $+$) of the force from any given particle depends on whether $z_a > z_b$ or vice versa. The force is therefore always attractive: if $z_a$ is to the right of $z_b$ ($z_a > z_b$) then particle $a$ will accelerate leftward toward $z_b$ (and $b$ rightward toward $a$) until they meet, after which time $z_a$ is to the left of $z_b$ ($z_a < z_b$) and the force changes sign, accelerating particle $a$ rightward. This scenario assumes that the particles can pass through each other without any influence, as would be appropriate for parallel sheets of particulate matter where collisions between the particles can be neglected. It is of course possible to include an additional structure—for example, modelling the particles as impenetrable points would make them bounce off of each other—but this would detract from the study of pure gravitational effects. Such an additional structure will not be considered in this article, apart from the inclusion of attractive and repulsive electromagnetic interactions.

Consider the case $N = 2$. If the particles are initially separated by some distance $d$, they will move toward each other with constant acceleration until they cross, after which the acceleration of each flips sign. The particles fly apart increasingly slowly until they reach a maximal separation $d$, after which they move toward each other again with increasing speed. After crossing a second time, the particles separate, moving with decreasing speed until they return to their original positions. Assuming no other interactions, the motion then repeats perpetually. Prior to crossing, the entire scenario is equivalent to that of a body of mass $m$ falling near the surface of the Earth.

For $N$ particles, every particle initially undergoes constant acceleration until the first two particles cross. This causes a sign flip in the force between each particle in the crossing pair, thereby changing the magnitudes (and perhaps signs) of the accelerations of each due to the presence of the other particles. As more particles cross, this changes the accelerations of more and more particles, generally yielding chaotic motion.

The simplest example of this occurs for $N = 3$. The 3-body OGS has been shown to exhibit a mild form of chaos [3]. This OGS can be mapped to a single-particle moving in two spatial dimensions in a hexagonal-well potential $V(x, y)$, whose sides are impenetrable flat sheets. If the masses of all particles are equal, then the shape is that of a regular hexagon; unequal masses distort this symmetry so that the sides are of unequal length [29,30]. Alternatively, the three-body OGS can be regarded as a single-particle under the influence of a constant gravitational field in two spatial dimensions that bounces off of a symmetric wedge of angle $2\theta$, where this angle parametrizes the relative inequality of the masses [3].

Constructing a relativistic of the Hamiltonian (1) is somewhat subtle. This is the subject of the next section.

## 3. Relativistic Gravity Coupled to Point Particles

In three spatial dimensions, a self-gravitating system would consist of a set of $N$ particles minimally coupled to Einstein gravity. Its action in $n$-dimensions is [31]

$$I_n = \int d^n x \left[ \frac{1}{2\kappa_n} \sqrt{-g}(R - 2\Lambda_n) - \sum_{a=1}^{N} m_a \int d\tau_a \left\{ -g_{\mu\nu}(x) \frac{dz_a^\mu}{d\tau_a} \frac{dz_a^\nu}{d\tau_a} \right\}^{1/2} \delta^{(n)}(x - z_a(\tau_a)) \right] \quad (6)$$

where $|\Lambda_n| = \frac{(n-2)(n-1)}{2\ell^2}$ is the cosmological constant (whose sign $+/-$, respectively, corresponds to asymptotically de Sitter/anti-de Sitter space–time), $\kappa_n = 8\pi G_n/c^4$ is the gravitational coupling, and $R$ is the Ricci scalar. Systems of astrophysical interest have $n = 4$. Notwithstanding issues connected with collisions between the particles, the field equations that follow from this action embody what we expect from a self-gravitating system: the curvature of space–time governs how the particles move along the trajectories $z_a(\tau_a)$, and the masses and motion of these particles in turn govern how space–time dynamically curves.

A ROGS that resembles a relativistic three-dimensional self-gravitating system as closely as possible should therefore have the following features.

1. The stress–energy of the particles generates a space–time curvature in as simple a manner as possible.
2. The curvature of space–time guides the motion of each particle in accordance with the equivalence principle, in the absence of any extraneous forces.
3. The dynamics of the system is self-consistent.

One might expect to obtain the ROGS by simply setting $n = 2$ in (6) (1 space dimension, 1 time dimension), but this will not work: the Ricci scalar is a total derivative in $(1 + 1)$ dimensions, in turn implying that the Einstein tensor is identically zero for any metric.

However, there is a way of taking the $n \to 2$ limit of the action (6) [28]. Consider the action

$$
\begin{aligned}
I_n^{EH} &= \frac{1}{2\kappa_n} \left( \int d^n x \sqrt{-\bar{g}} \bar{R} - \int d^n x \sqrt{-g} R \right) \\
&= \frac{1}{2\kappa_n} \int d^n x \left[ e^{\epsilon \Psi/2} \left[ \left( R - (\epsilon + 1) \Box \Psi \right) - \frac{1}{4} \epsilon(\epsilon + 1)(\partial \Psi)^2 \right] - R \right]
\end{aligned}
\quad (7)
$$

where $\bar{g} = e^\Psi g$ is a conformally rescaled metric and $\epsilon = n - 2$. Expanding in powers of $\epsilon$ yields

$$I_2^{EH} = \frac{n-2}{4\kappa_n} \int d^2 x \sqrt{-g} \left[ \Psi R - \Psi \Box \Psi - \frac{1}{2} (\nabla \Psi)^2 \right] \quad (8)$$

and so by setting $\kappa_2 = \frac{2\kappa_n}{n-2}$, we obtain

$$I_2^{EH} \equiv \lim_{n \to 2} I_n^{EH} = \frac{1}{2\kappa_2} \int d^2 x \sqrt{-g} \left[ \Psi R + \frac{1}{2} (\nabla \Psi)^2 \right] \quad (9)$$

upon discarding the total divergences. The limit $n \to 2$ in (6) is straightforward for the other two terms, yielding

$$
\begin{aligned}
I &= \int d^2x \sqrt{-g} \left[ \frac{1}{2\kappa} \left( \Psi R + \frac{1}{2} (\nabla \Psi)^2 - 2\Lambda \right) \right. \\
&\quad \left. - \sum_{a=1}^{N} m_a \int d\tau_a \left\{ -g_{\mu\nu}(x) \frac{dz_a^\mu}{d\tau_a} \frac{dz_a^\nu}{d\tau_a} \right\}^{1/2} \delta^{(2)}(x - z_a(\tau_a)) \right]
\end{aligned}
\tag{10}
$$

as the action for the ROGS, where $\kappa = \kappa_2$ and $\Lambda = \Lambda_2 = \pm \frac{1}{\ell^2}$.

Note that this procedure incorporates an additional field $\Psi$ in the gravitational action. This might seem to be in tension with the second desirable feature in the list above. However, the field equations derived from the variations $\delta\Psi$ and $\delta g_{\mu\nu}$ are

$$
R - g^{\mu\nu} \nabla_\mu \nabla_\nu \Psi = R - \frac{1}{\sqrt{-g}} \partial_\mu \left( \sqrt{-g} g^{\mu\nu} \partial_\nu \Psi \right) = 0
\tag{11}
$$

$$
\frac{1}{2} \nabla_\mu \Psi \nabla_\nu \Psi - \frac{1}{4} g_{\mu\nu} \nabla^\lambda \Psi \nabla_\lambda \Psi + g_{\mu\nu} \nabla^\lambda \nabla_\lambda \Psi - \nabla_\mu \nabla_\nu \Psi = \kappa T_{\mu\nu} - g_{\mu\nu} \Lambda
\tag{12}
$$

where

$$
T_{\mu\nu} = -\frac{2}{\sqrt{-g}} \frac{\delta \mathcal{L}_M}{\delta g^{\mu\nu}} = \sum_a m_a \int d\tau_a \frac{1}{\sqrt{-g}} g_{\mu\sigma} g_{\nu\rho} \frac{dz_a^\sigma}{d\tau_a} \frac{dz_a^\rho}{d\tau_a} \delta^{(2)}(x - z_a(\tau_a))
\tag{13}
$$

where the last term in (10) is the matter Lagrangian $\mathcal{L}_M$. Taking the trace of (12) and inserting the result into (11) gives

$$
R + 2\Lambda = \kappa T^\mu_{\ \mu} .
\tag{14}
$$

which shows that the evolution of the metric depends only on the stress-energy, and decouples from the evolution of $\Psi$. The variation $\delta z_a^\mu$ yields

$$
\frac{d}{d\tau_a} \left\{ g_{\mu\nu}(z_a) \frac{dz_a^\nu}{d\tau_a} \right\} - \frac{1}{2} g_{\nu\lambda,\mu}(z_a) \frac{dz_a^\nu}{d\tau_a} \frac{dz_a^\lambda}{d\tau_a} = 0
\tag{15}
$$

which is the geodesic equation, or rather equations since there is one per particle.

The system (14,15) forms a closed relativistic self-gravitating system of $N$ point particles. The space–time curvature is determined from the stress-energy of the particulate matter from (14); in turn, the evolution of the particles is determined by the space–time curvature via (15).

The theory given by the action (10) is known as $R = T$ theory [32–34] . Its classical properties, including gravitational collapse, black holes, cosmological solutions, solitonic properties, and thermodynamics have been extensively studied [35–69]. Its chief interest lies in the fact that it captures the essence of classical general relativity in two space–time dimensions, and has $(1 + 1)$–dimensional analogs of many of its properties [35,70,71]. Moreover, it has a well-defined Newtonian limit [33,35]. This is in contrast to generic scalar-tensor theories [72], where the dilaton does not decouple from the evolution of the gravitational field.

The quantum properties of $R = T$ theory have also received attention from a variety of perspectives [37,73–89]. There is also a supersymmetric version of the theory [90], which has supersymmetric black hole solutions [91–93]. Recently, experiments have been carried out [94,95] that test certain aspects of the theory, albeit in a simulated context.

The remainder of this article is concerned with the $N$-body problem and its solutions as determined from the equations that follow from the action (10). The canonical approach is the most useful way to obtain these equations. This is the subject of the next section.

## 4. Canonical Formalism for Particle Dynamics

To obtain the ROGS Hamiltonian, the canonical Arnowitt–Deser–Misner (ADM) formalism [96–98] can be employed, as with the ADM formalism in $(3+1)$-dimensional theory. The result of this procedure is that constraints are eliminated, coordinate conditions are imposed, and the reduced Hamiltonian is given as a spatial integral of the second derivative of $\Psi$, which is a function of the dynamical variables $(z_a, p_a)$ of the particles [99]. Consequently, the Hamiltonian is completely determined in terms of the coordinates and momenta of the particles.

### 4.1. Neutral Particles

Consider first the *N*-body system whose interactions are purely gravitational. Writing the metric as

$$ds^2 = -N_0^2 dt^2 + \gamma \left( dx + \frac{N_1}{\gamma} dt \right)^2 \tag{16}$$

its extrinsic curvature is

$$K = (2N_0\gamma)^{-1}(2\partial_1 N_1 - \gamma^{-1} N_1 \partial_1 \gamma - \partial_0 \gamma) = \sqrt{\gamma}\kappa(\pi - \Pi/\gamma) \tag{17}$$

where $\pi$ and $\Pi$ are the respective conjugate momenta to $\gamma$ and $\Psi$. Using this, the action integral (10) can be written as

$$I = \int d^2x \left\{ \sum_a p_a \dot{z}_a \delta(x - z_a(x^0)) + \pi\dot{\gamma} + \Pi\dot{\Psi} + N_0 R^0 + N_1 R^1 \right\} \tag{18}$$

where

$$
\begin{aligned}
R^0 &= -\kappa\sqrt{\gamma}\gamma\pi^2 + 2\kappa\sqrt{\gamma}\pi\Pi + \frac{1}{4\kappa\sqrt{\gamma}}(\Psi')^2 - \frac{1}{\kappa}\left(\frac{\Psi'}{\sqrt{\gamma}}\right)' \\
&\quad - \frac{\Lambda}{\kappa}\sqrt{\gamma} - \sum_a \sqrt{\frac{p_a^2}{\gamma} + m_a^2}\, \delta(x - z_a(x^0))
\end{aligned}
\tag{19}
$$

$$
R^1 = \frac{\gamma'}{\gamma}\pi - \frac{1}{\gamma}\Pi\Psi' + 2\pi' + \sum_a \frac{p_a}{\gamma}\delta(x - z_a(x^0))
\tag{20}
$$

where $\partial_0$ is denoted by $(\,\dot{}\,)$ and $\partial_1$ by a symbol $(\,'\,)$.

Performing the canonical reduction is a somewhat involved procedure, but is fully analogous to the $(3+1)$-dimensional case [100,101]. Suppressing the details, the coordinate conditions

$$\gamma = 1 \qquad \text{and} \qquad \Pi = 0 \tag{21}$$

can be consistently chosen, and subsequently, the action integral reduces to [99]

$$I_R = \int dx^2 \left\{ \sum_a p_a \dot{z}_a \delta(x - z_a) - \mathcal{H} \right\} \ . \tag{22}$$

where, writing $\triangle\Psi \equiv \Psi''$,

$$H = \int dx \mathcal{H} = -\frac{1}{\kappa}\int dx \triangle\Psi \tag{23}$$

which can be verified using a superpotential approach based on Noether's theorem [102]. The auxiliary field $\Psi$ is determined by solving [99,103,104]

$$\triangle\Psi - \frac{1}{4}(\Psi')^2 + \kappa^2\pi^2 + \Lambda + \kappa\sum_a \sqrt{p_a^2 + m_a^2}\delta(x - z_a) = 0 \ , \tag{24}$$

$$2\pi' + \sum_a p_a\delta(x - z_a) = 0 \tag{25}$$

which follow from reducing the constraint Equations (19) and (20) via the canonical reduction procedure. Note that $\Psi$ is a function of the dynamical variables $(z_a, p_a)$ of the particles, as stated above.

The expression (23) is analogous to the reduced Hamiltonian in $(3 + 1)$ dimensional general relativity [98,100,101]. Equation (24) can be shown to be an energy balance condition: the energy of the gravitational field (expressed in terms of $(\Psi, \pi)$ and the cosmological constant $\Lambda$) plus the relativistic energy of motion of the particles must add to zero. Likewise, (25) expresses the fact that the total momentum of the gravitational field and the particles must add to zero.

Two interesting approximations can be obtained from (23) and the constraint equations. One is an expansion in powers of the coupling $\kappa$—the weak-field expansion. The other is an expansion in inverse powers of $c$—the post-Newtonian expansion.

The $\kappa$-expansion can be carried out by solving (24) for $\triangle\Psi$ and inserting the result into (23). Setting $\Lambda = 0$ for simplicity, this yields

$$
\begin{aligned}
H &= \sum_a \sqrt{p_a^2 + m_a^2}\left\{1 - \frac{1}{4}\Psi(z_a) + \frac{1}{32}\Psi(z_a)\Psi(z_a)\right\} \\
&\quad + \frac{\kappa}{2}\sum_a p_a \chi(z_a)\left\{1 - \frac{1}{4}\Psi(z_a)\right\} + \frac{\kappa^2}{8}\sum_a \sqrt{p_a^2 + m_a^2}\,\chi(z_a)\chi(z_a) \\
&= \sum_a \sqrt{p_a^2 + m_a^2} + \frac{\kappa}{8}\sum_a\sum_b\left(\sqrt{p_a^2 + m_a^2}\sqrt{p_b^2 + m_b^2} - p_a p_b\right)r_{ab} \\
&\quad + \frac{\epsilon\kappa}{8}\sum_a\sum_b\left(\sqrt{p_a^2 + m_a^2}\,p_b - p_a\sqrt{p_b^2 + m_b^2}\right)(z_a - z_b) \\
&\quad + \frac{1}{4}\left(\frac{\kappa}{4}\right)^2\left\{\sum_a\sqrt{p_a^2 + m_a^2}\left[\sum_b p_b r_{ab} + \epsilon\sum_b\sqrt{p_b^2 + m_b^2}(z_a - z_b)\right]^2\right. \\
&\quad - \sum_a p_a\left[\sum_b p_b r_{ab} + \epsilon\sum_b\sqrt{p_b^2 + m_b^2}(z_a - z_b)\right]\left[\sum_c\sqrt{p_c^2 + m_c^2}\,r_{ac} + \epsilon\sum_c p_c(z_a - z_c)\right] \\
&\quad + \sum_a\sum_b\left[\sqrt{p_a^2 + m_a^2}\sqrt{p_b^2 + m_b^2}\,r_{ab} - \epsilon\,p_a\sqrt{p_b^2 + m_b^2}(z_a - z_b)\right] \\
&\quad\quad\quad\quad\quad\quad \times\left[\sum_c\sqrt{p_c^2 + m_c^2}\,r_{bc} + \epsilon\sum_c p_c(z_b - z_c)\right] \\
&\quad - \sum_{a,b}\left[\sqrt{p_a^2 + m_a^2}\,p_b r_{ab} - \epsilon\,p_a p_b(z_a - z_b)\right]\left[\sum_c p_c r_{bc} + \epsilon\sum_c\sqrt{p_c^2 + m_c^2}(z_b - z_c)\right]\right\}
\end{aligned}
\tag{26}
$$

to second order in $\kappa$, where $\chi$ is defined by $\chi' \equiv \pi$. In order to obtain (26), the boundary term

$$
S_b = \left[-\frac{1}{4\kappa}\Psi\Psi' + \kappa\chi\chi' - \frac{\kappa}{8}\left\{\Psi\left(\chi^2\right)' - \Psi'\chi^2\right\} + \frac{1}{32\kappa}\Psi^2\Psi'\right]_{-\infty}^{\infty}
\tag{27}
$$

must vanish. Here, there is a subtle problem in comparison to the $(3 + 1)$–dimensional setting because the dimensionless potential $Gm|x|/c^2$ becomes infinite at spatial infinity. It is straightforward to show that, if the function $f(x) \equiv \Psi^2 - 4\kappa^2\chi^2$ and its first derivative vanish for $| x | >> | z_a |$ for all $a$, then the surface term (27) vanishes, since it is a sum of the terms proportional to $f'$ and $(4 - \Psi)f' + f\Psi'$.

The $\kappa$-expansion is the successive approximation of the ROGS in the background of Minkowskian space–time. At each order in $\kappa$, the relativistic form is preserved, and so this approximation is appropriate for describing the relativistic motion of the particles in weak gravitational fields.

The post-Newtonian expansion is an approximation of the powers of $1/c$. Temporarily restoring $t \to ct$, we note that both $p_a^2/m_a^2$ and $\sqrt{\kappa}$ are of the order of $c^{-2}$. This yields

$$
\begin{aligned}
H = & \sum_a m_a c^2 + \sum_a \frac{\tilde{p}_a^2}{2m_a} + \frac{\kappa}{8} \sum_a \sum_b m_a m_b \tilde{r}_{ab} \\
& - \sum_a \frac{\tilde{p}_a^4}{8 m_a^3 c^2} + \frac{\kappa c^2}{8} \sum_a \sum_b m_a \frac{\tilde{p}_b^2}{m_b} \tilde{r}_{ab} - \frac{\kappa c^2}{8} \sum_a \sum_b \tilde{p}_a \tilde{p}_b \tilde{r}_{ab} \\
& + \frac{1}{4} \left( \frac{\kappa c^3}{4} \right)^2 \sum_a \sum_b \sum_c m_a m_b m_c \{ \tilde{r}_{ab} \tilde{r}_{ac} - (\tilde{z}_a - \tilde{z}_b)(\tilde{z}_a - \tilde{z}_c) \}
\end{aligned}
\tag{28}
$$

from (26), where the canonical variables $(z_a, p_a)$ have been redefined

$$
z_a \longrightarrow \tilde{z}_a = z_a \qquad p_a \longrightarrow \tilde{p}_a = p_a - \frac{\epsilon \kappa}{4} \sum_b m_a m_b r_{ab}
\tag{29}
$$

to eliminate a spurious term of order $1/c$, with $r_{ab} = (z_a - z_b)$. Under this redefinition, the Poisson brackets amongst the canonical variables remain unchanged. It is straightforward to show that the canonical equations of motion

$$
\dot{\tilde{z}}_a = \frac{\partial H}{\partial \tilde{p}_a} \qquad \dot{\tilde{p}}_a = -\frac{\partial H}{\partial \tilde{z}_a}
\tag{30}
$$

are equivalent to the geodesic Equation (15) [99].

For illustrative purposes, consider a single static source at the origin with $\Lambda = 0$. The constraint Equations (24) and (25) become

$$
-\frac{1}{\kappa} \triangle \Psi = \kappa \pi^2 - \frac{1}{4\kappa}(\Psi')^2 + M\delta(x)
\tag{31}
$$

$$
\pi' = 0
\tag{32}
$$

and the Hamiltonian equations that follow from (18) are

$$
\dot{\pi} + N_0 \left[ \frac{3\kappa}{2} \pi^2 + \frac{1}{8\kappa}(\Psi')^2 \right] + N_1 \pi' + \frac{1}{2\kappa} N_0' \Psi' + N_1' \pi = 0
\tag{33}
$$

$$
\dot{\Psi} + 2\kappa N_0 \pi - N_1 \Psi' = 0
\tag{34}
$$

$$
\kappa \pi N_0 + N_1' = 0
\tag{35}
$$

$$
\partial_1 \left( \frac{1}{2} N_0 \Psi' + N_0' \right) = 0
\tag{36}
$$

Solutions to these equations that ensure the boundary conditions $S_b = 0$ hold are

$$
\Psi = -\frac{\kappa M}{2} \mid x \mid + \frac{\epsilon \kappa M}{2} t
\tag{37}
$$

$$
\pi = \chi' = -\frac{\epsilon}{4} M \qquad \left( \chi = -\frac{\epsilon M}{4} x \right)
\tag{38}
$$

$$
N_0 = e^{\frac{\kappa M}{4} |x|}
\tag{39}
$$

$$
N_1 = \epsilon \frac{x}{\mid x \mid} \left( e^{\frac{\kappa M}{4} |x|} - 1 \right)
\tag{40}
$$

where $\epsilon$ is a constant of integration, with $\epsilon^2 = 1$. Solutions with $\epsilon = 1$ are related by time-reversal to those with $\epsilon = -1$. This factor guarantees the invariance of the whole theory under the time reversal. We also see from (37) that, even for a static source, the auxiliary field is not static.

The Hamiltonian is

$$H = -\frac{1}{\kappa} \int dx \triangle \Psi = M \tag{41}$$

showing that the total energy is the mass, and

$$g_{00} = -N_0^2 + N_1^2 = 1 - 2e^{\frac{\kappa M}{4}|x|} < -1$$

indicating that this static point–particle solution has no event horizon.

### 4.2. Charged Particles

It is also possible to include non-gravitational interactions into the ROGS. Electromagnetism is an obvious choice to consider, and in fact the derivation of the canonically reduced Hamiltonian for charged particles is parallel to that of the previous subsection for the uncharged case [105]. Here, I shall highlight aspects of the charged case that are distinct from the uncharged case.

The action integral for gravitational and electromagnetic fields coupled with *N*-charged point masses is

$$
\begin{aligned}
I = & \int d^2x \left[ \frac{1}{2\kappa} \sqrt{-g} g^{\mu\nu} \left\{ \Psi R_{\mu\nu} + \frac{1}{2} \nabla_\mu \Psi \nabla_\nu \Psi - g_{\mu\nu} \Lambda \right\} \right. \\
& - A_\mu \mathcal{F}^{\mu\nu}{}_{,\nu} + \frac{1}{4\sqrt{-g}} \mathcal{F}^{\mu\nu} \mathcal{F}^{\alpha\beta} g_{\mu\alpha} g_{\nu\beta} \\
& \left. + \sum_a \int d\tau_a \left\{ -m_a \left( -g_{\mu\nu}(x) \frac{dz_a^\mu}{d\tau_a} \frac{dz_a^\nu}{d\tau_a} \right)^{1/2} + e_a \frac{dz_a^\mu}{d\tau_a} A_\mu(x) \right\} \delta^2(x - z_a(\tau_a)) \right]
\end{aligned}
\tag{42}
$$

where $A_\mu$ and $\mathcal{F}^{\mu\nu}$ are the vector potential and the field strength, and $e_a$ and $\tau_a$ are the charge and the proper time of *a*-th particle, respectively. The field equations that follow from the action (42) are

$$R - g^{\mu\nu} \nabla_\mu \nabla_\nu \Psi = 0 \tag{43}$$

$$\frac{1}{2} \nabla_\mu \Psi \nabla_\nu \Psi - \frac{1}{4} g_{\mu\nu} \nabla^\lambda \Psi \nabla_\lambda \Psi + g_{\mu\nu} \nabla^\lambda \nabla_\lambda \Psi - \nabla_\mu \nabla_\nu \Psi = \kappa T_{\mu\nu} - g_{\mu\nu} \Lambda \tag{44}$$

$$\mathcal{F}^{\mu\nu}{}_{,\nu} = \sum_a e_a \int d\tau_a \frac{dz_a^\mu}{d\tau_a} \delta^2(x - z_a(\tau_a)) \tag{45}$$

$$\frac{1}{\sqrt{-g}} \mathcal{F}_{\mu\nu} = \partial_\mu A_\nu - \partial_\nu A_\mu \ , \tag{46}$$

$$m_a \left[ \frac{d}{d\tau_a} \left\{ g_{\mu\nu}(z_a) \frac{dz_a^\nu}{d\tau_a} \right\} - \frac{1}{2} g_{\nu\lambda,\mu}(z_a) \frac{dz_a^\nu}{d\tau_a} \frac{dz_a^\lambda}{d\tau_a} \right] = e_a \frac{dz_a^\nu}{d\tau_a} \left\{ A_{\nu,\mu}(z_a) - A_{\mu,\nu}(z_a) \right\} \tag{47}$$

where the stress–energy is

$$T_{\mu\nu} = \sum_a m_a \int d\tau_a \frac{1}{\sqrt{-g}} g_{\mu\sigma} g_{\nu\rho} \frac{dz_a^\sigma}{d\tau_a} \frac{dz_a^\rho}{d\tau_a} \delta^2(x - z_a(\tau_a)) + \frac{1}{(-g)} \left\{ \mathcal{F}_{\mu\alpha} \mathcal{F}_{\nu\beta} g^{\alpha\beta} - \frac{1}{4} g_{\mu\nu} \mathcal{F}_{\alpha\beta} \mathcal{F}^{\alpha\beta} \right\} \tag{48}$$

whose conservation is guaranteed from (44). Note that, in (1 + 1) dimensions, no magnetic component of the field exists. Inserting the trace of (44) into (43) yields (14), as before. This equation, along with (45)–(47), form a closed system of equations for gravity, electromagnetism, and charged *N*-body matter.

Using the form (16) of the metric, the variational principle yields the set of equations

$$
\begin{aligned}
\dot{\pi} \;+\; & N_0 \Bigg\{ \frac{3\kappa}{2}\sqrt{\gamma}\pi^2 - \frac{\kappa}{\sqrt{\gamma}}\pi\Pi + \frac{1}{8\kappa\sqrt{\gamma}\gamma}(\Psi')^2 + \frac{1}{4\sqrt{\gamma}}\Big(E^2 + 2\frac{\Lambda}{\kappa}\Big) \\
& \qquad\qquad\qquad\qquad - \sum_a \frac{p_a^2}{2\gamma^2\sqrt{\frac{p_a^2}{\gamma}+m_a^2}}\,\delta(x - z_a(t)) \Bigg\} \\
\;+\; & N_1 \Bigg\{ -\frac{1}{\gamma^2}\Pi\Psi' + \frac{\pi'}{\gamma} + \sum_a \frac{p_a}{\gamma^2}\,\delta(x - z_a(t)) \Bigg\} + N_0'\frac{1}{2\kappa\sqrt{\gamma}\gamma}\Psi' + N_1'\frac{\pi}{\gamma} = 0\;, \quad (49)
\end{aligned}
$$

$$
\dot{\gamma} - N_0(2\kappa\sqrt{\gamma}\gamma\pi - 2\kappa\sqrt{\gamma}\Pi) + N_1\frac{\gamma'}{\gamma} - 2N_1' = 0\;, \tag{50}
$$

$$
R^0 = 0\;, \tag{51}
$$

$$
R^1 = 0\;, \tag{52}
$$

$$
\dot{\Pi} + \partial_1\Big(-\frac{1}{\gamma}N_1\Pi + \frac{1}{2\kappa\sqrt{\gamma}}N_0\Psi' + \frac{1}{\kappa\sqrt{\gamma}}N_0'\Big) = 0\;, \tag{53}
$$

$$
\dot{\Psi} + N_0(2\kappa\sqrt{\gamma}\pi) - N_1\Big(\frac{1}{\gamma}\Psi'\Big) = 0\;, \tag{54}
$$

$$
\begin{aligned}
\dot{p}_a + \frac{\partial N_0}{\partial z_a}\sqrt{\frac{p_a^2}{\gamma}+m_a^2} &- \frac{N_0}{2\sqrt{\frac{p_a^2}{\gamma}+m_a^2}}\frac{p_a^2}{\gamma^2}\frac{\partial\gamma}{\partial z_a} - \frac{\partial N_1}{\partial z_a}\frac{p_a}{\gamma} \\
&+ N_1\frac{p_a}{\gamma^2}\frac{\partial\gamma}{\partial z_a} + \int dx\, N_0\sqrt{\gamma}E\frac{\partial E}{\partial z_a} = 0\;, \tag{55}
\end{aligned}
$$

$$
\dot{z_a} - N_0\frac{\frac{p_a}{\gamma}}{\sqrt{\frac{p_a^2}{\gamma}+m_a^2}} + \frac{N_1}{\gamma} = 0\;. \tag{56}
$$

$$
\frac{\partial E}{\partial t} + \sum_a e_a\frac{dz_a}{dt}\delta(x - z_a(t)) = 0\;, \tag{57}
$$

$$
\sqrt{-g}E = -\frac{\partial\varphi}{\partial x} - \frac{\partial A}{\partial t} \tag{58}
$$

where

$$
A_\mu = (-\varphi, A) \qquad E = \mathcal{F}^{01} \tag{59}
$$

and the quantities

$$
\begin{aligned}
R^0 &= -\kappa\sqrt{\gamma}\gamma\pi^2 + 2\kappa\sqrt{\gamma}\pi\Pi + \frac{1}{4\kappa\sqrt{\gamma}}(\Psi')^2 - \frac{1}{\kappa}\Big(\frac{\Psi'}{\sqrt{\gamma}}\Big)' - \frac{1}{2}\sqrt{\gamma}\Big(E^2 + 2\frac{\Lambda}{\kappa}\Big) \\
&\quad - \sum_a \sqrt{\frac{p_a^2}{\gamma}+m_a^2}\,\delta(x - z_a(t)) \tag{60}
\end{aligned}
$$

$$
R^1 = \frac{\gamma'}{\gamma}\pi - \frac{1}{\gamma}\Pi\Psi' + 2\pi' + \sum_a \frac{p_a}{\gamma}\delta(x - z_a(t)) \tag{61}
$$

$$
\frac{\partial E}{\partial x} = \sum_a e_a\delta(x - z_a(t)) \tag{62}
$$

vanish due to the constraints. The set of Equations (49)–(58) are equivalent to the set of Equations (43)–(47).

Note that, in (55) and (56), all metric components ($N_0$, $N_1$, $\gamma$) are evaluated at the point $x = z_a$ with

$$\frac{\partial f}{\partial z_a} \equiv \frac{\partial f(x)}{\partial x}\bigg|_{x=z_a}.$$

Furthermore, in $(1+1)$ dimensions, the electric field has no independent degrees of freedom. For charged particles moving within a finite region ($|x| < L$), the electric field in an outer region ($|x| > L$) is $E = \pm\frac{1}{2}\sum_a e_a$. $E = 0$ in the outer region for a system of zero total charge.

The canonical reduction of this system proceeds in a manner similar to the uncharged case. The coordinate conditions (21) can again be chosen and the reduced action is still given by (22). The reduced Hamiltonian is again given by (23), but now the constraints become

$$\triangle\Psi - \frac{1}{4}(\Psi')^2 + \kappa^2\pi^2 + \frac{1}{2}(\kappa E^2 + 2\Lambda) + \kappa\sum_a \sqrt{p_a^2 + m_a^2}\,\delta(x - z_a) = 0\,, \qquad (63)$$

$$2\pi' + \sum_a p_a\delta(x - z_a) = 0\,. \qquad (64)$$

The consistency of this canonical reduction is proven by showing that the canonical equations of motion derived from the reduced Hamiltonian (23) are identical with the Equations (55) and (56) [99,105].

## 5. The Two-Body Problem

As stated in the introduction, an exact solution to the two-body problem is known in Newtonian theory, but not in general relativity due to the dissipation of energy in the form of gravitational radiation. Hence, an analysis of a two-body system in general relativity (e.g., binary pulsars) necessarily involves resorting to approximation methods such as a post-Newtonian expansion [100,106]. Quite remarkably, there are a number of exact solutions [104,105,107–109] to the two-body problem in the ROGS theory described above. While this is in part understandable due to the lack of gravitational radiation (there are no gravitons in $(1+1)$ dimensions), the nonlinearity of the system does not obviously admit exact solutions, and in fact, a general dilaton theory of gravity will not have them.

### 5.1. Solution for Two Charged Particles

Here, we present the exact solution for the charged case [105], since from it, the electrically neutral case is recovered by setting $e_1 = e_2 = 0$. The standard approach for obtaining a solution is to find an explicit expression for the Hamiltonian (23) by solving (63) and (64). From this, the equations of motion can be derived, which in turn, can be solved to obtain the trajectories of the particles.

Noting that the electric field appears in the combination ($E^2 + 2\Lambda/\kappa$) in all equations, it is convenient to set

$$V(x) \equiv E^2 - C \quad \text{and} \quad \Lambda_e \equiv -2\Lambda - \kappa C \qquad (65)$$

with $C \equiv \frac{1}{4}(\sum_a e_a)^2$. The quantity $\Lambda_e$ is an effective cosmological constant that includes the contribution from the electric field. This situation arises because, analogous to the way in which a four-form behaves in $(3+1)$ dimensions [110], in $(1+1)$ dimensions, the electromagnetic field strength is a two-form, and so contributes to the stress–energy tensor in the same manner as a cosmological constant in compact spatial regions.

Defining $\phi$ by

$$\Psi = -4\ln|\phi|\,, \qquad (66)$$

the constraints (63) and (64) for a two-particle system become

$$\triangle\phi - \frac{1}{4}\left\{\kappa^2(\chi')^2 + \frac{\kappa}{2}V - \frac{1}{2}\Lambda_e\right\}\phi = \frac{\kappa}{4}\left\{\sqrt{p_1^2 + m_1^2}\,\phi(z_1)\delta(x - z_1) \right.$$
$$\left. + \sqrt{p_2^2 + m_2^2}\,\phi(z_2)\delta(x - z_2)\right\} \quad (67)$$

$$\triangle\chi = -\frac{1}{2}\{p_1\delta(x - z_1) + p_2\delta(x - z_2)\} \quad (68)$$

along with (62) for the electric field, whose solution is

$$E = \frac{1}{2}\partial_x \sum_a e_a |x - z_a(t)| . \quad (69)$$

It is also straightforward to solve (68); the result is

$$\chi = -\frac{1}{4}\{p_1 \mid x - z_1 \mid + p_2 \mid x - z_2 \mid\} - \epsilon X x + \epsilon C_\chi \quad (70)$$

where $X$ and $C_\chi$ are constants of integration and, as in (37), $\epsilon$ ($\epsilon^2 = 1$) ensures that the time reversal properties of $\chi$ are explicitly manifest. By definition, $\epsilon$ changes sign under time reversal and thus so does $\chi$.

Suppose that $z_2 < z_1$. We can divide space into three regions: $z_1 < x$ ((+) region), $z_2 < x < z_1$ ((0) region) and $x < z_2$ ((−) region). From (69) and (70), one sees that, in each region, $\chi'$ and $V$ are constant:

$$V = \begin{cases} 0 & \text{(+) region,} \\ -e_1 e_2 & \text{(0) region,} \\ 0 & \text{(−) region,} \end{cases} \quad (71)$$

$$\chi' = \begin{cases} -\epsilon X - \frac{1}{4}(p_1 + p_2) & \text{(+) region,} \\ -\epsilon X + \frac{1}{4}(p_1 - p_2) & \text{(0) region,} \\ -\epsilon X + \frac{1}{4}(p_1 + p_2) & \text{(−) region} \end{cases} \quad (72)$$

Obtaining the solution to (67) is somewhat more complicated [104], entailing a proper choice of boundary conditions that ensure the finiteness of the Hamiltonian and the vanishing of surface terms which arise in transforming the action. The result is [105]

$$\phi_+ = \left(\frac{K_1}{\mathcal{M}_1}\right)^{\frac{1}{2}} e^{-\frac{1}{4}(K_{01}z_1 - K_{02}z_2) + \frac{1}{2}K_+(x - z_1)}$$

$$\phi_0 = \frac{e^{-\frac{1}{4}(K_{01}z_1 - K_{02}z_2)}}{4K_0}\left\{(K_1\mathcal{M}_1)^{1/2}e^{-\frac{1}{2}K_0(x - z_1)} + (K_2\mathcal{M}_2)^{1/2}e^{\frac{1}{2}K_0(x - z_2)}\right\} \quad (73)$$

$$\phi_- = \left(\frac{K_2}{\mathcal{M}_2}\right)^{\frac{1}{2}} e^{-\frac{1}{4}(K_{01}z_1 - K_{02}z_2) - \frac{1}{2}K_-(x - z_2)}$$

where

$$\begin{cases} K_+ = \sqrt{\kappa^2\left(X + \frac{\epsilon}{4}(\tilde{p}_1 + \tilde{p}_2)\right)^2 - \frac{1}{2}\Lambda_e} & \text{(+) region} \\ K_0 = \sqrt{\kappa^2\left(X - \frac{\epsilon}{4}(\tilde{p}_1 - \tilde{p}_2)\right)^2 - \frac{\kappa}{2}e_1 e_2 - \frac{1}{2}\Lambda_e} & \text{(0) region} \\ K_- = \sqrt{\kappa^2\left(X - \frac{\epsilon}{4}(\tilde{p}_1 + \tilde{p}_2)\right)^2 - \frac{1}{2}\Lambda_e} & \text{(−) region} \end{cases} \quad (74)$$

with $\tilde{p}_i = p_i \, \text{sgn}(z_1 - z_2)$, and

$$
\begin{aligned}
K_1 &\equiv 2K_0 + 2K_- - \kappa\sqrt{p_2^2 + m_2^2} \\
K_2 &\equiv 2K_0 + 2K_+ - \kappa\sqrt{p_1^2 + m_1^2} \\
K_{01} &\equiv K_0 - K_+ + \frac{\kappa\epsilon}{2}\tilde{p}_1 \,, \\
K_{02} &\equiv K_0 - K_- - \frac{\kappa\epsilon}{2}\tilde{p}_2 \,, \\
\mathcal{M}_1 &\equiv \kappa\sqrt{p_1^2 + m_1^2} + 2K_0 - 2K_+ \\
\mathcal{M}_2 &\equiv \kappa\sqrt{p_2^2 + m_2^2} + 2K_0 - 2K_-
\end{aligned}
\tag{75}
$$

The Hamiltonian (23) becomes

$$
H = -\frac{1}{\kappa}\int dx \triangle\Psi = \frac{4}{\kappa}\left[\frac{\phi'}{\phi}\right]_{-\infty}^{\infty} = \frac{2(K_+ + K_-)}{\kappa}
\tag{76}
$$

and is explicitly determined in terms of the canonical variables once the solution $X$ to

$$
K_1 K_2 = \mathcal{M}_1 \mathcal{M}_2 \, e^{K_0 |z_1 - z_2|}
\tag{77}
$$

is obtained. This relation is a (not obvious) consequence of solving (67), and can more explicitly be written as

$$
\begin{aligned}
\left(4K_0^2 + [\kappa\sqrt{p_1^2 + m_1^2} - 2K_+][\kappa\sqrt{p_2^2 + m_2^2} - 2K_-]\right)\tanh\left(\frac{1}{2}K_0|z_1 - z_2|\right) \\
= -2K_0\left([\kappa\sqrt{p_1^2 + m_1^2} - 2K_+] + [\kappa\sqrt{p_2^2 + m_2^2} - 2K_-]\right)
\end{aligned}
\tag{78}
$$

which is the determining equation of the Hamiltonian.

Note that $K_{\pm}$ must both be real in order for the Hamiltonian (76) to have a definite meaning. This imposes a restriction on $X$ corresponding to a value of the effective cosmological constant $\Lambda_e$. But $K_0$ is not necessarily real; indeed, for a sufficiently strong electromagnetic repulsion (sufficiently large positive $e_1 e_2$), $K_0$ will be imaginary. In this case, in the (0) region, the solution for $\phi$ becomes

$$
\phi_0(x) = A_s \sin\frac{1}{2}\tilde{K}_0 x + A_c \cos\frac{1}{2}\tilde{K}_0 x \,,
\tag{79}
$$

instead of (73), where

$$
\tilde{K}_0 = -iK_0 = \sqrt{\frac{\kappa}{2}e_1 e_2 + \frac{1}{2}\Lambda_e - \kappa^2\left(X - \frac{\epsilon}{4}(\tilde{p}_1 - \tilde{p}_2)\right)^2}
\tag{80}
$$

yielding a new determining equation of the Hamiltonian

$$
\begin{aligned}
\left(4\tilde{K}_0^2 - [\kappa\sqrt{p_1^2 + m_1^2} - 2K_+][\kappa\sqrt{p_2^2 + m_2^2} - 2K_-]\right)\tan\left(\frac{1}{2}\tilde{K}_0|z_1 - z_2|\right) \\
= 2\tilde{K}_0\left([\kappa\sqrt{p_1^2 + m_1^2} - 2K_+] + [\kappa\sqrt{p_2^2 + m_2^2} - 2K_-]\right)
\end{aligned}
\tag{81}
$$

instead of (78). Indeed (81) and (79) can, respectively, be obtained from (78) and (73) by formally replacing $K_0$ with $i\tilde{K}_0$.

Consequently, (77) is applicable for all values of $K_0$. This transcendental equation determines $H$ in terms of the momenta and positions of the particles. Although (77) cannot be explicitly written in terms of known functions, it can be used to obtain exactly the canonical equations of motion for $z_i$ and $p_i$, which are [105]

$$\dot{p}_1 = -\frac{\partial H}{\partial z_1} = -\frac{2}{\kappa}\left(\frac{Y_+}{K_+} + \frac{Y_-}{K_-}\right)\frac{K_0 K_1 K_2}{J} \tag{82}$$

$$\dot{z}_1 = \frac{\partial H}{\partial p_1} = \epsilon\frac{Y_+}{K_+} + \frac{8}{J}\left(\frac{Y_+}{K_+} + \frac{Y_-}{K_-}\right)\frac{K_0 K_1}{\mathcal{M}_1}\left\{\frac{p_1}{\sqrt{p_1^2 + m_1^2}} - \epsilon\frac{Y_+}{K_+}\right\} \tag{83}$$

$$\dot{p}_2 = \frac{2}{\kappa}\left(\frac{Y_+}{K_+} + \frac{Y_-}{K_-}\right)\frac{K_0 K_1 K_2}{J} \tag{84}$$

$$\dot{z}_2 = -\epsilon\frac{Y_-}{K_-} + \frac{8}{J}\left(\frac{Y_+}{K_+} + \frac{Y_-}{K_-}\right)\frac{K_0 K_2}{\mathcal{M}_2}\left\{\frac{p_2}{\sqrt{p_2^2 + m_2^2}} + \epsilon\frac{Y_-}{K_-}\right\} \tag{85}$$

where

$$Y_\pm \equiv \kappa\left[X \pm \frac{\epsilon}{4}(p_1 + p_2)\right]$$
$$Y_0 \equiv \kappa\left[X - \frac{\epsilon}{4}(p_1 - p_2)\right] \tag{86}$$

and

$$J = 2\left\{\left(\frac{Y_0}{K_0} + \frac{Y_+}{K_+}\right)K_1 + \left(\frac{Y_0}{K_0} + \frac{Y_-}{K_-}\right)K_2\right\}$$
$$- 2\left\{\left(\frac{Y_0}{K_0} - \frac{Y_+}{K_+}\right)\frac{1}{\mathcal{M}_1} + \left(\frac{Y_0}{K_0} - \frac{Y_-}{K_-}\right)\frac{1}{\mathcal{M}_2}\right\}K_1 K_2 - \frac{Y_0}{K_0}K_1 K_2(z_1 - z_2). \tag{87}$$

These canonical equations ensure the conservation of the Hamiltonian and the total momentum $p_1 + p_2$. They can also be shown [105] to be equivalent to the geodesic equations (55) and (56), which become

$$\dot{p}_a = -\frac{\partial N_0}{\partial z_a}\sqrt{p_a^2 + m_a^2} + \frac{\partial N_1}{\partial z_a}p_a + \frac{1}{2}\sum_b e_a e_b N_0\frac{\partial}{\partial z_a}|z_a - z_b|, \tag{88}$$

$$\dot{z}_a = N_0\frac{p_a}{\sqrt{p_a^2 + m_a^2}} - N_1. \tag{89}$$

under the coordinate conditions (21), where

$$\frac{\partial N_{0,1}}{\partial z_i} \equiv \frac{1}{2}\left\{\frac{\partial N_{0,1}}{\partial x}\bigg|_{x=z_i+0} + \frac{\partial N_{0,1}}{\partial x}\bigg|_{x=z_i-0}\right\}. \tag{90}$$

defines the partial derivatives at $z_1, z_2$. Equations (49), (50), (53) and (54) yield the metric under the coordinate conditions (21) [105].

## 5.2. Test Particle Limit

To make sense of the solutions contained in (77), it is useful to first consider the test particle limit in which $m_1 = \mu << m = m_2$ and both particles have zero charge. Setting the latter to be a static source at the origin, with $z_2 = 0 = p_2$, the defining Equation (77) becomes

$$(\sqrt{p^2 + \mu^2} - H)(m + \epsilon\tilde{p} - H) = (\sqrt{p^2 + \mu^2} - \epsilon\tilde{p})m\ e^{\frac{\kappa}{4}(H - \epsilon\tilde{p})|z|} \tag{91}$$

where $z_1 = z$ and $p_1 = p$. The corresponding non-relativistic Hamiltonian is

$$H = m + \mu + \frac{\kappa m\mu}{4}\ |z| + \frac{p^2}{2\mu} \tag{92}$$

obtained by expanding (91) for $p << m$, $\kappa m|z| << 1$. Typically, the sum $mc^2 + \mu c^2$ of the rest-energies (where we have set $c = 1$ in the above) is subtracted from the non-relativistic Hamiltonian (92). However, for a proper comparison of the energy to the relativisitic case, these terms need to be retained.

Figure 1 shows a comparison between the relativistic case (91) and non-relativistic case (92) in the phase space of the test particle. Even for the relatively small values of the total energy, the non-relativistic motion (red) is notably distinct. For the given value of $H = 1.125m$ and an initial condition of $z = 0$, the initial momentum $p = 0.075m$ in the relativistic case as compared to $p = 0.071m$ in the non-relativisitic case. In both cases, the static mass $m$ attracts the test particle, but in the relativistic case, its momentum initially decreases less rapidly. However, this situation changes once the particle reaches its maximal separation from $m$. This occurs at $\kappa m|z| = 1$ in the non-relativistic case, but at a shorter separation $\kappa m|z| = 0.89$ in the relativistic case—relativistic gravitational attraction is stronger for the same total energy. One can observe the rather counter-intuitive feature that the particle is then attracted back to the source even though it has positive momentum in the relativistic case! The loss of momentum is more rapid than in the non-relativistic case, and continues to be so until the test particle returns to the origin, after which (in both cases) the motion repeats.

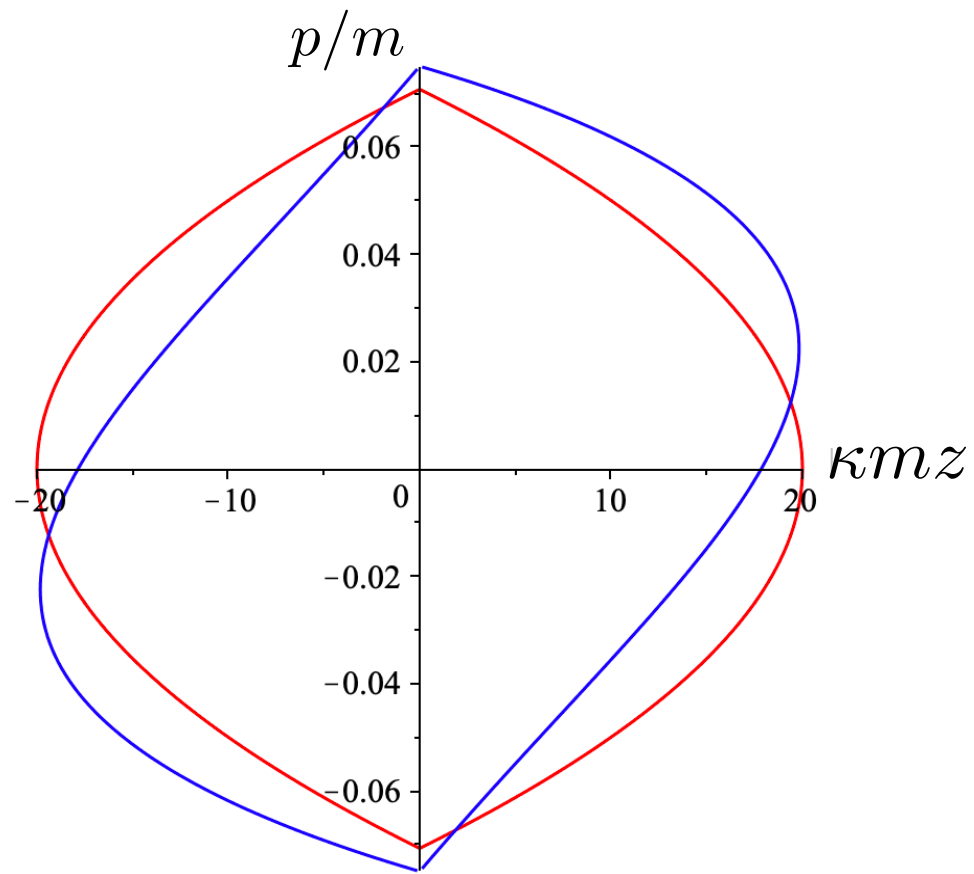

**Figure 1.** A plot of the relativistic trajectory (blue) of a test particle in comparison to the non-relativistic case (red). The Hamiltonian for each case is set to $H = 1.125m$ and $\mu = 0.1m$.

### 5.3. Exact Equal Mass Two-Body Motion for $\Lambda = 0$

For neutral bodies of equal mass, the determining Equation (77) can be solved explicitly for $\Lambda = 0$, yielding [107,108]

$$H = \sqrt{p^2 + m^2} + \epsilon p \; \text{sgn}(r) - \frac{8W\left[-\frac{\kappa}{8}(|r|\sqrt{p^2 + m^2} - \epsilon pr)\exp\left(\frac{\kappa}{8}(|r|\sqrt{p^2 + m^2} - \epsilon pr)\right)\right]}{\kappa|r|} \quad (93)$$

in the centre of inertia system $p_1 = -p_2 = p$, where $r = z_1 - z_2$. Here, $W(x)$ is the Lambert $W$ function [111]

$$y \cdot e^y = x \quad \Rightarrow \quad y = W(x) \tag{94}$$

which has two real branches $W_0$ and $W_{-1}$ for real $x$.

Since the Hamiltonian (93) is exact for arbitrary values of $m$ and $p$ to infinite order in $\kappa$, the whole structure of the theory can be studied, from the weak field to the strong field limits. As in the test particle case, a phase space trajectory in $(r, p)$ space can be obtained by setting $H = H_0$. Alternatively, the equations of motion (82)–(85) yield

$$\dot{p} = -\frac{\frac{\kappa}{4}(H - 2\epsilon\tilde{p})(H - \epsilon\tilde{p} - \sqrt{p^2 + m^2})}{2 - \frac{\kappa r}{4}(H - \epsilon\tilde{p} - \sqrt{p^2 + m^2})} \, \mathrm{sgn}(r) \,, \tag{95}$$

$$\dot{r} = 2\epsilon \left\{ 1 - \frac{H - 2\epsilon\tilde{p}}{2 - \frac{\kappa r}{4}(H - \epsilon\tilde{p} - \sqrt{p^2 + m^2})} \cdot \frac{1}{\sqrt{p^2 + m^2}} \right\} \mathrm{sgn}(r) \,, \tag{96}$$

and can be solved numerically. However, superficial coordinate singularities appear where the denominator $\left\{ 2 - \kappa r(H - \epsilon\tilde{p} - \sqrt{p^2 + m^2})/4 \right\}$ vanishes, corresponding to $W(x) = -1$. This is the transit point between two branches $W_0$ and $W_{-1}$ and are a consequence of $t$ being a coordinate time.

This problem can be dealt with by describing the trajectories in terms of the proper time $\tau_a$ of each particle. Using (89), this is

$$d\tau_a^2 = dt^2 \left\{ N_0(z_a)^2 - (N_1(z_a) + \dot{z}_a)^2 \right\} = dt^2 N_0(z_a)^2 \frac{m_a^2}{p_a^2 + m_a^2} \qquad (a = 1, 2) \tag{97}$$

becoming

$$d\tau = d\tau_1 = d\tau_2 = \frac{(H - 2\epsilon\tilde{p})m}{\left\{ 2 - \frac{\kappa r}{4}(H - 2\epsilon\tilde{p}) \right\} \left( \sqrt{p^2 + m^2} - \epsilon\tilde{p} \right) \sqrt{p^2 + m^2}} \, dt \tag{98}$$

for both particles in the equal mass case. The canonical Equations (95) and (96) then become

$$\frac{dp}{d\tau} = -\frac{\kappa}{4m} \sqrt{p^2 + m^2} \left\{ H \left( \sqrt{p^2 + m^2} - \epsilon\tilde{p} \right) - m^2 \right\} \mathrm{sgn}(r) \tag{99}$$

$$\frac{dr}{d\tau} = \frac{2\epsilon}{m} \left( \sqrt{p^2 + m^2} - \epsilon\tilde{p} \right) \left\{ \frac{\left[ 2 - \frac{\kappa r}{4}(H - \epsilon\tilde{p} - \sqrt{p^2 + m^2}) \right] \sqrt{p^2 + m^2}}{H - 2\epsilon\tilde{p}} - 1 \right\} \mathrm{sgn}(r) \tag{100}$$

which remarkably have an exact solution.

This is procured as follows. Noting that $H$ is constant, (99) is an ordinary differential equation that can be solved for $p(\tau)$. The trajectory $r(\tau)$ can then be obtained by solving (77) or by directly solving (100) after substituting the solution for $p$, yielding an exact expression for the proper separation of the two bodies as a function of their mutual proper time. The result is [108]

$$p(\tau) = \frac{\epsilon m}{2} \left( f_0(\tau) - \frac{1}{f_0(\tau)} \right) \mathrm{sgn}(r) \tag{101}$$

$$r(\tau) = \frac{16 \, \mathrm{sgn}(r)}{\kappa \left\{ H - m \left( f_0 - \frac{1}{f_0} \right) \right\}} \tanh^{-1} \left[ \frac{H - m \left( f_0 + \frac{1}{f_0} \right)}{H - m \left( f_0 - \frac{1}{f_0} \right)} \right] \,, \tag{102}$$

with

$$f_0(\tau) = \frac{H}{m} \left[ 1 - \frac{\sqrt{p_0^2 + m^2} - \epsilon p_0 \, \mathrm{sgn}(r) - \frac{m^2}{H}}{\sqrt{p_0^2 + m^2} - \epsilon p_0 \, \mathrm{sgn}(r)} \, e^{\frac{\epsilon \kappa m}{4}(\tau - \tau_0)} \right] \,, \tag{103}$$

where $p_0$ is the initial momentum at $\tau = \tau_0$.

In Figure 2, the trajectory $r$ is plotted for various values of $H_0$ as a function of the mutual proper time of the two equal mass particles, with the corresponding phase space trajectories plotted underneath. For a small $H_0$, we see the distorted oval-shaped trajectory similar to that in Figure 1. The corresponding motion (shown as the solid curve in the upper diagram) is oscillatory, corresponding to the two particles flying apart to some maximal separation and then merging together to repeat the motion with their positions interchanged. As $H_0$ increases, we observe considerable distortion in the phase-space trajectory. The motion becomes increasingly asymmetric over a given half-period, with the maximal separation occurring at relatively smaller values of $\tau$. For most of the motion, $p > 0$; once $p < 0$, the particles rapidly merge together to then repeat the motion with their positions switched.

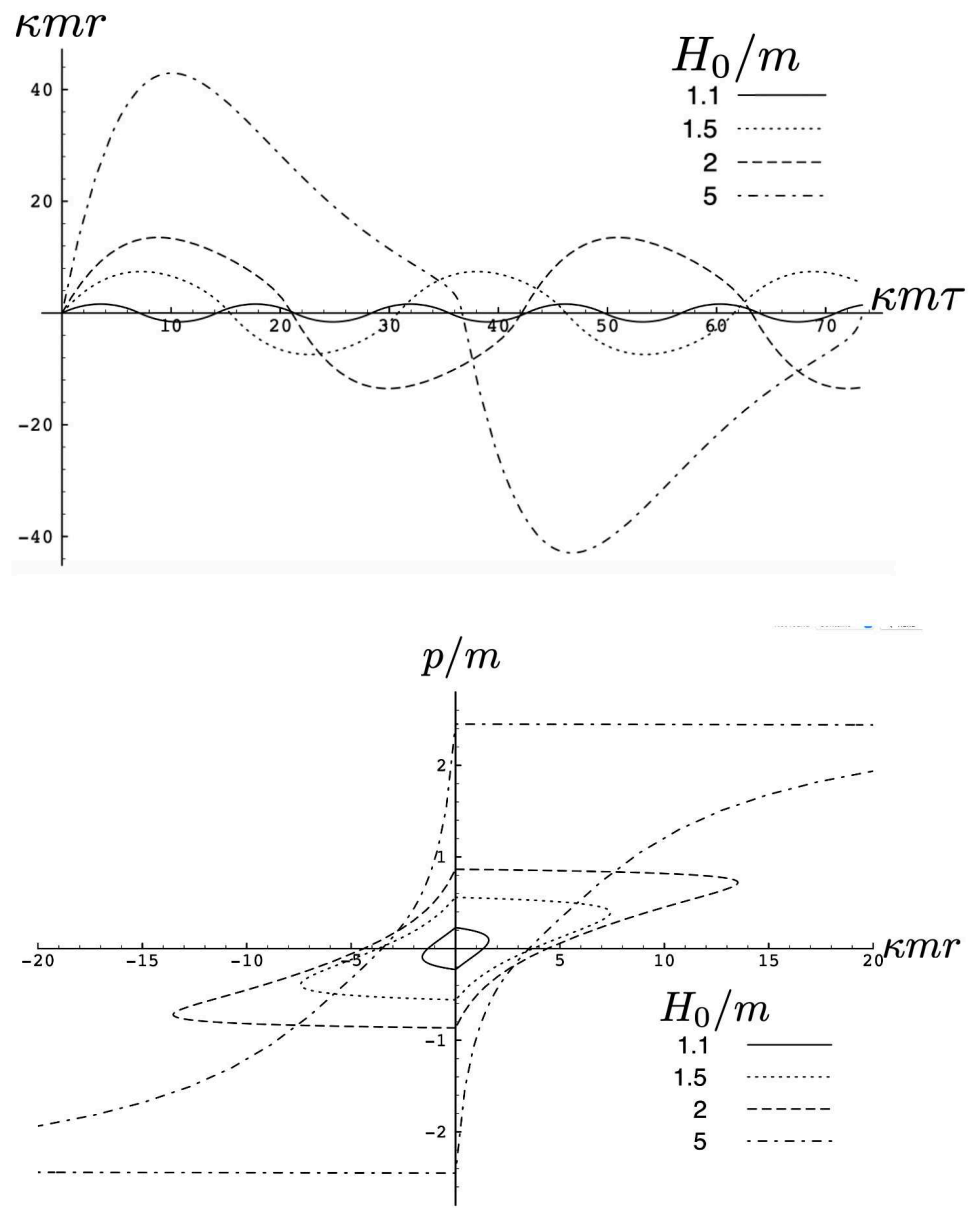

**Figure 2. Top**: A plot of the relativistic trajectories of neutral particles of equal mass as a function of their mutual proper time for various values of the conserved energy $H_0$. All motions begin at $r = 0$ with an initial momentum given by solving (93). **Bottom**: The corresponding phase-space plots for the top diagram. Note that the curves for the largest value of $H_0$ extend outside the range of the figure.

As the energy $H_0$ of the two-body system increases, the departure from non-relativistic motion is quite striking. This is illustrated in Figure 3 for $H_0 = 25m$. We see that the period of the relativistic motion is only slightly longer than a half-period of its non-relativistic counterpart. Furthermore, the relativistic system is much more tightly bound, with the maximal separation approximately half that of the non-relativistic case.

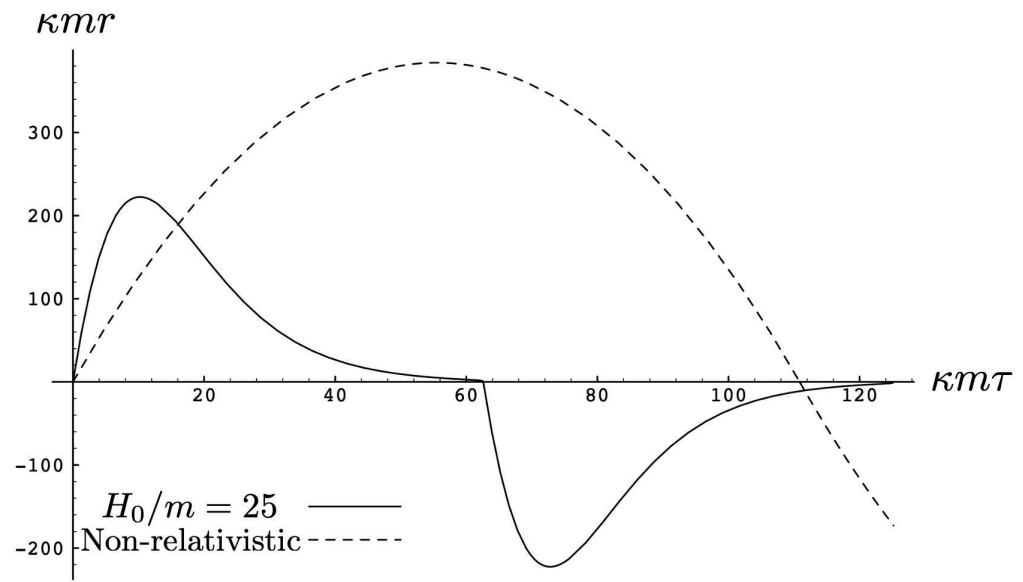

**Figure 3.** A comparison of the relativistic and non-relativistic trajectories of the neutral particles of equal mass as a function of their mutual proper time at a large value of the conserved energy $H_0$. Both motions begin at $r = 0$ with an initial momentum given by solving (93) (solid line) and its corresponding counterpart (92) with $z = r$ and $\mu = m$.

### 5.4. Exact Two-Body Motion with Equal Masses and Arbitrary Charges

Turning now to the charged system, in the centre of inertia frame $p_1 = -p_2 = p$, the determining Equations (78) and (81) become

$$(\mathcal{J}_\Lambda^2 + B^2) \tanh\left(\frac{\kappa}{8}\mathcal{J}_\Lambda |r|\right) = 2\mathcal{J}_\Lambda B , \tag{104}$$

and

$$(\tilde{\mathcal{J}}_\Lambda^2 - B^2) \tan\left(\frac{\kappa}{8}\tilde{\mathcal{J}}_\Lambda |r|\right) = -2\tilde{\mathcal{J}}_\Lambda B , \tag{105}$$

respectively, where

$$\mathcal{J}_\Lambda = \sqrt{\left(\sqrt{H^2 + \frac{8\Lambda_e}{\kappa^2} - 2\epsilon\tilde{p}}\,\right)^2 - \frac{8e_1 e_2}{\kappa} - \frac{8\Lambda_e}{\kappa^2}}$$

$$\tilde{\mathcal{J}}_\Lambda = \sqrt{\frac{8e_1 e_2}{\kappa} + \frac{8\Lambda_e}{\kappa^2} - \left(\sqrt{H^2 + \frac{8\Lambda_e}{\kappa^2} - 2\epsilon\tilde{p}}\,\right)^2}$$

$$B = H - 2\sqrt{p^2 + m^2} . \tag{106}$$

Each case further subdivides into two parts:

$$\tanh\left(\frac{\kappa}{16}\mathcal{J}_\Lambda |r|\right) = \frac{B}{\mathcal{J}_\Lambda} \qquad \text{(tanh-type A)} \tag{107}$$

$$\tanh\left(\frac{\kappa}{16}\mathcal{J}_\Lambda |r|\right) = \frac{\mathcal{J}_\Lambda}{B} \qquad \text{(tanh-type B)} \tag{108}$$

for (104) and

$$\tan\left(\frac{\kappa}{16}\tilde{\mathcal{J}}_\Lambda|r|\right) = -\frac{B}{\tilde{\mathcal{J}}_\Lambda} \qquad \text{(tan-type A)} \tag{109}$$

$$\tan\left(\frac{\kappa}{16}\tilde{\mathcal{J}}_\Lambda|r|\right) = \frac{\tilde{\mathcal{J}}_\Lambda}{B} \qquad \text{(tan-type B)} \tag{110}$$

for (105). If both $\Lambda_e = 0$ and $e_a = 0$, then (108) has no solutions since $\mathcal{J}_\Lambda/B$ exceeds unity.

The canonical equations of motion

$$\dot{p} = -\frac{\kappa \mathcal{J}_\Lambda^2 (\mathcal{J}_\Lambda^2 - B^2)}{16\mathcal{C}}\,\mathrm{sgn}(r)\,, \tag{111}$$

$$\dot{r} = 2\epsilon\sqrt{1 + \frac{8\Lambda_e}{\kappa^2 H^2}}\left(1 - \frac{\mathcal{J}_\Lambda^2}{\mathcal{C}}\right)\mathrm{sgn}(r) + \frac{2\mathcal{J}_\Lambda^2}{\mathcal{C}}\frac{p}{\sqrt{p^2 + m^2}}\,. \tag{112}$$

where

$$\mathcal{C} = \frac{1}{\sqrt{1 + \frac{8\Lambda_e}{\kappa^2 H^2}}}\left\{\sqrt{1 + \frac{8\Lambda_e}{\kappa^2 H^2}}\,\mathcal{J}_\Lambda^2 - \left(\sqrt{H^2 + \frac{8\Lambda_e}{\kappa^2}} - 2\epsilon\bar{p}\right)\left(B + \frac{\kappa}{16}(\mathcal{J}_\Lambda^2 - B^2)\,r\right)\right\} \tag{113}$$

are the same for all four types. Coordinate singularities are present here as before, and these can again be dealt with by choosing the time coordinate to be the common proper time (97), which in this case is

$$d\tau = d\tau_1 = d\tau_2 = \frac{m}{\sqrt{p^2 + m^2}}\frac{\mathcal{J}_\Lambda^2}{\mathcal{C}}dt \tag{114}$$

yielding

$$\frac{dp}{d\tau} = -\frac{\kappa\sqrt{p^2 + m^2}(\mathcal{J}_\Lambda^2 - B^2)}{16m}\,\mathrm{sgn}(r) \tag{115}$$

$$\frac{dr}{d\tau} = \frac{2\epsilon}{m}\left\{\frac{\sqrt{p^2 + m^2}\,\mathcal{C}}{\mathcal{J}_\Lambda^2} - (\sqrt{p^2 + m^2} - \epsilon\bar{p})\right\}\mathrm{sgn}(r) \tag{116}$$

for the canonical equations of motion (111) and (112).

Solving (115) gives a solution of the form (101), but where

$$f(\tau) = \begin{cases} \dfrac{\frac{H}{m}(1 + \sqrt{\gamma_H})\left\{1 - \eta\,e^{\frac{\epsilon\kappa m}{4}\sqrt{\gamma_m}(\tau - \tau_0)}\right\}}{\gamma_e + \sqrt{\gamma_m} + (\sqrt{\gamma_m} - \gamma_e)\,\eta\,e^{\frac{\epsilon\kappa m}{4}\sqrt{\gamma_m}(\tau - \tau_0)}} & \gamma_m > 0 \\[4mm] \dfrac{\frac{m}{H}\frac{1 + \sqrt{\gamma_H}}{\gamma_e + \frac{v}{m - \sigma}\frac{\epsilon\kappa H}{8}(\tau - \tau_0)}}{} & \gamma_m = 0 \\[4mm] \dfrac{\frac{H}{m}(1 + \sqrt{\gamma_H})}{\gamma_e + \sqrt{-\gamma_m}\frac{\frac{m^2}{H}\sqrt{-\gamma_m} - \sigma\tan\left[\frac{\epsilon\kappa m}{8}\sqrt{-\gamma_m}(\tau - \tau_0)\right]}{\frac{m^2}{H}\sqrt{-\gamma_m} - \sigma\tan\left[\frac{\epsilon\kappa m}{8}\sqrt{-\gamma_m}(\tau - \tau_0)\right]}} & \gamma_m < 0 \end{cases} \tag{117}$$

with

$$\gamma_H = 1 + \frac{8\Lambda_e}{\kappa^2 H^2} \qquad \gamma_e = 1 + \frac{2e_1 e_2}{\kappa m^2} \qquad \gamma_m = \gamma_e^2 + \frac{8\Lambda_e}{\kappa^2 m^2} \tag{118}$$

$$\sigma = (1 + \sqrt{\gamma_H})(\sqrt{p_0^2 + m^2} - \epsilon p_0 \mathrm{sgn}(r)) - \frac{m^2}{H}\gamma_e \qquad \eta = \frac{\sigma - \frac{m^2}{H}\sqrt{\gamma_m}}{\sigma + \frac{m^2}{H}\sqrt{\gamma_m}} \tag{119}$$

and $p_0$ is the initial momentum at $\tau = \tau_0$. These solutions are valid provided

$$1 + \frac{8\Lambda_e}{\kappa^2 H^2} \geq 0 \tag{120}$$

which is satisfied for all $\Lambda_e > 0$, and for $\Lambda_e < 0$ imposes the constraint

$$H \geq \sqrt{-\frac{8\Lambda_e}{\kappa^2}} \tag{121}$$

on the Hamiltonian.

The solution for $r(\tau)$ is then

$$r(\tau) = \begin{cases} \dfrac{16\,\mathrm{sgn}(r)\tanh^{-1}\left[\dfrac{\kappa\left(H-m\left|f(\tau)+\frac{1}{f(\tau)}\right|\right)}{\sqrt{\left(\sqrt{\kappa^2 H^2 + 8\Lambda_e} - m\kappa(f(\tau)-\frac{1}{f(\tau)})\right)^2 - 8\kappa e_1 e_2 - 8\Lambda_e}}\right]}{\sqrt{\left(\sqrt{\kappa^2 H^2 + 8\Lambda_e} - m\kappa(f(\tau)-\frac{1}{f(\tau)})\right)^2 - 8\kappa e_1 e_2 - 8\Lambda_e}} & \text{(tanh-type A)} \\[4em] \dfrac{16\,\mathrm{sgn}(r)\tanh^{-1}\left[\dfrac{\sqrt{\left(\sqrt{\kappa^2 H^2 + 8\Lambda_e} - m\kappa(f(\tau)-\frac{1}{f(\tau)})\right)^2 - 8\kappa e_1 e_2 - 8\Lambda_e}}{\kappa\left(H-m\left|f(\tau)+\frac{1}{f(\tau)}\right|\right)}\right]}{\sqrt{\left(\sqrt{\kappa^2 H^2 + 8\Lambda_e} - m\kappa(f(\tau)-\frac{1}{f(\tau)})\right)^2 - 8\kappa e_1 e_2 - 8\Lambda_e}} & \text{(tanh-type B)} \end{cases} \tag{122}$$

for the tanh-type solutions, and

$$r(\tau) = \begin{cases} \dfrac{16\,\mathrm{sgn}(r)\left(\tan^{-1}\left[\dfrac{\kappa\left(m\left|f(\tau)+\frac{1}{f(\tau)}\right|-H\right)}{\sqrt{8\Lambda_e + 8\kappa e_1 e_2 - \left(\sqrt{\kappa^2 H^2 + 8\Lambda_e} - m\kappa(f(\tau)-\frac{1}{f(\tau)})\right)^2}}\right]+n\pi\right)}{\sqrt{8\Lambda_e + 8\kappa e_1 e_2 - \left(\sqrt{\kappa^2 H^2 + 8\Lambda_e} - m\kappa(f(\tau)-\frac{1}{f(\tau)})\right)^2}} & \text{(tan-type A)} \\[4em] \dfrac{16\,\mathrm{sgn}(r)\left(\tan^{-1}\left[\dfrac{\sqrt{8\Lambda_e + 8\kappa e_1 e_2 - \left(\sqrt{\kappa^2 H^2 + 8\Lambda_e} - m\kappa(f(\tau)-\frac{1}{f(\tau)})\right)^2}}{\kappa\left(H-m\left|f(\tau)+\frac{1}{f(\tau)}\right|\right)}\right]+n\pi\right)}{\sqrt{8\Lambda_e + 8\kappa e_1 e_2 - \left(\sqrt{\kappa^2 H^2 + 8\Lambda_e} - m\kappa(f(\tau)-\frac{1}{f(\tau)})\right)^2}} & \text{(tan-type B)} \end{cases} \tag{123}$$

for the tan-type solutions.

### 5.4.1. Neutral Particle Motion

For electrically neutral particles, with $e_1 = e_2 = 0$, the quantity $\Lambda_e = -2\Lambda$. In this case, the gravitational attraction of the masses competes with cosmic expansion or attraction, depending on the sign of $\Lambda_e$.

It is instructive to examine the motion for a range of masses. To this end, an arbitrary fiducial mass $m_0$ can be chosen, which sets a calibration scale for all other quantities. Typical scenarios are illustrated in Figures 4 and 5.

For $\Lambda_e < 0$, the curvature $R < 0$ in the absence of stress–energy (from (14)) corresponding to cosmic attraction is depicted in Figure 4. The particles always remain bound, and for large values of $m$, undergo oscillation about their centre of inertia in a manner not too different from that shown in Figure 2 for small $H_0$. However, for very small $m$, a rather surprising situation arises: a sharp extremum develops early in each half-period followed by a second, broader extremum. The first extremum is visible in Figure 4 for the $m = 0.05m_0$ curve, but also appears for smaller values of $m$; it is not visible for $m = 0.001m_0$ due to plotting precision.

In all cases, in Figure 4, the two particles start at $r = 0$ with equal opposing momentum and depart in opposite directions. For a small enough $m$, they quickly reach a maximum separation and then go back toward one another for a certain period of proper time. At some point, they each reverse direction, moving apart more slowly until they attain a second maximal separation. They then reverse direction again, returning to their starting point where the motion then repeats itself with the particles interchanged.

This peculiar behavior, first observed in [103,104], is due to a subtle interplay between the gravitational attraction and relativistic motion of the particles in a space–time with cosmic attraction. It is demonstrative of the unexpected relativistic behaviour that self-gravitating systems can exhibit.

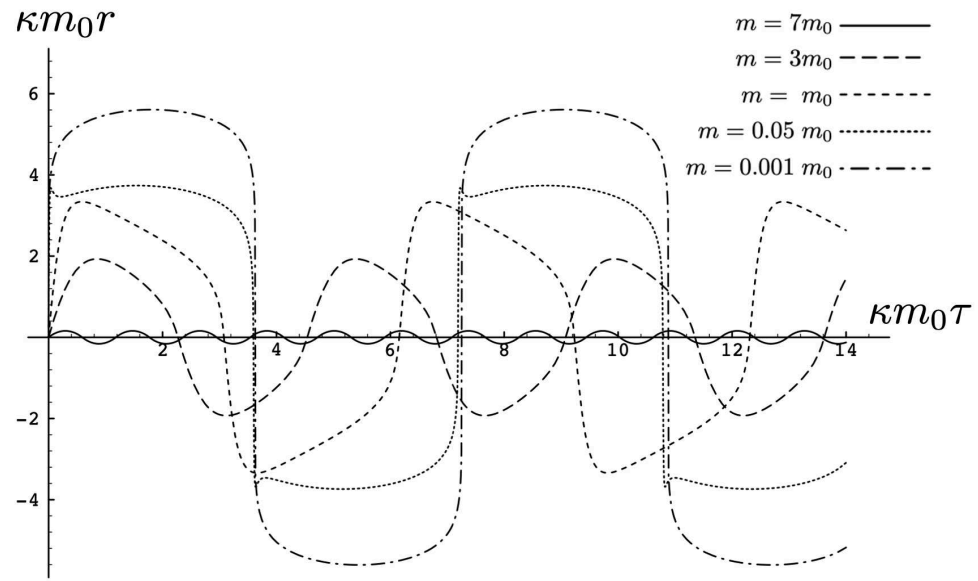

**Figure 4.** A sequence of equal mass curves in the cosmic-attractive case for $\frac{\Lambda_e}{\kappa^2 m_0^2} = -1.5$ and $\frac{H_0}{m_0} = 16$. There is a second extremum in each half-period for $m = 0.05 m_0$.

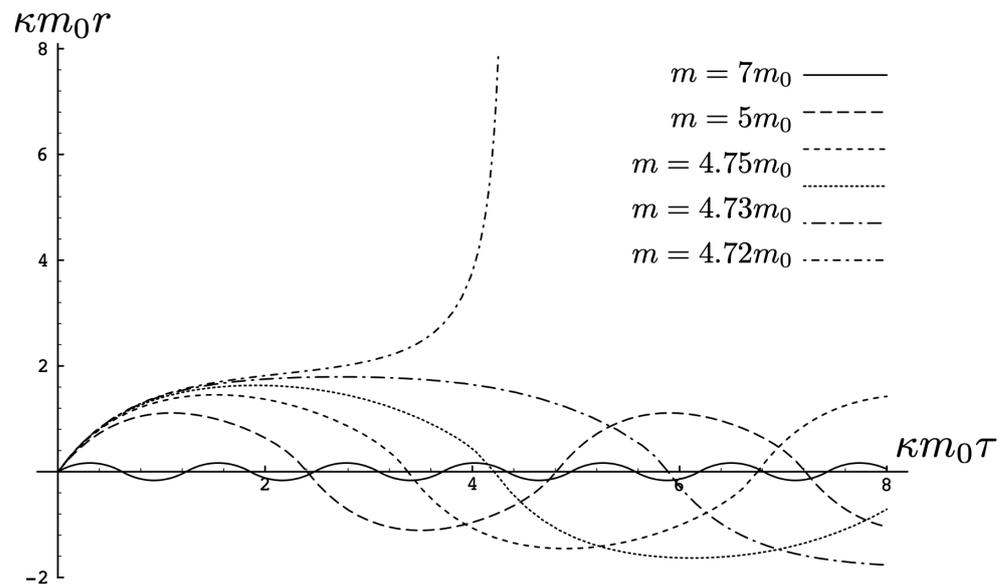

**Figure 5.** A sequence of equal mass curves in the cosmic-expansive case for $\frac{\Lambda_e}{\kappa^2 m_0^2} = 1.5$ and $\frac{H_0}{m_0} = 16$. The motion becomes unbound for masses $m < 4.73 m_0$.

If $\Lambda_e > 0$, $R > 0$ in the absence of stress–energy, and a de Sitter-like cosmic expansion ensues. The situation for various values of $m$ is illustrated in Figure 5. For a large enough $m$, the motion is bounded. As $m$ decreases, the period of the motion becomes larger, until at a critical value of $m \lesssim 4.729$, the period becomes infinite. Here, the cosmic expansion balances the gravitational attraction of the particles, and dominates for smaller values of $m$. The bound behaviour is fully analogous to that of an overdense universe that expands to some maximal value of the scale factor and then recontracts; the unbound behaviour is analogous to that of an underdense universe whose deceleration parameter is insufficient to prevent full cosmic dilution.

All of the above behaviour is described by the tanh-type solutions. However, there is a countably infinite set of unbounded motions described by the tan-type solutions. Some examples are shown in Figure 6 for the phase-space. For $0 < \Lambda_e < \Lambda_c$, both bounded and unbounded motions exist for a fixed value of $H = H_0$, shown in the left panel of Figure 6. The bounded motion consists of the distorted oscillations noted previously, whereas the unbounded motion consists of two particles simply approaching each other at some minimal value of $|r|$ after which they reverse direction and proceed toward infinity. The dotted curves come from the upper expression in (123) and the dashed curves from the lower one. As $\Lambda_e \to \Lambda_c$, the bulges of the solid tanh-type A curve and the dotted tan-type A curve come close, making contact at the critical value $\Lambda_e = \Lambda_c$. For $\Lambda_e > \Lambda_c$, these two curves bifurcate into the solid curves shown in the right panel of Figure 6: the particles cross one another before receding toward infinity. The upper solid curve represents the motion in which

$$p \to p_\pm = \frac{\left( \pm \sqrt{\kappa^2 H^2 + 8\Lambda_e} + \sqrt{8\Lambda_e} \right)}{2\kappa} \tag{124}$$

as $r \to \pm \infty$.

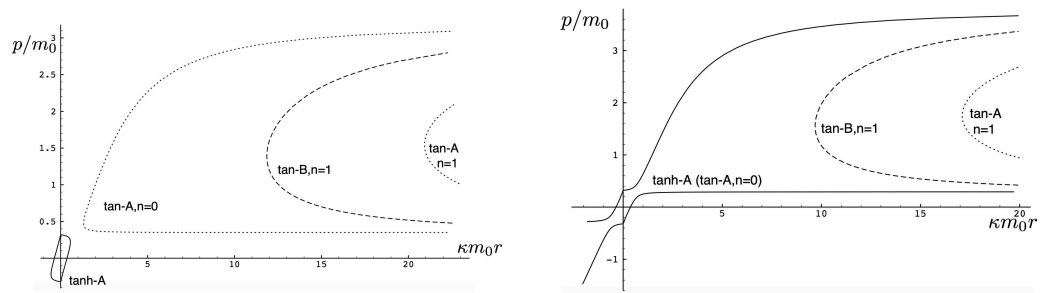

**Figure 6.** Phase space trajectories of bounded and the unbounded motions for $H_0 = 2.1\ m_0$ and $\frac{\Lambda_e}{\kappa^2 m_0^2} = 1.0$ (**left**) and $\frac{\Lambda_e}{\kappa^2 m_0^2} = 1.5$ (**right**).

One peculiar feature of the unbounded motion is that the two particles diverge to infinite separation at finite proper time. The trajectory $x(t)$ of a null ray emitted from particle 2 at time $T$ is governed by $d\tau = 0$, yielding

$$\frac{dx}{dt} = \pm N_0(x(t), z_1(t), z_2(t), p(t)) - N_1(x(t), z_1(t), z_2(t), p(t)) \tag{125}$$

for the equation of the light signal directed toward (+) or away from (−) particle 1. Numerically solving (125) yields the trajectories shown in Figure 7, which plots the trajectories of null rays emitted from particle 2 at various times $T$. For small $\kappa m_0 T \lesssim 3.4$), the particles are in causal contact, as shown by the dotted curve of positive slope in the left panel of Figure 7. At $\kappa m_0 T \approx 3.4$, the light ray just barely catches up to particle 1, which itself is almost in light-like motion; this is shown by the dashed curve of positive slope. Beyond this time, the particles are no longer in causal contact. For a larger $T$, the null ray of positive slope quickly asymptotes to a line parallel to the asymptotic worldline of particle 1, as shown by the dashed curve. The light rays emitted away from particle 1 all asymptote to curves parallel to the trajectory of particle 2. As $T$ becomes even larger, the null ray emitted in the direction of particle 1 experiences a repulsive effect, ultimately reversing its direction, as shown in the right panel of Figure 7. For $\kappa m_0 T \approx 4.82$ the null ray emitted toward particle 1 asymptotes to $r = 0$. For larger values of $T$, null rays emitted in the positive direction eventually reverse their motion and asymptote towards curves that are parallel to the asymptotic trajectory of particle 2, as shown by the dashed and dot-dashed lines in the left panel of Figure 7. The trajectories of the light signals emitted away from particle 1 cannot be distinguished in the plot from the trajectory of particle 2.

This behaviour can be captured in a flat-space model as follows. Consider the two-velocity

$$u^\mu = (h(\sigma\tau), \sqrt{h^2(\sigma\tau) - 1}) \tag{126}$$

where $h(\sigma\tau)$ is some function and

$$ds^2 = -dt^2 + dx^2 \tag{127}$$

is the flat metric. The general expression for the two-acceleration is

$$a^\mu = \frac{du^\mu}{d\tau} = \sigma h'(1, \frac{h}{\sqrt{h^2(\sigma\tau) - 1}}) \tag{128}$$

whose magnitude is

$$a \cdot a = \frac{(\sigma h')^2}{h^2(\sigma\tau) - 1} \tag{129}$$

where $h' = dh(\tau)/d\tau$. Noting that $u \cdot u = 1$ and $a \cdot u = 0$, we have the following possible scenarios:

1. As $\tau \to \infty$, $h \to h_0$ where $h_0$ is finite. In this case, the particle never becomes light-like.
2. As $\tau \to \infty$, $h \to \infty$. In this case, the particle becomes light-like, but this happens in an infinite amount of proper time (and coordinate time). The standard example is $h = \cosh(\sigma\tau)$, the constant acceleration example.
3. The function $h \to \infty$ as $\tau \to \tau_0$, where $\tau_0$ is finite. In this case, the particle asymptotes to a light-like trajectory in a finite amount of proper time, but an infinite amount of coordinate time; an example is $h = \sec(\sigma\tau)$. The acceleration increases as a function of proper time, diverging at $\tau = \tau_0$. This last situation is realized by the exact solutions (123) with $\Lambda > \Lambda_c$.

The proper time $\tau_\infty$ at which the particles attain infinite separation $r \longrightarrow \infty$ is

$$\tau_\infty = \frac{4}{\kappa m \sqrt{\gamma_m}} \log\left( \frac{H_0(1 + \sqrt{\gamma_H}) - (p_+ + \sqrt{p_+^2 + m^2})(1 + \sqrt{\gamma_m})}{\eta\left[ H_0(1 + \sqrt{\gamma_H}) - (p_+ + \sqrt{p_+^2 + m^2})(\sqrt{\gamma_m} - 1) \right]} \right) \tag{130}$$

for any given $m$, where $p_+$ is given by (124).

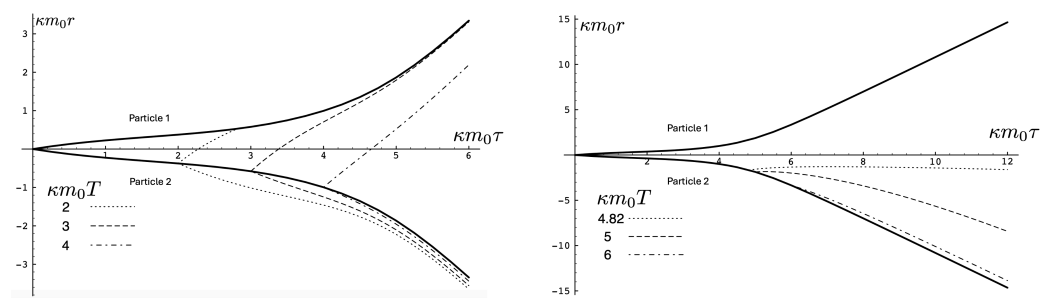

**Figure 7.** Trajectories of unbounded motion for the same parameters as in Figure 6. At early times, the two particles are in casual constant (**left** panel), but null rays emitted from particle 2 for proper times $\kappa m_0 T \geq 3.4$ will never reach particle 1, as shown by the dashed and dot-dashed lines of positive slope in the **left** panel. Null rays emitted in the opposite direction asymptote to curves that are parallel to the asymptotic trajectory of particle 2. As the emission time $T$ from particle 2 increases, the null rays experience increasing effective repulsion, with those emitted toward particle 1 eventually reversing their directions to asymptote to curves that are parallel to the asymptotic trajectory of particle 2, as shown in the **right** panel. The null rays emitted in the opposite direction follow too close to the trajectory of particle 2 to be distinguished in the figure in the **right** panel.

### 5.4.2. Charged Particle Motion with $\Lambda_e = 0$

There are two qualitatively distinct cases of charged particle motion: attractive, when the particles have opposite sign charges, and repulsive, when the charges have the same sign. The situation becomes further complicated for nonzero $\Lambda_e$, and so it is more instructive to first consider $\Lambda_e = 0$. Since the charges appear in all solutions as the product $e_1 e_2$, it is sufficient to set $e_1 = -e_2 = q$ for the attractive case and $e_1 = e_2 = q$ for the repulsive case.

For $\Lambda_e = 0$, (117) simplifies to

$$f_e(\tau) = \frac{H}{m\gamma_e}\left\{1 - \eta_e\, e^{\frac{\epsilon\kappa m}{4}\gamma_e(\tau - \tau_0)}\right\} \tag{131}$$

where

$$\eta_e = \frac{\sqrt{p_0^2 + m^2} - \epsilon p_0 \mathrm{sgn}(r) - \frac{m^2}{H}\gamma_e}{\sqrt{p_0^2 + m^2} - \epsilon p_0 \mathrm{sgn}(r)}$$

from (118) and (119) and $p(\tau)$ has the form (101) with $f_0 \to f_e$; the relative distance $r(\tau)$ is likewise obtained from (107)–(110).

**The attractive case: $e_1 = -e_2 = q$**

The quantity $\kappa^2\left(H - mf(\tau) + \frac{m}{f(\tau)}\right)^2 - 8\kappa e_1 e_2 > 0$ in the attractive case and the two particles always remain bound, with the motion described by the tanh-type solution (107). As expected, as $|q|$ increases, the particles are more tightly bound. This is clear from Figure 8, where the maximal separation of the particles decreases as $|q|$ increases, due to the additional electromagnetic attraction. The frequency of the motion correspondingly increases.

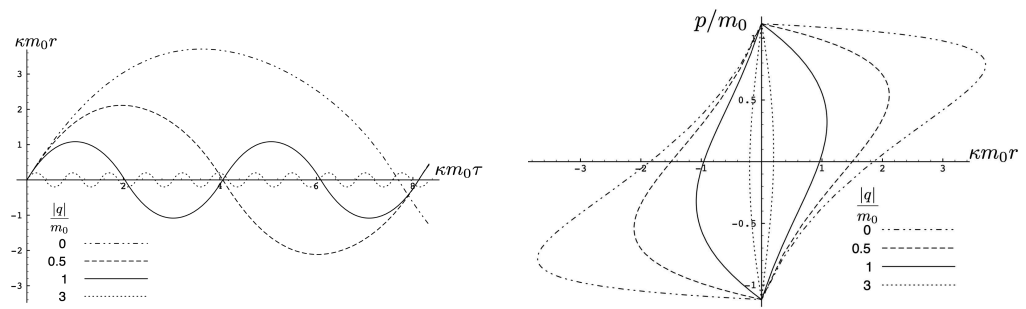

**Figure 8.** **Left**: The exact $r(\tau)$ plots for $H_0 = 3m_0$ and four different values of $|q|/m_0$. **Right**: Phase-space trajectories corresponding to the $r(\tau)$ plots at the left.

The period is determined from the initial value of $p_0 = \sqrt{(H/2)^2 - m^2}$ at $r = 0$:

$$T = \frac{16}{\kappa m\gamma_e}\tanh^{-1}\left(\frac{\gamma_e\sqrt{H^2 - 4m^2}}{(2 - \gamma_e)H}\right). \tag{132}$$

Although the above expression diverges when $\gamma_e = \frac{2H}{2p_0 + H}$, this situation is never realized in the attractive case, since $\gamma_e < 1$ whereas $\frac{2H}{2p_0 + H} > 1$.

It is instructive to compare the exact relativistic motion to the motion in three approximations:

1. The non-relativistic motion described by the Hamiltonian

$$H = 2m + \frac{p^2}{m} + \frac{q^2}{2}|r| + \frac{\kappa m^2}{4}|r|, \tag{133}$$

2.  The linear approximation in $\kappa$, whose Hamiltonian is

$$
\begin{aligned}
H \;=\; & 2\sqrt{p^2+m^2}+\frac{1}{2}q^2|r|+\frac{\kappa}{4}\Big\{(2p^2+m^2-2\epsilon\tilde{p}\sqrt{p^2+m^2})|r| \\
& +\frac{1}{2}q^2(\sqrt{p^2+m^2}-\epsilon\tilde{p})r^2+\frac{1}{24}q^4|r|^3\Big\}\,,
\end{aligned}
\tag{134}
$$

3.  The $\kappa \to 0$ limit, which is special-relativistic electrodynamics in $(1+1)$-dimensional flat space–time; its Hamiltonian is

$$
H=2\sqrt{p^2+m^2}+\frac{q^2}{2}\,|r|\,.
\tag{135}
$$

This comparison is illustrated in Figure 9 for $H_0=3m_0$ for $q/m_0=0.5,1,5,$ and 10. For small $q=0.5$, both the exact solution (solid curve) and the linear approximation (dashed curve) follow $S$-shaped trajectories, whereas the non-relativistic (dotted curve) and flat electrodynamics (dot-dashed curve) trajectories have symmetric oval shapes. As $|q|$, the effect of gravity becomes relatively weak and all trajectories tend to coincide with the trajectory of flat electrodynamics.

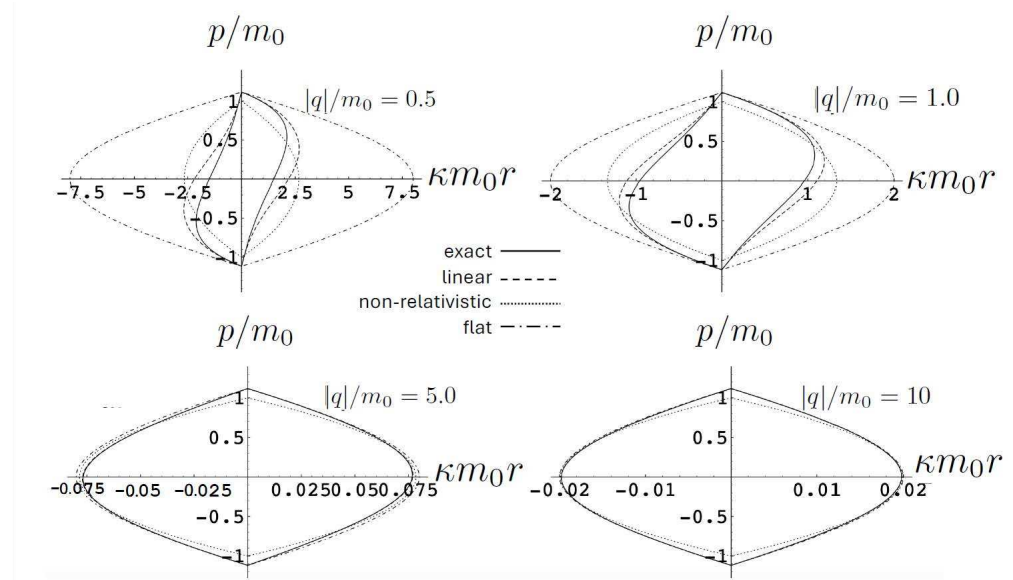

**Figure 9.** A comparison of the phase–space trajectories of $H_0=3m_0$ for exact, linear, non-relativistic, and flat electrodynamics for various values of $|q|$.

The repulsive case: $e_1=e_2=q$, $|q|\le q_c$

In this case, the electromagnetic force is repulsive and competes with the attractive gravitational force. There is a critical value

$$
q_c=\sqrt{\frac{\kappa}{8}}\left(H-\sqrt{H^2-4m^2}\right),
\tag{136}
$$

of the charge, obtained from setting $\mathcal{J}^2=(H-2\epsilon\tilde{p})^2-8q^2/\kappa=0$. If $|q|<q_c$, bounded and unbounded motions can occur, described, respectively, by the tanh-type ($\mathcal{J}^2>0$) and tan-type solutions ($\mathcal{J}^2<0$) for $r(\tau)$. Alternatively, (136) gives the critical value of $H$ for fixed $\kappa$ and $q$, or the critical value of $\kappa$ for fixed $H$ and $q$, both corresponding to the transition from bounded to unbounded motion.

The two possibilities are illustrated in Figure 10, which plots $r(\tau)$ plots for fixed $H_0=3m_0$ (left) or fixed $|q|/m_0=0.1$ (right). We see for fixed $H_0$ that the period becomes larger as $|q|$ increases; once the critical value $q_c/m_0=0.2700907567$ is exceeded, the motion becomes unbounded and the separation of the two particles diverges at finite $\tau$. Likewise,

if $|q|$ is fixed, then a similar transition takes place at a threshold value of $H_0 = H_c$; for the particular choice given here, $H_c/m_0 = 7.21249$.

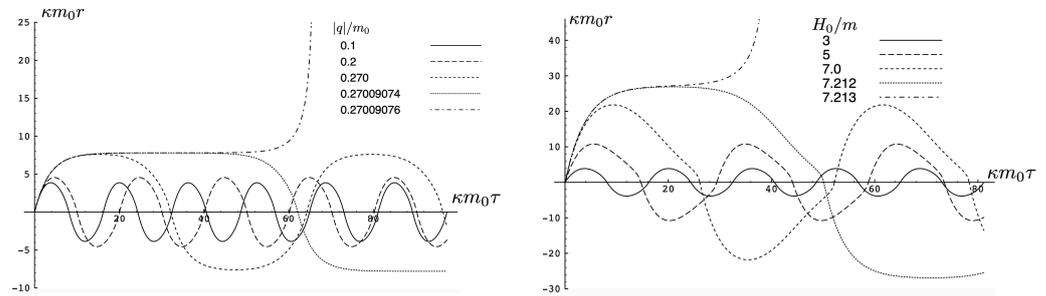

**Figure 10.** A comparison of sub-critical repulsive motion from two perspectives. Left: Plots for various values of $|q|/m_0$ for $H_0 = 3m_0$—the threshold value for escape is $|q|/m_0 = 0.2700907567$. Right: Plots for various values of $H_0/m_0$ for $|q|/m_0 = 0.1$—the threshold value for escape is $H_0/m_0 = 7.21249$.

The existence of two types of motion for fixed $H$ and $q$ has no non-relativistic analogue, and is a qualitatively new aspect of relativistic gravitation. An illustration is provided in Figure 11, showing $r(\tau)$ for $H_0 = 3m_0$ and $|q|/m_0 = 0.25 < q_c/m_0$. Both bounded motion and a sequence of possible unbounded motions can exist. The tanh-type A and B motions have the specific feature that $r(\tau)$ diverges (the particles attain infinite separation) at finite proper time (but at infinite coordinate time).

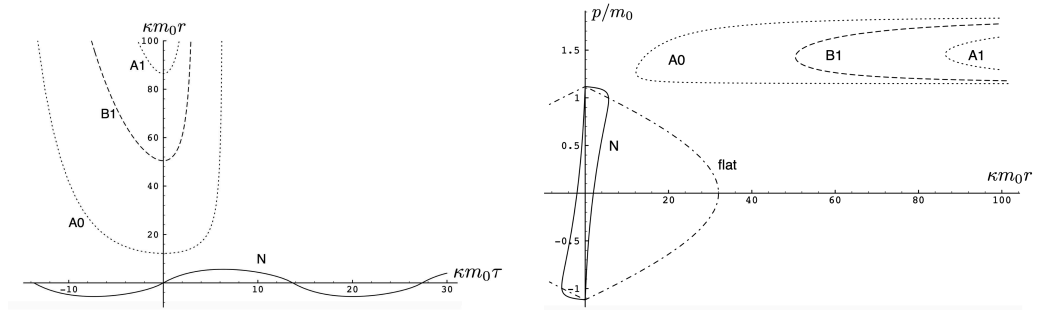

**Figure 11.** An illustration of the possible bounded motion (solid) and unbounded motions (dotted and dashed) for the same values of $H_0 = 3m_0$ and $|q|/m_0 = 0.25 < q_c/m_0$ depicted in position space (**left**) and phase space (**right**). The initial value $r(0)$ determined which of these motions is realized. The flat-space ($\kappa = 0$) electrodynamic phase space trajectory is shown in the right panel for comparison.

The repulsive case: $e_1 = e_2 = q$, $|q| > q_c$

Only unbounded motion is permitted in this case since the electromagnetic repulsive force overwhelms the attractive gravitational force if $|q| > q_c$. The $r(\tau)$ trajectories are qualitatively similar to the unbounded motions depicted in Figure 11; to understand the distinctions between unbounded motion for this case and that for the $|q| \leq q_c$ case, it is instructive to consider the phase space plots.

First, consider the physical region of the $(|q|, p)$ parameter space, as shown in Figure 12 for $H_0 = 3m_0$. From (106), the shaded area is the region $\mathcal{J}^2 > 0$ and $B > 0$, where the tanh-type A and tanh-type B give the actual trajectories. The boundary of this area is fixed by $p = -\sqrt{2/\kappa} \, |q| + H_0/2$ and $p = \pm p_0 = \pm\sqrt{(H_0/2)^2 - m^2}$ ($= \pm\sqrt{5}m_0/2$ for the specific choice in Figure 12). The values of $|q|$ at the intersections of these boundary lines are $q_d$ (for the negative solution) and $q_c$ (the critical value (136)) for the positive one. The tanh-type B solution is realized in a very narrow region between the dashed ($\mathcal{J} - B = 0$) and solid ($p = -\sqrt{2/\kappa} \, |q| + H_0/2$) curves in the shaded region. The tan-type A and B trajectories are realized in the unshaded $\mathcal{J}^2 < 0$ region.

The dotted line in Figure 12 denotes a constant $|q|/m_0$ line whose intercepts with the diagonal lines $p = \pm\sqrt{2/\kappa} \, |q| + H_0/2$ are $p_1/m_0$ and $p_2/m_0$. In the subcritical region $0 < |q| < q_c$, this dotted line would be to the left of $q_c/m_0$. The tanh-type A solutions are realized in the shaded region with $-p_0 < p < p_0$, and the associated $r(\tau)$ trajectory is given by the solid line in Figure 11. In the unshaded region $p_2 < p < p_1$, between the diagonal lines and to the left of $q_c/m_0$, both tan-type A and B are realized; the associated $r(\tau)$ trajectories are, respectively, given by the dotted and dashed lines in Figure 11. The region $p_0 < p < p_2$ is forbidden.

In the supercritical region $q_c < |q| < q_d$, all values of $p$ in the range $-p_0 < p < p_1$ are allowed, and the associated phase-space trajectories are shown in the left panel of Figure 13. The single-bounded motion $N$ in the right panel of Figure 11 merges with the unbounded motion $A0$ at $|q| = q_c$ to yield two new unbounded trajectories $N1$ (corresponding to the unshaded region $p_2 < p < p_1$), and $N2$ (corresponding to the shaded region $-p_0 < p < p_2$) in Figure 13. The remaining unbounded motions $A_n$ and $B_n$ ($n = 1, 2, \ldots$) are all described by, respectively, by the tan-type A and B solutions. In both Figures 11 and 13, the analogous trajectories in flat space electrodynamics (the dot-dashed curves) are shown to illustrate the strongly deforming effects of gravity on the trajectories.

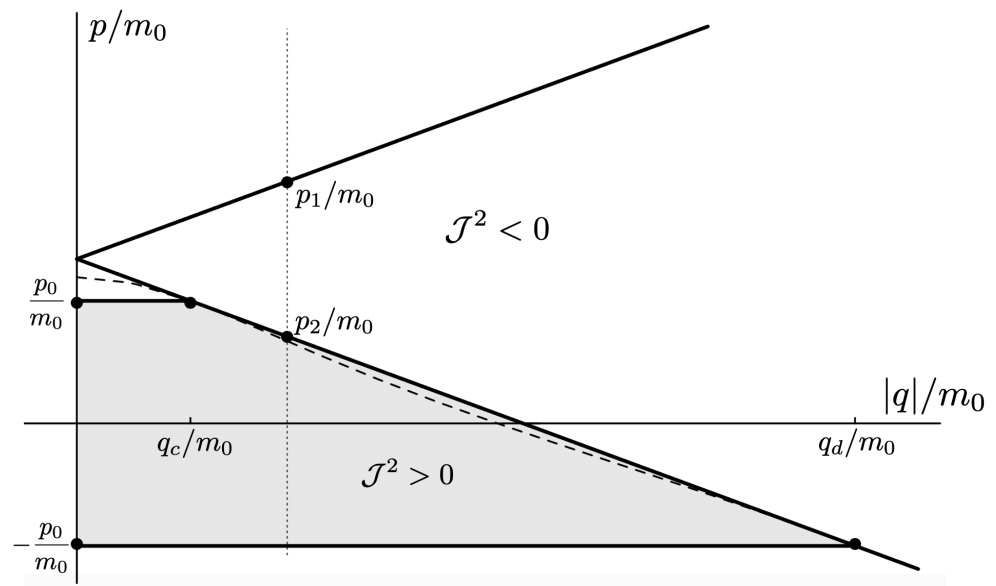

**Figure 12.** The physical region of the $(|q|, p)$ parameter space for $H_0/m_0 = 3$ in the repulsive charged case.

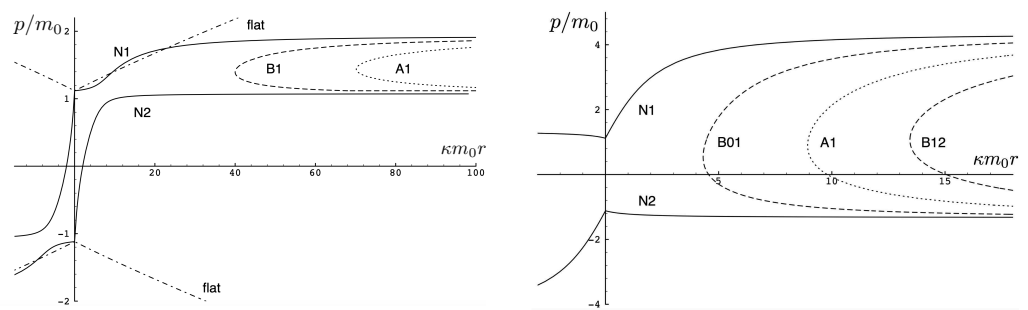

**Figure 13.** Phase-space trajectories for $|q| > q_c$ and $H_0/m_0 = 3$ **Left**: Unbounded motions for $|q|/m_0 = 0.3$. **Right**: Unbounded motions for $|q|/m_0 = 2$. A comparison to the motion in flat space electrodynamics ($\kappa = 0$) is given in the left panel.

Finally, for $|q| > q_d$, all solutions are tan-type A and B since $p_2 < -p_0$. Sample-phase space trajectories are shown in the right panel of Figure 13. A characteristic cusp appears at $r = 0$ in the $N1$ and $N2$ trajectories. For the B trajectories, the motion switches between $B0$ (for $-p_0 \leq p \leq p_0$) and $B1$ (for $p_2 \leq p \leq -p_0$ and $p_0 \leq p \leq p_1$); the full motion is denoted $B01$. $B12$ is likewise composed of a combination of $B1$ and $B2$.

Another interesting aspect of the supercritical $|q| > q_c$ case is that $H < 2m$ is possible—the total energy can be less than the rest energy of the particles! For $H < 2m$, only $\mathcal{J}^2 < 0$ is possible and the shaded region in Figure 12 is absent. Consequently, only unbounded motion is possible; some trajectories for this case are shown in Figure 14. All types of unbounded motions $A_n$ and $B_n$ are realized. Note that all trajectories curve more toward the $r$ axis (due to the additional effect of gravitational attraction) relative to that of flat space electrodynamics. They are also shifted in the positive $p$ direction due to the $p$-dependence of the gravitational potential.

In $1 + 1$ flat-space electrodynamics, the total energy of the system is restricted to $H \geq 2m$ for attractive charges, but no such restriction on $H$ exists for repulsive charges. We see that, in a general relativistic theory, the same situation prevails.

The time-reversed trajectories corresponding to these cases can all be obtained by setting $\epsilon = -1$, and trajectories in the $r < 0$ region are obtained from those in $r > 0$ by reversing the signs of both $r$ and $p$.

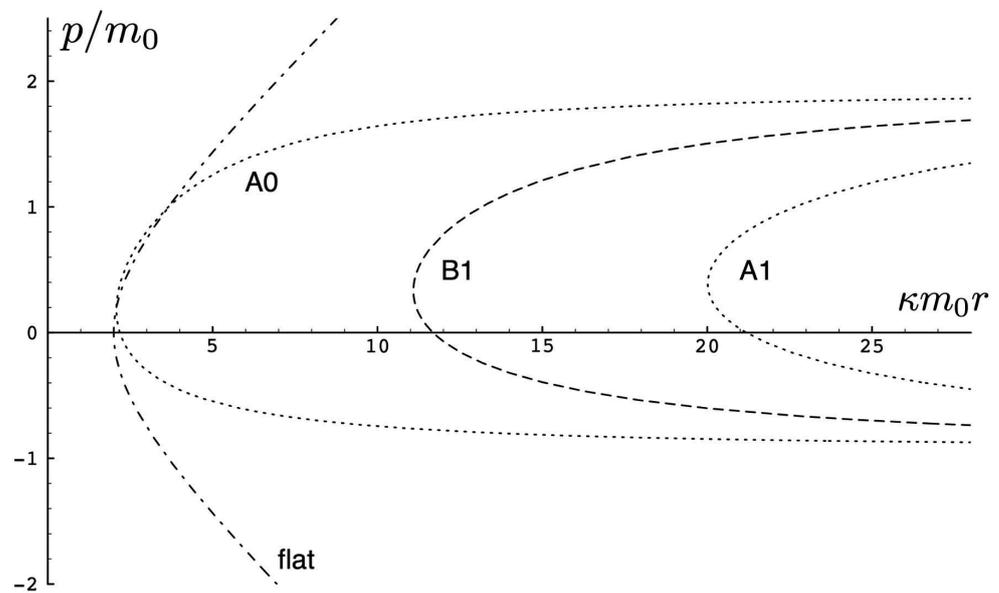

**Figure 14.** Phase-space trajectories of unbounded motions for $H_0/m_0 = 1 < 2$ and $|q|/m_0 = 1$.

### 5.4.3. Charged Particle Motion with $\Lambda_e \neq 0$

This is the most general case for two-body motion in the electrovacuum. The dynamics is governed by a combination of gravitational attraction, electromagnetic attraction/repulsion, and the cosmological constant. The signs of $\gamma_m$ and $\kappa^2 \left( H_0 - mf(\tau) + \frac{m}{f(\tau)} \right)^2 - 8\kappa e_1 e_2 - 8\Lambda_e$ characterize the solutions.

The possible range of motions is now rich and complicated [105], and a set of sample trajectories is presented in Figure 15. One of the noteworthy special cases occurs particular range of negative $\Lambda_e$ and small mass (or large $H/m_0$), where the trajectories have a double-peaked structure. This is shown in the top panel of Figure 15 and is particularly noticeable for $\Lambda_e/(\kappa m_0)^2 = -0.5$, shown as the solid curve. The particles begin at $r = 0$ with equal and opposite initial momenta, reach a maximum separation, and then reverse direction due to the combined attractive electromagentic and graviational forces. But they soon reverse direction again, moving apart until reaching a second maximum after which they return to the starting point. For small values of $|q|$, the double peak structure is present, but for sufficiently large $|q|$, it disappears.

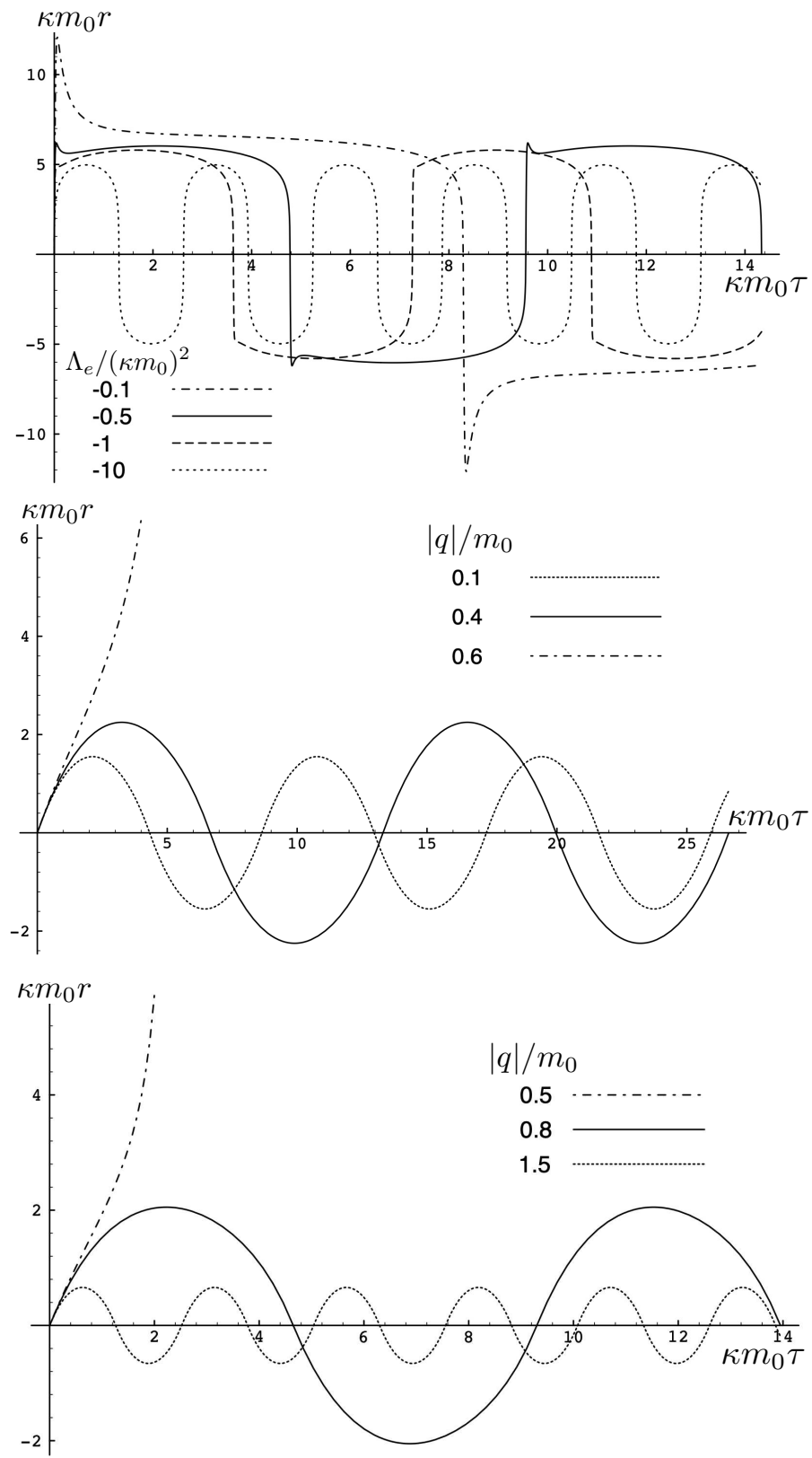

**Figure 15.** Trajectories of equal mass charged particles for $\Lambda_e \neq 0$. **Top**: $r(\tau)$ for $H/m_0 = 500$, $|q|/m_0 = 5$ (attractive), and various values of $\Lambda_e < 0$. **Middle**: $r(\tau)$ for $H/m_0 = 2.5$, $\Lambda_e/(\kappa m_0)^2 = -0.694$ for various values of $|q|$ (repulsive). **Bottom**: $r(\tau)$ for $H/m_0 = 3.0$, $\Lambda_e/(\kappa m_0)^2 = 1$ for various values of $|q|$ (attractive). Cases with $\Lambda_e > 0$ with electromagnetic repulsion have trajectories similar to those in Figure 11.

This peculiar behaviour takes place due to the induced momentum-dependent potential in conjunction with the dual force attraction in the cosmological vacuum. An expansion of the Hamiltonian in terms of $\kappa$ and $\Lambda_e/\kappa^2$

$$
\begin{aligned}
H &= 2\sqrt{p^2+m^2} + \frac{\kappa}{4}(\sqrt{p^2+m^2}-\epsilon\tilde{p})^2\,|r| + \frac{\kappa^2}{4^2}(\sqrt{p^2+m^2}-\epsilon\tilde{p})^3\,r^2 \\
&\quad + \frac{7\kappa^3}{6\times4^3}(\sqrt{p^2+m^2}-\epsilon\tilde{p})^4\,|r|^3 - \frac{\Lambda_e}{2\kappa}\cdot\frac{\epsilon\tilde{p}}{\sqrt{p^2+m^2}}\,|r| - \frac{\Lambda_e}{16}\cdot\frac{\epsilon\tilde{p}\,m^2}{p^2+m^2}\,r^2 \\
&\quad + \frac{\Lambda_e^2}{4\kappa^3}\cdot\frac{\epsilon\tilde{p}}{(p^2+m^2)^{3/2}}\,|r| + \cdots
\end{aligned}
\tag{137}
$$

illustrates that negative $\Lambda_e$ acts effectively as an attractive force leading to bounded (periodic) motions, whereas positive $\Lambda_e$ acts effectively as a repulsive force.

For $\Lambda_e < 0$, there can be both bounded and unbounded motion in the electromagnetically repulsive case depending on the size of $|q|$. For small $|q|$, cosmological attraction is smaller, whereas for large $|q|$, the opposite situation prevails. There is an intermediate range of $|q|$ for a given $\Lambda_e$, where both bounded and unbounded motions are possible [105]. Sample trajectories are shown in the middle panel of Figure 15.

In the electromagnetically attractive case, for $\Lambda_e > 0$, there can again be both bounded and unbounded motion depending on the size of $|q|$. In this case, attraction due to a sufficiently large value of $|q|$ will overwhelm cosmological repulsion. These effects depend on the relative size of $H$ as compared to $\sqrt{\kappa^2 m^2/2\Lambda_e + 4m^2}$; if $H$ is larger than this latter quantity, then there is again an intermediate range of $|q|$ where both bounded and unbounded motions exist [105]. Sample trajectories are shown in the bottom panel of Figure 15.

Finally, the case of joint electromagnetic and cosmological repulsion yields trajectories similar to those in Figure 11.

### 5.5. Exact Two-Body Motion with Unequal Masses

For unequal masses, the proper time (97) of each particle

$$
d\tau_1 = dt\,\frac{16YK_0K_1m_1}{JKM_1\sqrt{p^2+m_1^2}} \qquad d\tau_2 = dt\,\frac{16YK_0K_2m_2}{JKM_2\sqrt{p^2+m_2^2}}
\tag{138}
$$

differs, where $K \equiv K_+ = K_-$ and $Y \equiv Y_+ = Y_-$. While there are a variety of choices of time parameters one could use to describe the motion, the most natural one appears to be [104,105]

$$
d\tilde{\tau} \equiv dt\,\frac{16YK_0}{JK}\left(\frac{K_1K_2m_1m_2}{M_1M_2\sqrt{p^2+m_1^2}\sqrt{p^2+m_2^2}}\right)^{1/2},
\tag{139}
$$

which is symmetric with respect to $1 \leftrightarrow 2$ and reduces to the proper time (114) when $m_1 = m_2$.

The parameter space now has an additional variable and the analysis of the various types of motions is correspondingly more detailed. However, the general rubric of the four types of motion depending on cosmological attraction/repulsion and electromagnetic attraction/repulsion continues to hold.

A number of more interesting sample cases are illustrated in Figures 16 and 17. When $m_1$ is large, $m_2$ behaves like a test body and the two particles oscillate about their centre of inertia for all values of the parameters shown. The gravitational attraction is stronger and both the period and proper maximal separation between the particles becomes shorter.

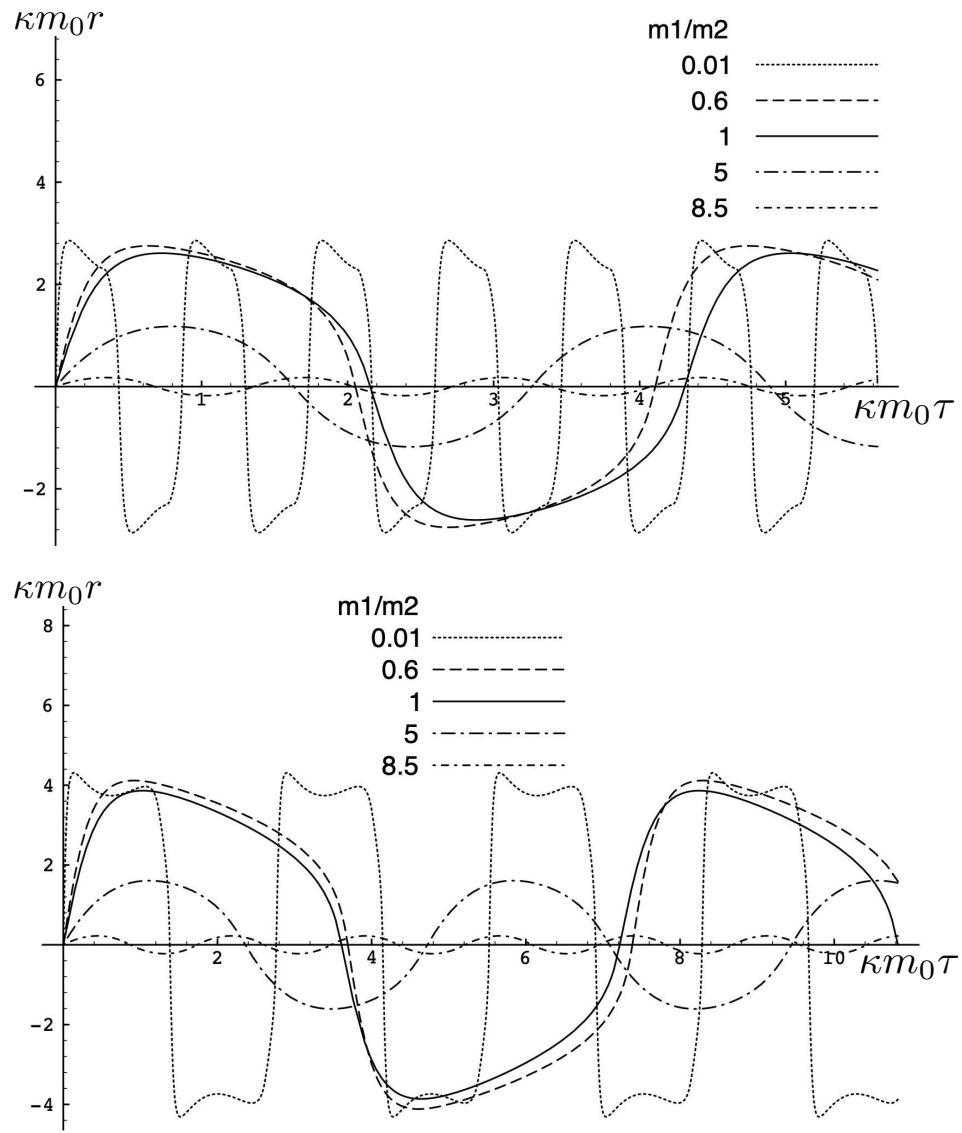

**Figure 16.** Trajectories $r(\tau)$ of unequal mass charged particles in the electromagnetically attractive case for a variety of mass ratios with $H/m_0 = 10$, $\Lambda_e/(\kappa m_0)^2 = -1$, and $m_2 = m_0$. **Top**: $|q|/m_0 = 1$. **Bottom**: $|q|/m_0 = 0.1$.

For $\Lambda_e < 0$, as the mass ratio decreases, the maximal separation and period both become larger, as shown in Figure 16. However, eventually, the attractive effect due to the cosmological constant begins to dominate, and for quite small mass ratios, the period and proper maximal separation decrease and the double-peak structures emerge. For larger $|q|$ (top panel), the electromagnetic attraction is stronger, yielding correspondingly shorter periods and maximal separations as compared to the smaller $|q|$ case (bottom panel).

If there is either cosmological or electromagnetic repulsion (or both), then unbounded motion is possible, as shown in Figure 17. Large $m_1$ yields bounded motion as before, but for a small enough mass ratio, the repulsive effect dominates and the particles fly apart to infinity. Maximal separations and periods are larger for cosmological and electromagnetic repulsion, as shown in the bottom panel of Figure 17.

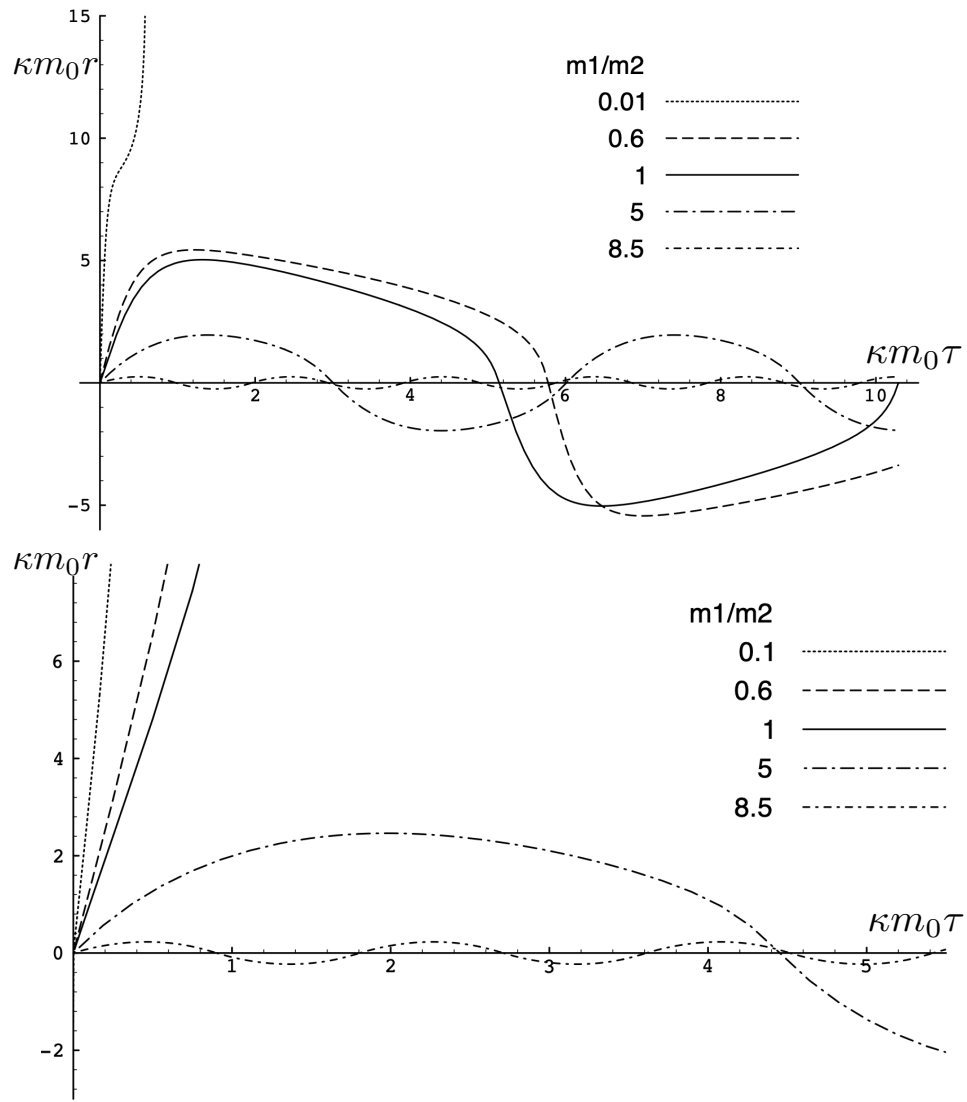

**Figure 17.** Trajectories $r(\tau)$ of unequal mass-charged particles in the electromagnetically repulsive case for a variety of mass ratios with $H/m_0 = 10$ and $m_2 = m_0$. **Top**: $|q|/m_0 = 0.6$ and $\Lambda_e/(\kappa m_0)^2 = -1$ (cosmological attraction) **Bottom**: $|q|/m_0 = 0.1$ and $\Lambda_e/(\kappa m_0)^2 = 0.1$ (cosmological repulsion).

## 5.6. Static Balance

In $(3 + 1)$ dimensions, exact solutions to the two-body problem in general relativity have been unobtainable, primarily because gravitational radiation carries energy away from the system. However, a condition of static balance—in which gravitational attraction is exactly balanced by electromagnetic repulsion—was attained for two bodies [112,113] and later for $N$ bodies on a line [114]. This condition is

$$e_i = \pm\sqrt{4\pi G}\, m_i\,,\tag{140}$$

which is more stringent than the corresponding condition

$$Gm_1m_2 - \frac{e_1e_2}{4\pi} = 0\tag{141}$$

in Newtonian theory. Whether or not (140) is a is a necessary condition for static balance remains an open question.

The corresponding static balance condition in $(1+1)$ dimensions can be straightforwardly obtained from the determining Equation (77): it is simply the extremum of $H$ with respect to $r$. Setting $\partial H / \partial r = 0$ yields, after some algebra,

$$\frac{\kappa}{2}\left(\sqrt{p^2 + m_1^2} - \epsilon \tilde{p}\right)\left(\sqrt{p^2 + m_2^2} - \epsilon \tilde{p}\right) - e_1 e_2 = 0 \tag{142}$$

which is the force–balance condition. If $p = 0$, then

$$\frac{\kappa}{2} m_1 m_2 - e_1 e_2 = 0 \tag{143}$$

which is the static balance condition [109]. This condition is identical to that of (141) in Newtonian theory in $(1+1)$ dimensions.

However, there is a more general solution to (142), which is

$$p_c = \pm \frac{|(\frac{\kappa}{2})^2 m_1^2 m_2^2 - e_1^2 e_2^2|}{\sqrt{2 \kappa e_1 e_2} \sqrt{(\frac{\kappa}{2} m_1^2 + e_1 e_2)(\frac{\kappa}{2} m_2^2 + e_1 e_2)}} \ . \tag{144}$$

provided $e_1 e_2 > 0$, where $p = p_c$ is the (constant) value of the momentum for which the repulsive electromagnetic and attractive gravitational forces between both particles are the same.

Non-relativistically, (143) is the force–balance condition that includes both the static case and uniform motion. However, in the relativistic case, these two are different. In general the force–balance condition (142) depends on the momentum and allows for uniform motion in the centre of inertia system in which both particles either approach or recede with constant momentum (144). The static balance condition (143) is the special case for which $p_c = 0$.

## 6. The Three-Body Problem

The 3-body problem for a relativistic self-gravitating system is an interesting subject of study, in large part because its non-relativistic counterpart models several interesting physical systems. These include two elastically colliding billiard balls in a uniform gravitational field [14], with the perfectly elastic collisions of a particle with a wedge in a uniform gravitational field [3], and a "linear baryon" (a bound state of three quarks along a line) [15]. These systems can even be tested experimentally [115].

The first study of three-body motion in a fully relativistic context was carried out over 20 years ago [29,30], with extensions to include unequal masses [116], a cosmological constant [117], and charge [118] subsequently undertaken. The most effective means to study the dynamics of the 3-body ROGS is to work in the canonical formalism [99]. This approach yields an exact expression for the Hamiltonian in terms of the four physical degrees of freedom of the system: the two proper separations of the point particles and their conjugate momenta.

The non-relativistic system can be shown to be equivalent to that of a single-particle moving in a hexagonal-well potential in two spatial dimensions. The three-body ROGS has an exact relativistic generalization of this hexagonal-well problem, providing insights into intrinsically non-perturbative relativistic effects. For equal mass particles, the cross-sectional shape of the well is that of a regular hexagon in the non-relativistic case. Unequal masses distort this symmetry to that of a 'squashed' hexagon, whose sides have differing lengths. The hexagonal symmetry is maintained relativistically, with the sides of the hexagon curved outward.

Poincaré sections can be used to examine the global structure of the phase space. The relativistic plots are qualitatively similar to their non-relativistic counterparts, but distorted toward the lower-right of the phase plane. This is due the relativistic gravitational momentum that continuously distorts the basic structure of the Poincaré plot.

### 6.1. Three-Body Constraint Equations

The constraints (63) and (64) for a three-body system become

$$
\begin{aligned}
\triangle\phi - \frac{\kappa^2}{4}(\chi')^2\phi \;=\;& \frac{\kappa}{4}\Big\{\sqrt{p_1^2 + m_1^2}\,\phi(z_1)\delta(x - z_1) + \sqrt{p_2^2 + m_2^2}\,\phi(z_2)\delta(x - z_2) \\
& + \sqrt{p_3^2 + m_3^2}\,\phi(z_3)\delta(x - z_3)\Big\}
\end{aligned}
\tag{145}
$$

$$
\triangle\chi \;=\; -\frac{1}{2}\{p_1\delta(x - z_1) + p_2\delta(x - z_2) + p_3\delta(x - z_3)\}
\tag{146}
$$

where the general solution to this latter equation is

$$
\chi = -\frac{1}{4}\{p_1|x - z_1| + p_2|x - z_2| + p_3|x - z_3|\} - \epsilon X x + \epsilon C_\chi \;.
\tag{147}
$$

As before, $\epsilon$ ($\epsilon^2 = 1$) flips sign under time-reversal, and $X$ and $C_\chi$ are constants of integration.

The strategy for solving (145) is similar to the two-body case: space–time is divided into four regions

$$
\begin{aligned}
z_1 < x && \text{(+) region} \\
z_2 < x < z_1 && \text{(1) region} \\
z_3 < x < z_2 && \text{(2) region} \\
x < z_3 && \text{(−) region}
\end{aligned}
$$

assuming $z_3 < z_2 < z_1$. Within each of these regions, $\chi'$ is constant

$$
\chi' = \begin{cases}
-\epsilon X - \frac{1}{4}(p_1 + p_2 + p_3) & \text{(+) region} \\
-\epsilon X + \frac{1}{4}(p_1 - p_2 - p_3) & \text{(1) region} \\
-\epsilon X + \frac{1}{4}(p_1 + p_2 - p_3) & \text{(2) region} \\
-\epsilon X + \frac{1}{4}(p_1 + p_2 + p_3) & \text{(−) region}
\end{cases}
\tag{148}
$$

and (145) has the solution

$$
\begin{cases}
\phi_+(x) = A_+ e^{\frac{\kappa}{2}K_+ x} + B_+ e^{-\frac{\kappa}{2}K_+ x} \\
\phi_1(x) = A_1 e^{\frac{\kappa}{2}K_1 x} + B_1 e^{-\frac{\kappa}{2}K_1 x} \\
\phi_2(x) = A_2 e^{\frac{\kappa}{2}K_2 x} + B_2 e^{-\frac{\kappa}{2}K_2 x} \\
\phi_-(x) = A_- e^{\frac{\kappa}{2}K_- x} + B_- e^{-\frac{\kappa}{2}K_- x}
\end{cases}
\tag{149}
$$

in each region, where

$$
K_+ \equiv X + \frac{\epsilon}{4}(p_1 + p_2 + p_3) \qquad K_- \equiv X - \frac{\epsilon}{4}(p_1 + p_2 + p_3)
$$

$$
K_1 \equiv X - \frac{\epsilon}{4}(p_1 - p_2 - p_3) \qquad K_2 \equiv X - \frac{\epsilon}{4}(p_1 + p_2 - p_3)
\tag{150}
$$

In order for a full solution to (145) respect boundary conditions that ensure the finiteness of the Hamiltonian, the function $\phi$ must satisfy the following matching conditions

$$
\phi_+(z_1) = \phi_1(z_1) \qquad \phi_1(z_2) = \phi_2(z_2) \qquad \phi_-(z_3) = \phi_2(z_3)
\tag{151a}
$$

$$
\phi'_+(z_1) - \phi'_1(z_1) = \frac{\kappa}{4}\sqrt{p_1^2 + m_1^2}\phi(z_1)
\tag{151b}
$$

$$
\phi'_1(z_2) - \phi'_2(z_2) = \frac{\kappa}{4}\sqrt{p_2^2 + m_2^2}\phi(z_2)
\tag{151c}
$$

$$
\phi'_2(z_3) - \phi'_-(z_3) = \frac{\kappa}{4}\sqrt{p_3^2 + m_3^2}\phi(z_3) \;.
\tag{151d}
$$

at the locations of the particles $x = z_1, z_2, z_3$. Satisfying these conditions is a straightforward but tedious exercise, yielding the result

$$
\begin{aligned}
L_1 L_2 L_3 =\ & \mathfrak{M}_{12}\mathfrak{M}_{21}L_3^* e^{\frac{\kappa}{4}s_{12}[(L_1+\mathfrak{M}_{12})z_{13}-(L_2+\mathfrak{M}_{21})z_{23}]} \\
& + \mathfrak{M}_{23}\mathfrak{M}_{32}L_1^* e^{\frac{\kappa}{4}s_{23}[(L_2+\mathfrak{M}_{23})z_{21}-(L_3+\mathfrak{M}_{32})z_{31}]} \\
& + \mathfrak{M}_{31}\mathfrak{M}_{13}L_2^* e^{\frac{\kappa}{4}s_{31}[(L_3+\mathfrak{M}_{31})z_{32}-(L_1+\mathfrak{M}_{13})z_{12}]}
\end{aligned}
\tag{152}
$$

or more compactly

$$
L_1 L_2 L_3 = \sum_{ijk}\left|\epsilon^{ijk}\right|\mathfrak{M}_{ij}\mathfrak{M}_{ji}L_k^* e^{\frac{\kappa}{4}s_{ij}[(L_i+\mathfrak{M}_{ij})z_{ik}-(L_j+\mathfrak{M}_{ji})z_{jk}]}
\tag{153}
$$

for the full determining equation of the Hamiltonian, where

$$
\mathfrak{M}_{ij} = M_i - \epsilon p_i s_{ij}, \qquad\qquad M_i = \sqrt{p_i^2 + m_i^2}
\tag{154}
$$

$$
L_i = H - M_i - \epsilon(\sum_j p_j s_{ji}) \qquad L_i^* = (1 - \prod_{j<k\neq i} s_{ij}s_{ik})M_i + L_i
\tag{155}
$$

with $z_{ij} = (z_i - z_j)$, $s_{ij} = \text{sgn}(z_{ij})$, and $\epsilon^{ijk}$ the three-dimensional Levi–Civita tensor. Equation (153) reproduces the correct determining equation for any permutation of the particles.

The components of the metric can be determined from (49)–(54). Of greater relevance are the canonical equations of motion (30), which become

$$
\begin{aligned}
\dot{z}_1 \{ & L_2 L_3 + L_1 L_3 + L_1 L_2 \\
& - [M_2 - \epsilon p_2 s_{21}][M_1 - \epsilon p_1 s_{12}][1 + \frac{\kappa}{4}L_3^*|z_{12}|]e^{\frac{\kappa}{4}s_{12}[(L_1+\mathfrak{M}_{12})z_{13}-(L_2+\mathfrak{M}_{21})z_{23}]} \\
& - [M_3 - \epsilon p_3 s_{31}][M_1 - \epsilon p_1 s_{13}][1 + \frac{\kappa}{4}L_2^*|z_{13}|]e^{\frac{\kappa}{4}s_{13}[(L_1+\mathfrak{M}_{13})z_{12}+(L_3+\mathfrak{M}_{32})z_{23}]} \\
& - [M_2 - \epsilon p_2 s_{23}][M_3 - \epsilon p_3 s_{32}][1 + \frac{\kappa}{4}L_1^*|z_{23}|]e^{\frac{\kappa}{4}s_{23}[(L_3+\mathfrak{M}_{31})z_{13}-(L_2+\mathfrak{M}_{13})z_{12}]} \} \\
= & [M_2 - \epsilon p_2 s_{21}][(\frac{\partial M_1}{\partial p_1} - \epsilon s_{12})L_3^* - (M_1 - \epsilon p_1 s_{12})(\epsilon s_{13} + \frac{\kappa}{4}L_3^*(\epsilon z_{12}))] \\
& \qquad\qquad\qquad\qquad\qquad\qquad\qquad\qquad\qquad \times e^{\frac{\kappa}{4}s_{12}[(L_1+\mathfrak{M}_{12})z_{13}-(L_2+\mathfrak{M}_{21})z_{23}]} \\
& + [M_3 - \epsilon p_3 s_{31}][(\frac{\partial M_1}{\partial p_1} - \epsilon s_{13})L_2^* - (M_1 - \epsilon p_1 s_{13})\{\epsilon s_{12} + \frac{\kappa}{4}L_2^*(\epsilon z_{13})\}] \\
& \qquad\qquad\qquad\qquad\qquad\qquad\qquad\qquad\qquad \times e^{\frac{\kappa}{4}s_{13}[(L_1+\mathfrak{M}_{23})z_{12}+(L_3+\mathfrak{M}_{32})z_{23}]} \\
& + [M_2 - \epsilon p_2 s_{23}][M_3 - \epsilon p_3 s_{32}][-s_{12}s_{13}\frac{\partial M_1}{\partial p_1} + \frac{\kappa}{4}s_{23}L_1^*[\epsilon|z_{12}| - \epsilon|z_{13}|]] \\
& \qquad\qquad\qquad\qquad\qquad\qquad\qquad\qquad\qquad \times e^{\frac{\kappa}{4}s_{23}[(L_3+\mathfrak{M}_{31})z_{13}-(L_2+\mathfrak{M}_{13})z_{12}]} \\
& + \frac{\partial M_1}{\partial p_1}L_2 L_3 + \epsilon(s_{12}L_1 L_3 + s_{13}L_2 L_1)
\end{aligned}
\tag{156}
$$

for $\dot{z}_1$, with similar expressions for $\dot{z}_2$ and $\dot{z}_3$. Differentiating (152) with respect to $z_1$ yields

$$
\dot{p}_1 \{ L_2 L_3 + L_1 L_3 + L_1 L_2
$$

$$
- [M_2 - \epsilon p_2 s_{21}][M_1 - \epsilon p_1 s_{12}][1 + \tfrac{\kappa}{4} L_3^* |z_{12}|] e^{\frac{\kappa}{4} s_{12}[(L_1 + \mathfrak{M}_{12}) z_{13} - (L_2 + \mathfrak{M}_{21}) z_{23}]}
$$

$$
- [M_3 - \epsilon p_3 s_{31}][M_1 - \epsilon p_1 s_{13}][1 + \tfrac{\kappa}{4} L_2^* |z_{13}|] e^{\frac{\kappa}{4} s_{13}[(L_1 + \mathfrak{M}_{13}) z_{12} + (L_3 + \mathfrak{M}_{32}) z_{23}]}
$$

$$
- [M_2 - \epsilon p_2 s_{23}][M_3 - \epsilon p_3 s_{32}][1 + \tfrac{\kappa}{4} L_1^* |z_{23}|] e^{\kappa/4 s_{23}[(L_3 + M_3 - \epsilon p_3 s_{32}) z_{13} - (L_2 + M_2 - \epsilon p_2 s_{23}) z_{12}]} \}
$$

$$
= [M_2 - \epsilon p_2 s_{21}][M_1 - \epsilon p_1 s_{12}][\tfrac{\kappa}{4} s_{12} L_3^* [H + \epsilon (p_2 - p_1) s_{12} + \epsilon p_3 s_{13}]]
$$
$$
\times e^{\frac{\kappa}{4} s_{12}[(L_1 + \mathfrak{M}_{12}) z_{13} - (L_2 + \mathfrak{M}_{21}) z_{23}]}
$$
$$
+ [M_3 - \epsilon p_3 s_{31}][M_1 - \epsilon p_1 s_{13}][\tfrac{\kappa}{4} s_{13} L_2^* [H + \epsilon p_2 s_{12} + \epsilon (p_3 - p_1) s_{13}]]
$$
$$
\times e^{\frac{\kappa}{4} s_{13}[(L_1 + \mathfrak{M}_{23}) z_{12} + (L_3 + \mathfrak{M}_{32}) z_{23}]}
$$
$$
+ [M_2 - \epsilon p_2 s_{23}][M_3 - \epsilon p_3 s_{32}][\tfrac{\kappa}{4} s_{23} L_1^* p_1 (s_{12} - s_{13})] e^{\frac{\kappa}{4} s_{23}[(L_3 + \mathfrak{M}_{31}) z_{13} - (L_2 + \mathfrak{M}_{13}) z_{12}]}
\tag{157}
$$

for $\dot{p}_1$, again with similar expressions for $\dot{p}_2$ and $\dot{p}_3$.

### 6.2. Effective Potential

The determining Equation (152) indicates that the Hamiltonian is a function of only four independent variables: the two separations between the particles and their conjugate momenta. Writing

$$
z_1 - z_2 = \sqrt{2} \rho \tag{158}
$$

$$
z_1 + z_2 - 2 z_3 = \sqrt{6} \lambda \tag{159}
$$

$$
z_1 + z_2 + z_3 = Z \tag{160}
$$

in turn implies the conjugate momenta

$$
p_\rho = \frac{1}{\sqrt{2}} (p_1 - p_2) \tag{161}
$$

$$
p_\lambda = \frac{1}{\sqrt{6}} (p_1 + p_2 - 2 p_3) \tag{162}
$$

$$
p_Z = \frac{1}{3} (p_1 + p_2 + p_3) \tag{163}
$$

obtained by imposing the requirement

$$
\{ q_A, p_B \} = \delta_{AB} \tag{164}
$$

where A, B $= \rho, \lambda, Z$.

The coordinates $\rho$ and $\lambda$ describe the motions of the three particles about their centre of mass. The remaining conjugate pair $(Z, p_Z)$, respectively, correspond to the centre of mass and its conjugate momenta in the non-relativistic limit. The Hamiltonian is independent of these variables and so $p_Z$ can be set to zero without loss of generality; similarly, the origin can be chosen to be $Z = 0$.

The Hamiltonian can be therefore be regarded as a function $H = H(\rho, \lambda, p_\rho, p_\lambda)$ of the four canonical degrees of freedom. The determining equation can be rewritten as

$$
L_+ L_- L_0 = \mathfrak{M}_{+-} \mathfrak{M}_{-+} L_0^* e^{\frac{\kappa}{4} s_0 H_0} + \mathfrak{M}_{-0} \mathfrak{M}_{01} L_+^* e^{\frac{\kappa}{4} s_- H_-} + \mathfrak{M}_{0+} \mathfrak{M}_{+0} L_-^* e^{\frac{\kappa}{4} s_+ H_+} \tag{165}
$$

where

$$
\begin{aligned}
H_0 &\equiv (L_1 + \mathfrak{M}_{12})z_{13} - (L_2 + \mathfrak{M}_{21})z_{23} \\
&= \sqrt{2}H\rho - \epsilon\left(2|\rho|p_\rho + \left[\left|\lambda + \frac{\rho}{\sqrt{3}}\right| - \left|\lambda - \frac{\rho}{\sqrt{3}}\right|\right]p_\lambda\right)
\end{aligned}
\tag{166}
$$

$$
\begin{aligned}
H_- &\equiv (L_2 + \mathfrak{M}_{23})z_{21} - (L_3 + \mathfrak{M}_{32})z_{31} \tag{167}\\
&= H\left(\lambda - \frac{\rho}{\sqrt{3}}\right) - \epsilon\left(\left[|\rho| - \frac{\sqrt{3}}{2}\left|\lambda + \frac{\rho}{\sqrt{3}}\right|\right]\left(\frac{p_\lambda}{\sqrt{3}} + p_\rho\right) + \frac{3}{2}\left(p_\lambda - \frac{p_\rho}{\sqrt{3}}\right)\left|\lambda - \frac{\rho}{\sqrt{3}}\right|\right)
\end{aligned}
$$

$$
\begin{aligned}
H_+ &\equiv (L_3 + \mathfrak{M}_{31})z_{32} - (L_1 + \mathfrak{M}_{13})z_{12} \tag{168}\\
&= -H\left(\lambda + \frac{\rho}{\sqrt{3}}\right) - \epsilon\left(\left[|\rho| - \frac{\sqrt{3}}{2}\left|\lambda - \frac{\rho}{\sqrt{3}}\right|\right]\left(\frac{p_\lambda}{\sqrt{3}} - p_\rho\right) - \frac{3}{2}\left(p_\lambda + \frac{p_\rho}{\sqrt{3}}\right)\left|\lambda + \frac{\rho}{\sqrt{3}}\right|\right)
\end{aligned}
$$

and

$$
M_0 = \sqrt{\frac{2}{3}}\sqrt{p_\lambda^2 + m_0^2} \qquad M_\pm = \frac{1}{\sqrt{2}}\sqrt{\left(\frac{p_\lambda}{\sqrt{3}} \pm p_\rho\right)^2 + 2m_\pm^2}
\tag{169}
$$

$$
L_\pm = H - M_\pm \pm \frac{\epsilon}{\sqrt{2}}\left[\left(\frac{p_\lambda}{\sqrt{3}} \mp p_\rho\right)s_0 \pm \frac{2}{\sqrt{3}}p_\lambda s_\pm\right]
\tag{170}
$$

$$
L_0 = H - M_0 - \frac{\epsilon}{\sqrt{2}}\left[\left(\frac{p_\lambda}{\sqrt{3}} + p_\rho\right)s_+ + \left(\frac{p_\lambda}{\sqrt{3}} - p_\rho\right)s_-\right]
\tag{171}
$$

$$
\mathfrak{M}_{\pm\mp} = M_\pm \mp \frac{\epsilon}{\sqrt{2}}\left(\frac{p_\lambda}{\sqrt{3}} \pm p_\rho\right)s_0, \quad \mathfrak{M}_{\pm 0} = M_\pm - \frac{\epsilon}{\sqrt{2}}\left(\frac{p_\lambda}{\sqrt{3}} \pm p_\rho\right)s_\pm
\tag{172}
$$

$$
\mathfrak{M}_{0\pm} = M_o + \frac{\epsilon\sqrt{2}}{\sqrt{3}}p_\lambda s_\pm \quad L_0^* = (1 - s_+s_-)M_0 + L_0
\tag{173}
$$

$$
L_\pm^* = (1 \mp s_0 s_\pm)M_\pm + L_\pm
\tag{174}
$$

It is instructive to carry out an expansion of (165) in the inverse powers of the speed of light $c$, which is the post-Newtonian (pN) expansion. This gives

$$
\begin{aligned}
H_{pN} &= 3mc^2 + \frac{p_\rho^2 + p_\lambda^2}{2m} + \frac{\kappa m^2 c^4}{\sqrt{8}}\left[|\rho| + \frac{\sqrt{3}}{2}\left(\left|\lambda + \frac{\rho}{\sqrt{3}}\right| + \left|\lambda - \frac{\rho}{\sqrt{3}}\right|\right)\right] - \frac{(p_\rho^2 + p_\lambda^2)^2}{16m^3c^2} + \frac{\kappa c^2}{\sqrt{8}}|\rho|p_\rho^2 \\
&\quad + \frac{\kappa c^2}{16\sqrt{2}}\left[3\left(\left|\lambda + \frac{\rho}{\sqrt{3}}\right| + \left|\lambda - \frac{\rho}{\sqrt{3}}\right|\right)\left(\sqrt{3}p_\lambda^2 + p_\rho^2\right) + 6\left(\left|\lambda + \frac{\rho}{\sqrt{3}}\right| - \left|\lambda - \frac{\rho}{\sqrt{3}}\right|\right)p_\rho p_\lambda\right] \\
&\quad + \frac{\kappa^2 m^3 c^6}{16}\left[\frac{|\rho|\sqrt{3}}{2}\left(\left|\lambda + \frac{\rho}{\sqrt{3}}\right| + \left|\lambda - \frac{\rho}{\sqrt{3}}\right|\right) + \frac{3}{4}\left|\lambda + \frac{\rho}{\sqrt{3}}\right|\left|\lambda - \frac{\rho}{\sqrt{3}}\right| - \frac{3}{4}\left(\lambda^2 + \rho^2\right)\right]
\end{aligned}
\tag{175}
$$

where the factors of the speed of light $c$ have been restored (recall $\kappa = \frac{8\pi G}{c^4}$). To leading order in $1/c$

$$
H = H_N \equiv 3mc^2 + \frac{p_\rho^2 + p_\lambda^2}{2m} + \frac{\kappa m^2 c^4}{\sqrt{8}}\left[|\rho| + \frac{\sqrt{3}}{2}\left(\left|\lambda + \frac{\rho}{\sqrt{3}}\right| + \left|\lambda - \frac{\rho}{\sqrt{3}}\right|\right)\right] + \cdots
\tag{176}
$$

which is the non-relativstic hexagonal-well Hamiltonian of a single particle [15] plus the total rest mass of the system. This latter quantity, while non-relativistically irrelevant, is useful to retain in order to straightforwardly compare the energies and motions and energies of the relativistic (R) and non-relativistic (N) systems.

The Hamiltonian (176) describes the non-relativistic motion of a single particle of mass $m$ (referred to as the hex-particle) in a linearly increasing potential well in the $(\rho, \lambda)$ plane, whose cross-sectional shape is that of a regular hexagon. To extend this to the pN and R cases, the potential can be defined by the relation $V(\rho, \lambda) = H(p_\rho = 0, p_\lambda = 0)$. From (165), we then obtain

$$\left(V_{\mathrm{R}} - m_1 c^2\right)\left(V_{\mathrm{R}} - m_2 c^2\right)\left(V_{\mathrm{R}} - m_3 c^2\right)$$

$$= \left(V_{\mathrm{R}} - s_+ s_- m_3 c^2\right) m_1 m_2 c^4 \exp\left[\frac{\sqrt{2}\kappa R}{4} V_{\mathrm{R}} |\sin\theta|\right]$$

$$+ \left(V_{\mathrm{R}} - s_\rho s_+ m_1 c^2\right) m_2 m_3 c^4 \exp\left[\frac{\sqrt{2}\kappa R}{4} V_{\mathrm{R}} \left|\sin\left(\theta - \frac{\pi}{3}\right)\right|\right]$$

$$+ \left(V_{\mathrm{R}} + s_\rho s_- m_2 c^2\right) m_3 m_1 c^4 \exp\left[\frac{\sqrt{2}\kappa R}{4} V_{\mathrm{R}} \left|\sin\left(\theta + \frac{\pi}{3}\right)\right|\right] \quad (177)$$

for the R potential, where

$$\rho = R\sin\theta \qquad \lambda = R\cos\theta \tag{178}$$

has been used to render the hexagonal symmetry manifest, and $s_\pm = \mathrm{sgn}\left(\sqrt{3}\lambda \pm \rho\right)$, $s_\rho = \mathrm{sgn}(\rho)$. The corresponding pN potential is

$$\tilde{V}_{pN} = (m_1 + m_2 + m_3)c^2 + \frac{\kappa c^4}{4\sqrt{2}}\left(2m_1 m_2 |\tilde{\rho}| + m_1 m_3 \left|\sqrt{3}\tilde{\lambda} + \tilde{\rho}\right| + m_2 m_3 \left|\sqrt{3}\tilde{\lambda} - \tilde{\rho}\right|\right)$$

$$+ \frac{1}{2c^2}\left(\frac{\kappa c^4}{4}\right)^2 m_1 m_2 m_3 \left((1 - \tilde{s}_\rho \tilde{s}_1)|\tilde{\rho}|\left|\sqrt{3}\tilde{\lambda} + \tilde{\rho}\right| + (1 + \tilde{s}_\rho \tilde{s}_2)|\tilde{\rho}|\left|\sqrt{3}\tilde{\lambda} - \tilde{\rho}\right|\right.$$

$$\left. + \frac{1}{2}(1 - \tilde{s}_1 \tilde{s}_2)\left|\sqrt{3}\tilde{\lambda} + \tilde{\rho}\right|\left|\sqrt{3}\tilde{\lambda} - \tilde{\rho}\right|\right) \tag{179}$$

where $\tilde{\rho}$ and $\tilde{\lambda}$ are defined as in (158) and (159) using the $\tilde{z}_a$ coordinates of (29). As $c \to \infty$, $\kappa \to 0$, and we recover

$$V_N = (m_1 + m_2 + m_3)c^2 + \frac{2\pi G}{\sqrt{2}}\left(2m_1 m_2 |\rho| + m_1 m_3 \left|\sqrt{3}\lambda + \rho\right| + m_2 m_3 \left|\sqrt{3}\lambda - \rho\right|\right) \tag{180}$$

which is the hexagonal well potential of the N-system.

A comparison of $V_N$ and $V_R$ is given in Figure 18. At very low energies, they are indistinguishable, but striking differences emerge with increasing energy. For all energies, equipotential lines of $V_N$ form the shape of a regular hexagon in the $(\rho, \lambda)$ plane, with the sides rising linearly in all directions, forming the hexagonal-well potential noted earlier. The post-Newtonian potential $V_{pN}$ retains the hexagonal symmetry, but distorts the sides to be parabolically concave. The relativistic potential $V_R$ also retains the hexagonal symmetry, but the sides of the hexagon become convex, even at energies only slightly larger than the rest mass. The overall size of the hexagon at a given value of $V_R$ is considerably smaller since its growth is extremely rapid compared to the other two cases. Its cross-sectional size reaches a maximum at $V_R = V_{Rc} = 6.711968022mc^2$, after which it decreases in diameter like $\ln(V_R)/V_R$ with an increasing $V_R$. The relativistic potential is therefore an hexagonal carafe, whose neck narrows as $V_R$ increases. The part of the potential for which $V_R > V_{Rc}$ is in an intrinsically non-perturbative relativistic regime: the motion for values of $V_R$ larger than this cannot be understood as a perturbation from some classical limit of the motion. A comparison of the equipotential lines for each case in Figure 19 highlights these distinctions.

Furthermore, since there are couplings between the momentum and position of the hex-particle, the potential does not fully govern the motion in both the pN and R systems. In the pN system, there is a momentum-dependent steepening of the walls of the hexagon to leading-order in $1/c^2$.

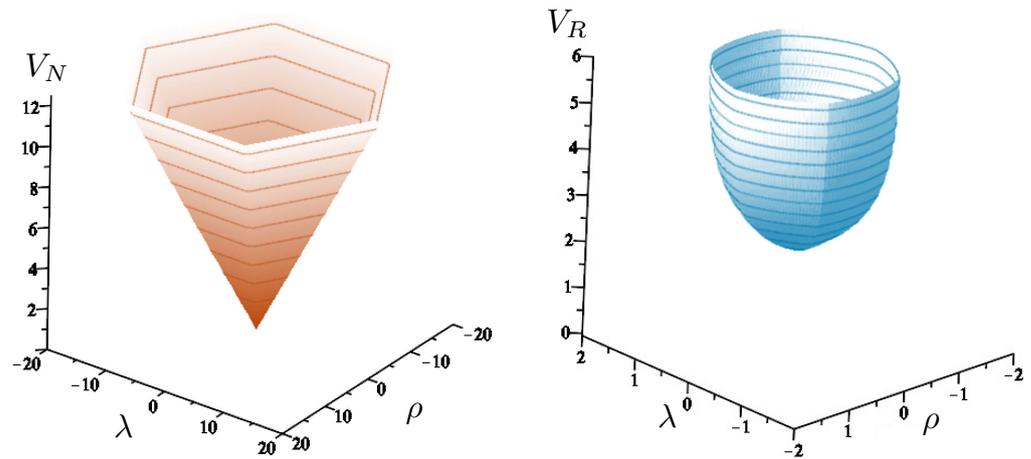

**Figure 18.** The shape of the non-relativistic potential (**left**) and relativistic potential (**right**) of the hex-particle in the equal mass case, using units in (181) and with potentials in units of $mc^2$.

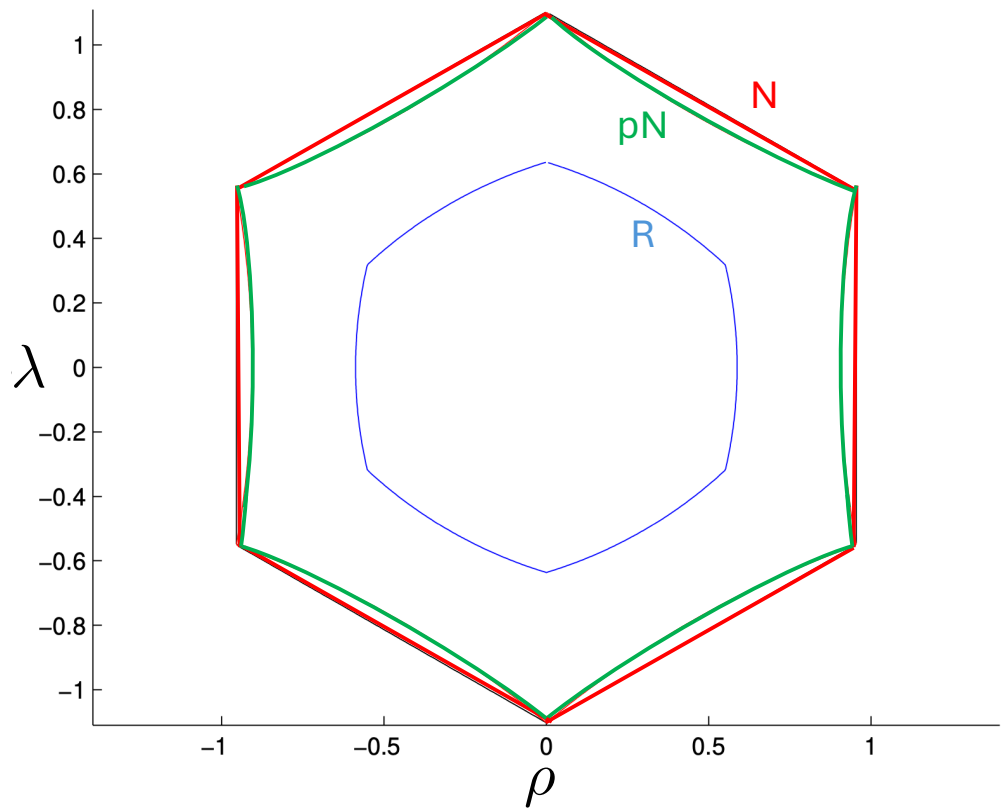

**Figure 19.** Equipotential lines at $V = 4mc^2$ for the non-relativistic potential (red) and post-Newtonian potential (green), and relativistic potential (blue) in the equal mass case.

### 6.3. Relativistic Equal Mass Three-Body Trajectories

An analysis of the 3-body system is best carried out by considering the motion of the hex-particle in the $(\rho, \lambda)$ plane. In the Hamiltonian formalism, the motion is given by two conjugate pairs of differential equations for $(\rho, p_\rho)$ and $(\lambda, p_\lambda)$ that are continuous everywhere except across the three hexagonal bisectors $\rho = 0$, $\rho - \sqrt{3}\lambda = 0$, and $\rho + \sqrt{3}\lambda = 0$. These bisectors, respectively, correspond to the crossings of particles 1 and 2, 2 and 3, or 1 and 3, and divide the hexagon into sextants.

The non-relativistic analogue of this system is that of a ball that elastically collides with a wedge whilst experiencing a constant gravitational force [3,17]. A bounce at one of the edges of the wedge corresponds to a crossing one of the three hexagonal bisectors, and a discrete mapping can be constructed that describes the particle's angular and radial velocities each time it collides with one of the edges. The systems are nearly identical in the equal mass case since a crossing between the two equal mass particles cannot be distinguished from an elastic collision between them.

The interesting dynamics of the system arises due to these crossings, or alternatively, due to the non-smoothness of the potential along the bisectors. Two types of motion can be distinguished [3]: *A*-motion, in which the hex-particle crosses a single bisector twice in succession (the same pair of particles cross twice in a row) and *B*-motion, corresponding to the hex-particle crossing two successive sextant boundaries (one particle crosses each of its compatriots in succession). Any given motion is characterized by a *symbol sequence*, a sequence of letters *A* and *B*, with a finite exponent *n* denoting *n*-repetitions and an overbar denoting an infinite repeated sequence. For example, the expression $A^2 B^3$ denotes three *A*-motions (two adjacent particles cross twice in a row twice in succession) followed by three *B*-motions (one particle crosses the other two in succession, which then cross each other). The expression $(A^n B^m)^p$ denotes the motion $A^n B^m$ repeated *p* times, and $\overline{(A^n B^m)}$ denotes infinitely many repeats of this motion. There is an ambiguity in classifying either the final or the initial crossing since whether or not a motion is of *A*-type or *B*-type is contingent upon the previous crossing; this ambiguity can be resolved by taking the initial crossing of any sequence of motions as being unlabeled.

Since the same initial conditions for the N, pN, and R systems do not yield the same conserved energy *H*, comparison of these cases necessitates a choice: one can either compare at fixed values of the energy (FE conditions), modifying the initial conditions as appropriate (as required by the conservation laws for each system), or else fix the initial conditions, comparing trajectories at differing values of *H* (fixed-momenta (FM) conditions). Numerically, it is useful to rescale

$$p_i = M_{tot} c \hat{p}_i \qquad z_i = \frac{4}{\kappa M_{tot} c^2} \hat{z}_i \tag{181}$$

in which case, the equations of motion become

$$\frac{\partial \eta}{\partial \hat{p}_i} = \frac{1}{c} \frac{\partial H}{\partial p_i} = \frac{4}{\kappa M_{tot} c^3} \frac{d \hat{z}_i}{dt} = \frac{d \hat{z}_i}{d \hat{t}} \tag{182}$$

$$\frac{\partial \eta}{\partial \hat{z}_i} = \left( \frac{4}{\kappa M_{tot}^2 c^4} \right) \frac{\partial H}{\partial z_i} = -\frac{4}{\kappa M_{tot} c^3} \frac{d \hat{p}_i}{dt} = -\frac{d \hat{p}_i}{d \hat{t}} \tag{183}$$

where $M_{tot} = 3m$ is the total mass of the system, $t = \frac{4}{\kappa M_{tot} c^3} \hat{t}$, and $\hat{z}_i$ and $\hat{p}_i$ are the respective dimensionless positions and momenta. The quantities $M_i$ and $L_i$ likewise rescale

$$M_i = M_{tot} c^2 \left( \sqrt{\hat{p}_i^2 + \hat{m}_i^2} + \hat{p}_i \right) = M_{tot} c^2 \hat{M}_i \tag{184}$$

$$L_i = M_{tot} c^2 \left( \eta + 1 - \sqrt{\hat{p}_i^2 + \hat{m}_i^2} - \epsilon (\sum_j \hat{p}_j s_{ji}) \right) = M_{tot} c^2 \hat{L}_i \tag{185}$$

where $\eta + 1 = H / M_{tot} c^2$, $\hat{m}_i = \left( \frac{m_i}{M_{tot}} \right)$.

One consideration in describing the motion is that the proper time (97) differs for each particle, even in the equal mass case. The simplest choice (but not the only one) is to work with the coordinate time *t*.

The plots in this section were obtained numerically [30,116], with a time step in the numerical code having a value $\hat{t} = 1$. Absolute and relative error tolerances of $\epsilon_{abs} = \epsilon_{rel} = 10^{-8}$ were imposed so that the estimated error in each of the dynamical variables $\rho(i), \lambda(i), p_\rho(i)$, and $p_\lambda(i)$ at each step *i* in the numerical integration is $\epsilon(i) \leq \max(\epsilon_{rel} |y(i)|, \epsilon_{abs})$.

### 6.3.1. General Features of the Motion

For each of the N, pN, and R systems, the motions fall into one of three principal classes: annuli, pretzel, and chaotic. Within each class, the orbits either (i) eventually densely cover the portion of $(\rho, \lambda)$ space they occupy, or (ii) do not. A symbol sequence consisting of a finite sequence repeated infinitely many times would be in case (ii) whereas all chaotic orbits (by definition) are in case (i). Quasi-regular orbits are also in case (i); for these, the symbol sequence consists of repeats of the same finite sequence, but with an $A$-motion added or removed at irregular intervals. In phase space, the two types of orbits are separated by separatrixes (trajectories joining a pair of hyperbolic fixed points). Regular orbits lie inside the 'elliptical' region surrounding an elliptical fixed point, whereas quasi-regular orbits lie outside such a region.

Quasi-periodic trajectories closely resemble their related periodic counterparts, except that the orbit does not exactly repeat itself. Consequently, a quasi-periodic orbit eventually densely covers some region of phase space despite its high degree of regularity, as manifested by its periodic symbol sequence. A particle moving on a torus $S^1 \times S^1$ is a textbook example. Its motion is characterized by its angular velocity around each copy of $S^1$; if the ratio of these is rational, the motion will be periodic, whereas the motion will be quasi-periodic if the ratio is irrational. For the three-body case, non-periodic orbits with fixed symbol sequences are quasi-periodic, and appear as a collection of closed circles, ovals, or crescents in the Poincaré section (discussed in the next subsection). Quasi-regular orbits have symbol sequences that are not fixed.

### 6.3.2. Annulus Orbits

Annulus orbits have the symbol sequence $\overline{B}$ and consist of an annulus encircling the origin in the $\rho - \lambda$ plane. In these orbits, the hex-particle never crosses the same bisector twice in succession.

Most annulus orbits are quasi-periodic and fill in the entire ring. However, a few repeat themselves after some number of rotations about the origin, and a wide variety of patterns are possible contingent upon the initial conditions for the N, pN, and R systems. No qualitative distinctions between N and pN annuli were observed within numerically attainable values of $\eta$ [29,30].

An example of annulus orbits for the N and R systems (for FE conditions), along with the positions of each of the three bodies as a function of time, is shown in Figure 20 (periodic) and Figure 21 (near-chaotic). Periodic orbits are numerically difficult to find, so the orbits in Figure 20 are actually very close to periodic orbits; this allows the pattern of the periodic orbit to be visible. At similar energies, the R hex-particle covers the $(\rho, \lambda)$ plane more densely than its N counterpart, and has a higher frequency of oscillation. The higher frequency for the R system was seen in the previous section for two bodies and appears to generally hold for the three-body system as well. The increased trajectory density for FE conditions in the R system consequently follows, since the same number of time steps were used for both.

These features of higher frequency and trajectory density are more apparent in Figure 22, which provides a comparison of the orbits using FM initial conditions. For these conditions, the R system (blue) has a slightly higher energy than its N counterpart (red). The N system is less dense, covers a smaller region of the $(\rho, \lambda)$ plane, and does not venture as close to the origin, characteristics that becoming increasingly pronounced for increasing $\eta$, provided that the R energy remains larger than its N counterpart. This is not guaranteed for FM conditions, as the bottom diagram in Figure 22 illustrates: here, the N system has about 14% more energy than its R counterpart, and so covers a larger region of the $(\rho, \lambda)$ plane.

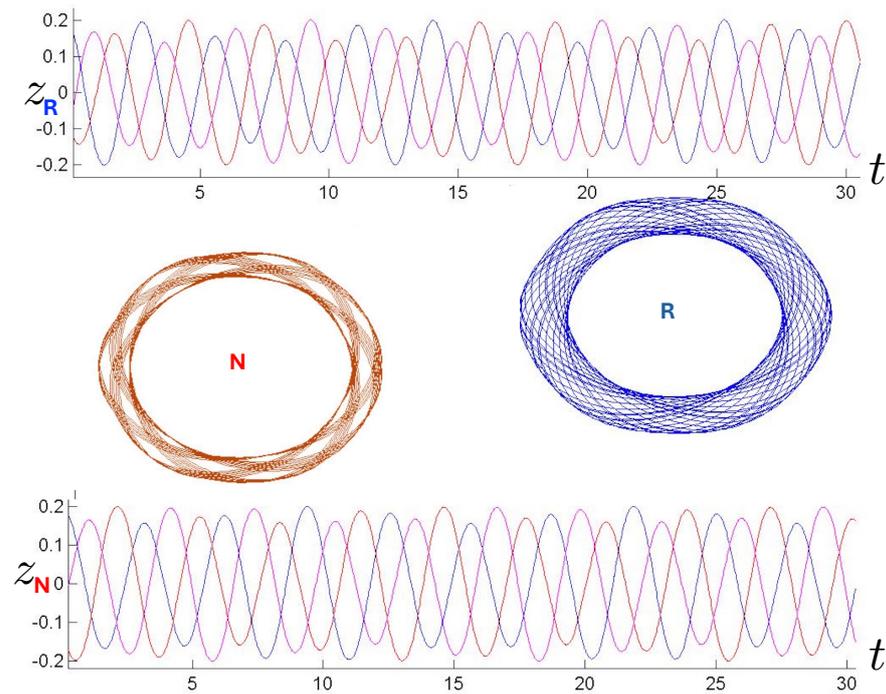

**Figure 20.** Annulus orbits (N-red, R-blue) shown in conjunction with their corresponding 3-particle trajectories $z(t)$ (blue, red, magenta) for 30 time steps (**top**: relativistic, **bottom**: non-relativistic). The quasi-regular annulus orbits are for the FE initial conditions with $\eta = 1.1$ and run for 200 time steps. They are far from being chaotic. The R motion is further from periodicity, leaving far fewer open regions in the $(\rho, \lambda)$ plane.

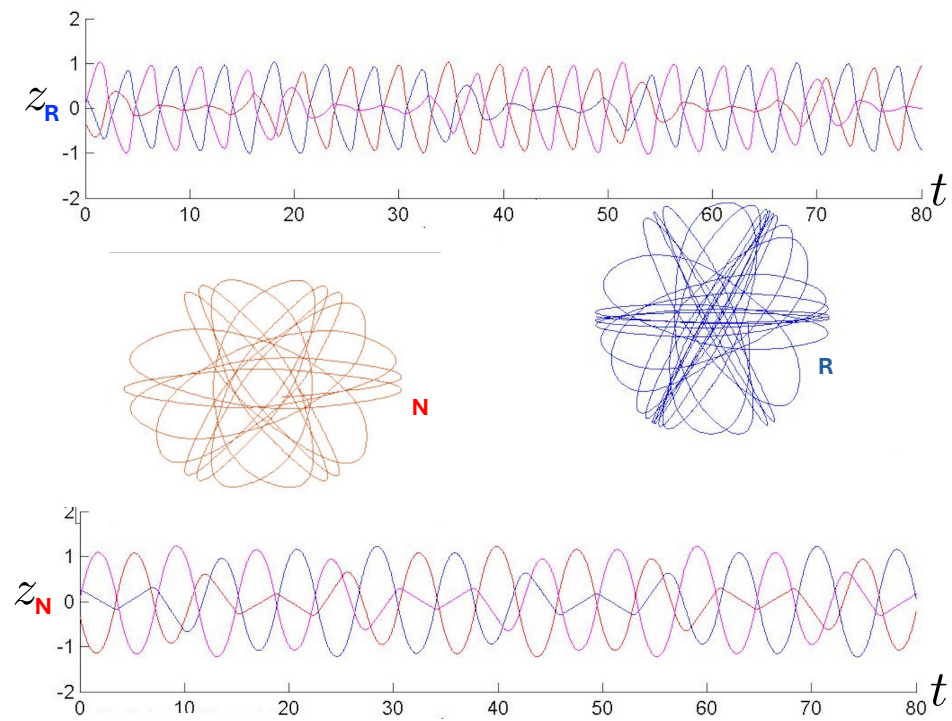

**Figure 21.** Near-chaotic annulus orbits (N-red; R-blue) shown in conjunction with their corresponding three-particle trajectories $z(t)$ (blue, red, magenta) for 80 time steps (**top**: relativistic, **bottom**: non-relativistic). These near-chaotic orbits were run for 200 time-steps using FE initial conditions with $\eta = 1.5$. The R trajectory is closer to chaos than the N trajectory.

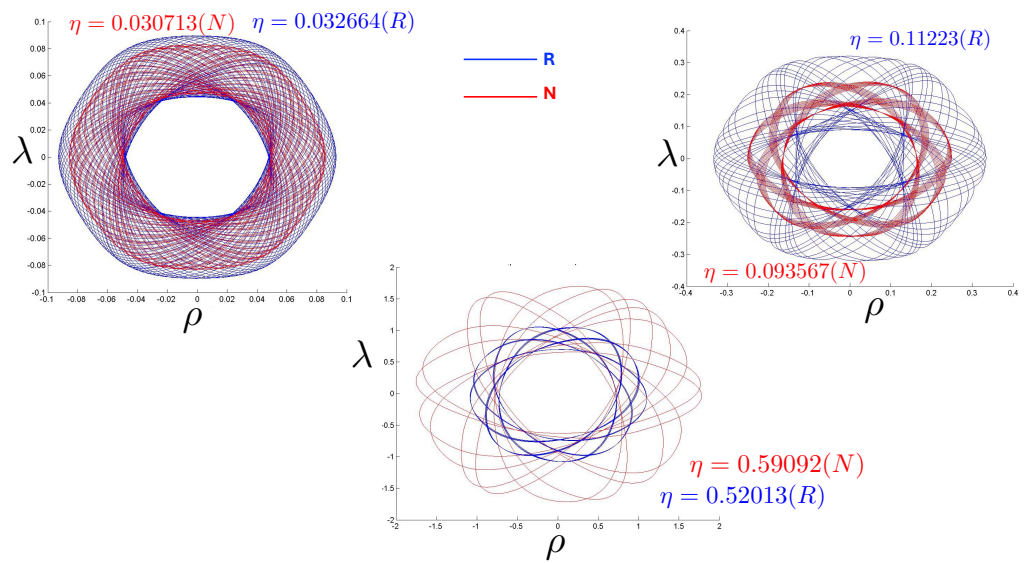

**Figure 22.** A comparison of the annulus orbits at identical FM conditions, for three similar values of $\eta$, for all 200 time steps. N trajectories (red) typically have less energy than R trajectories (blue) and so cover a smaller region of the $(\rho, \lambda)$ plane. However, for some initial conditions, the N system has a larger energy (bottom figure) and so covers a correspondingly larger region.

### 6.3.3. Pretzel Orbits

Pretzel orbits have the generic symbol sequence $\prod_{ijk}\left(A^{n_i}B^{3m_j}\right)^{l_k}$, where $n_i, m_j, l_k \in Z^+$, with some $l_k$ possibly infinite, and consist of orbits in which the hex-particle essentially oscillates back and forth about one of the three bisectors for some segments of its motion. A typical example is shown in Figure 23. For both the N and R systems, we see that two of the three bodies form a stable bound subsystem, which in turn orbits the third analogous to a 2-body system. The N system exhibits parabolic regularity for both the two-body subsystem and the full system, whereas the R system has shoulder-like distortions observed previously in the two-body case.

This formation of a stable (or quasi-stable) bound subsystem is characteristic of pretzel orbits, and the range of possible trajectories is extremely diverse. Many families of regular orbits exist. These generally have one base element in their symbol sequence (e.g., $AB^3$) and a sequence of elements formed by appending an $A$ to each existing sequence of $A$'s (for example, $\{AB^3, A^2B^3, A^3B^3, \ldots\}$. The $B^3$ sequence corresponds to a 180-degree rotation of the hex-particle about the origin, yielding a broad variety of twisted, pretzel-like figures. This is a notable distinction from the wedge system [3,17], for which $B$ and $B^2$ sequences are also observed; only $B^3$ sequences are present in all pretzel orbits.

Distinctions between the R, pN, and N systems are the strongest for pretzel orbits. Both regular orbits (with an infinitely repeating symbol sequence) and irregular orbits that densely fill a cylindrical tube in the $(\rho, \lambda)$ plane occur. Orbits in the R system generally have kinks about the $\lambda = 0$ line that are absent in their N and pN counterparts; a cylindrical-shaped trajectory in the N system looks like an hourglass in the R system, for example. Furthermore, periodic and quasi-periodic orbits in the N system have counterparts with the same symbol sequence in the R system but not in the pN system, which exhibits chaotic behaviour not seen in the N and R systems.

A comparison of the time–evolution of trajectories in the N and R systems is shown in Figure 24 for FE conditions at small and large values of $\eta$. For small $\eta$ ($\eta = 0.05$), there is very little distinction between the N and R motions, consistent with the smooth non-relativistic limit of (152). Significantly different trajectories occur for larger values $\eta$ ($\eta = 0.85$). In the R system, the oscillation frequency is higher and the pattern 'weaves' relative to the near-cylindrical shape in the N-system once enough time steps occurred.

The R trajectory is more tightly confined, commensurate with the 2-body motion seen in the previous section.

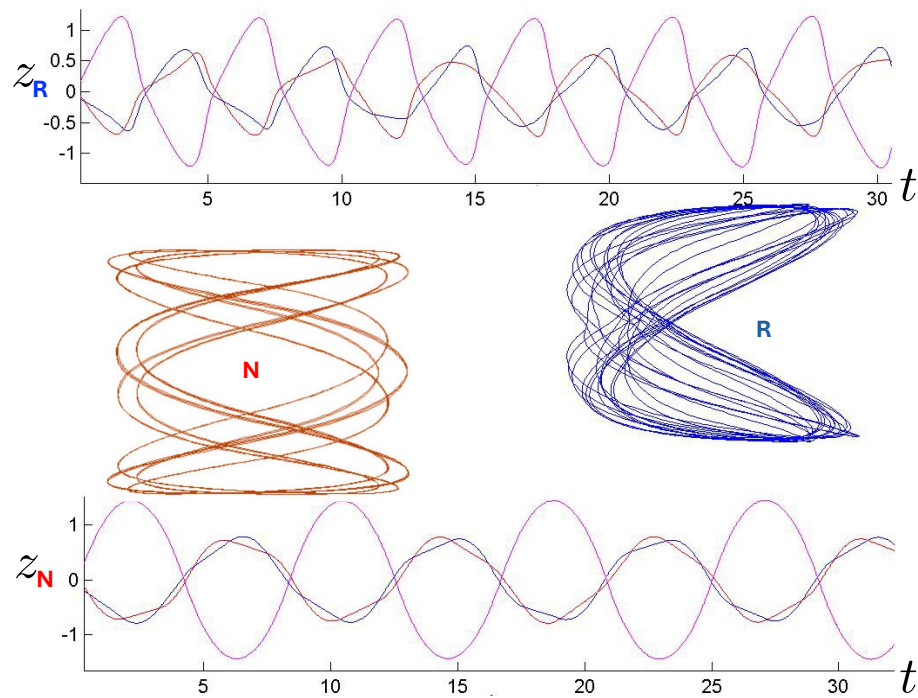

**Figure 23.** Regular pretzel orbits for FE conditions for the R (blue) and N (red) systems, each run for 120 time steps, with their corresponding 3-particle trajectories $z(t)$ (blue, red, magenta) truncated at 35 time steps (**top**: relativistic, **bottom**: non-relativistic). The collision sequences are $AB^3$ (R) and $A^2B^3$ (N), and differ due to FE initial conditions.

In Figure 25, we display the sensitivity of trajectories to initial conditions. The fish-like diagram has an $AB^6$ symbol sequence: two of the particles oscillate quasi-regularly about each other (shown in the upper $z(t)$ plot), with this pair undergoing larger-amplitude and lower-frequency oscillations with the third. A slight change in the initial FE conditions yields the strudel-like figure on the left. Now, one particle alternates its oscillations with the other two, maintaining a near-constant amplitude throughout.

By controlling the FM initial conditions, interesting sequences of hex-particle orbits can be obtained. An example is given in Figure 26, which compares snake-like orbits in the N (red) and R (blue) systems. These quasi-regular orbits have symbol sequences $\left\{ \overline{A^i B^3} \right\}$. In both systems, the orbits have two sharp turning points separated by some number $n$ of bumps. In the N system, these have been shown to exist for arbitrary $n$ [3], and it was conjectured that the same is true for the R and pN systems [30]. The figures in the R system develop an hourglass shape narrowing about $\lambda = 0$, and cover a much narrower region in the $\rho$ direction (note the scale in the bottom sequence of plots). The N orbits, by contrast, are circumscribed by a cylinder.

The symbol sequence $AB^3$ results in boomerang-like figures, shown in Figure 27, where the qualitatively different physics due to relativistic effects is manifest. At low energies ($\eta = 0.2$), the N (red) and R (blue) systems have similar boomerang shapes. However, as $\eta$ increases, orbits in the R system develop two distinct turning points at different distances from the $\rho = 0$ axis for $\lambda > 0$, with symmetric counterparts for $\lambda < 0$. This feature is particularly evident for $\eta = 0.75$. A kink at the right-hand-side of the boomerang emerges, becoming increasingly pronounced with an increasing $\eta$. The underlying reason behind the development of this structure is not clear.

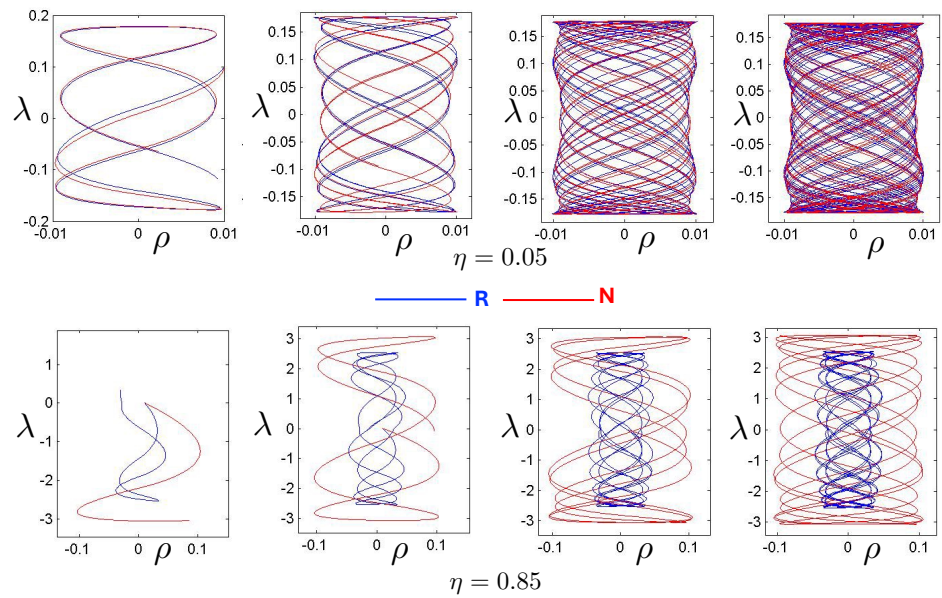

$\eta = 0.05$

$\eta = 0.85$

**Figure 24.** Time evolution of the hex-particle for a pretzel orbit, shown simultaneously in the N (red) and R (blue) systems at t = 3, 11, 25, and 35 time steps (moving left to right on both rows) at FE conditions. For low energies ($\eta = 0.05$, **top**), the trajectories in the two systems are very similar, but at high energies ($\eta = 0.85$, **bottom**), they differ significantly. In the latter case, the R orbit evolves with a higher collision frequency and stabilizes into a quasi-periodic cylindrical pattern. In contrast to this, the N trajectory extends considerably further from the origin and will eventually form a densely filled cylinder.

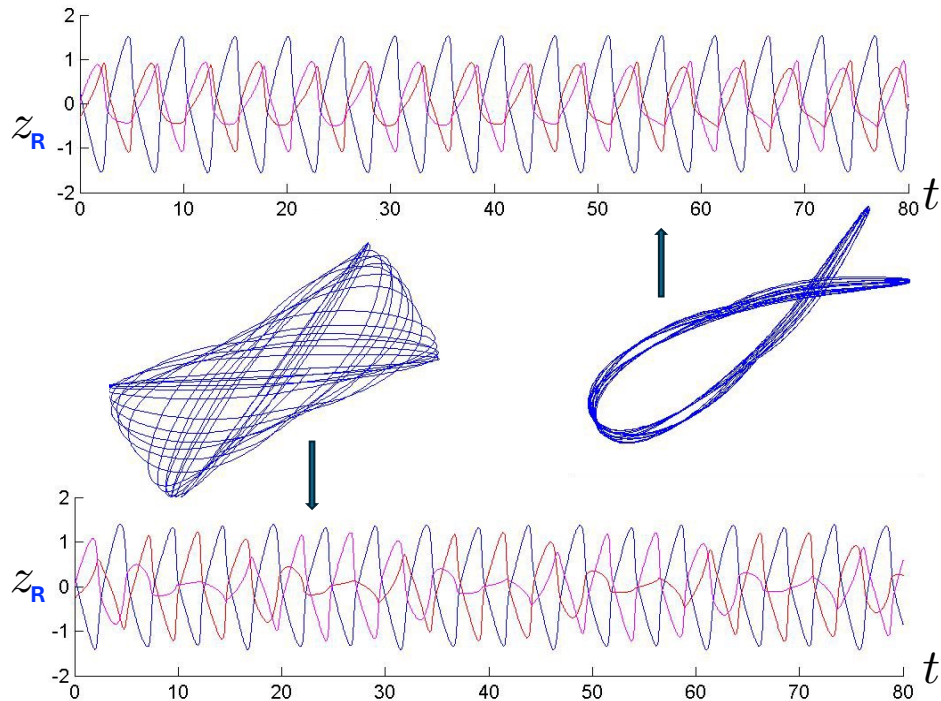

**Figure 25.** A comparison of pretzel orbits of the relativistic system for slightly different FE conditions, each run for 200 time steps, with their corresponding 3-particle trajectories $z(t)$ (blue, red, magenta) truncated at 80 time steps. A regular $AB^6$ orbit pattern (**top**) yields the fish-like structure at the right, whereas slightly different initial conditions (**bottom**) result in the Studel-like figure at the left. Here, two particles are in a large-amplitude bound state, with the particle undergoing lower-amplitude irregular oscillations with this pair.

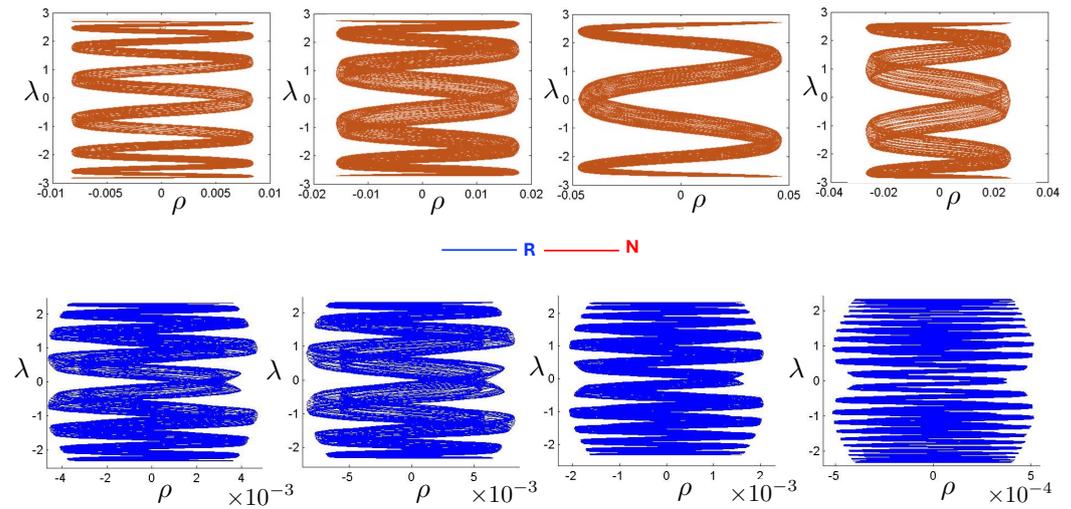

**Figure 26.** A comparison of the quasi-regular snake-like orbits for the N (red) and R (blue) systems, run for 200 time steps. These orbits have the symbol sequence $A^m B^3$ for $m$ odd, and in both systems, each trajectory has two sharp turning points separated by some number $n$ of bumps. The value of $n$ increases with decreasing initial angular momentum in the $(\rho, \lambda)$ plane. For the N system, FM initial conditions were used, with the square (barely visible near the top of each figure) indicating the starting point. In the R system, FE initial conditions were used with $\eta = 0.75$. The R orbits have a narrow hourglass shape, whereas the N orbits in the upper row lie in a cylinder notably larger in size in the $\rho$ direction.

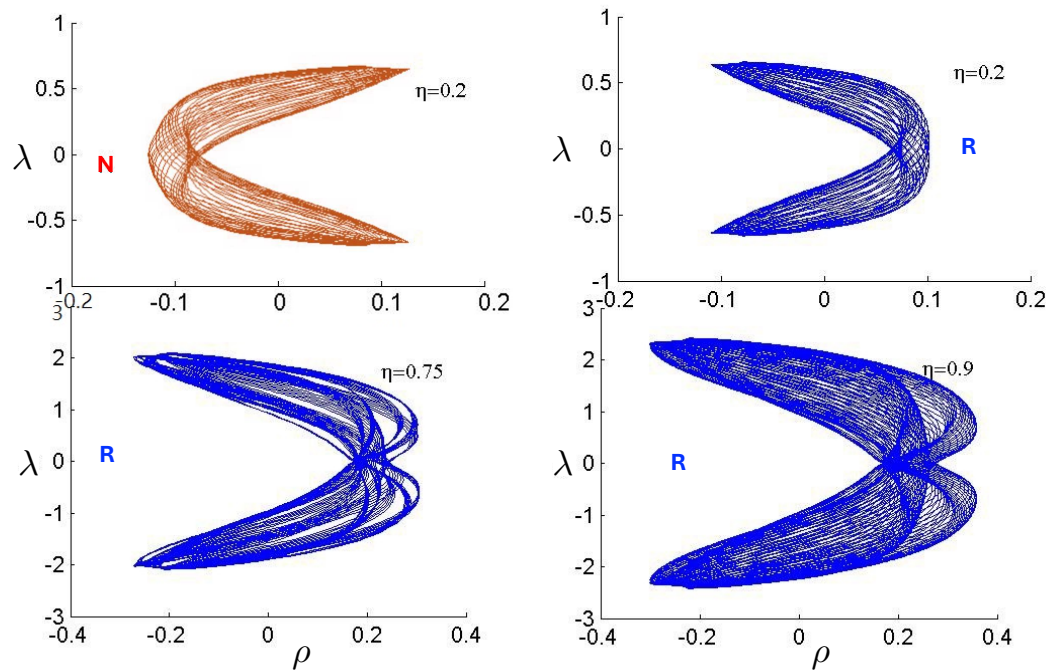

**Figure 27.** A comparison of orbits with the symbol sequence $AB^3$ for 200 time steps with FM initial conditions. The N system (red) is shown at the upper left and the R system (blue) in the remaining three plots for different values of $\eta$. As $\eta$ increases, the R trajectories develop a kink along the $\lambda = 0$ axis, displaying a double-banding pattern with two turning points at two distinct distances from the $\rho$ axis about $\lambda = 0$.

### 6.3.4. Chaotic Orbits

Chaotic orbits are those for which the hex-particle wanders between $A$-motions and $B$-motions in an apparently irregular fashion. Unlike the annuli and pretzel orbits, chaotic orbits eventually wander into all areas of the $\rho - \lambda$ plane. Chaos emerges at the transition between annulus and pretzel orbits, where the hex-particle passes very close to the origin, for each system.

Figure 28 illustrates a typical case for the N and R systems. Two particles undergo a large-amplitude oscillation with the third one (the middle 'm' particle) mildly oscillating near the centre of momentum. At irregular intervals, this third particle switches places with one of the other two, and the pattern repeats. The m-particle alternates in an irregular fashion, leading to chaos.

A comparison of the time development of chaotic trajectories in both the N and R systems is shown in Figure 29. The upper sequence shows how a chaotic trajectory can develop in the R system (blue) whilst the N system (red) forms a densely filled annulus for the same FE initial conditions with $\eta = 0.5$. The lower sequence shows how a chaotic trajectory can develop in the N system (red), whilst the R system (blue) forms a densely filled pretzel, for the same FE initial conditions, again for $\eta = 0.5$. In both cases, the R trajectory attains its final pattern much more rapidly than its N counterpart, a manifestation of the difference in frequencies noted earlier.

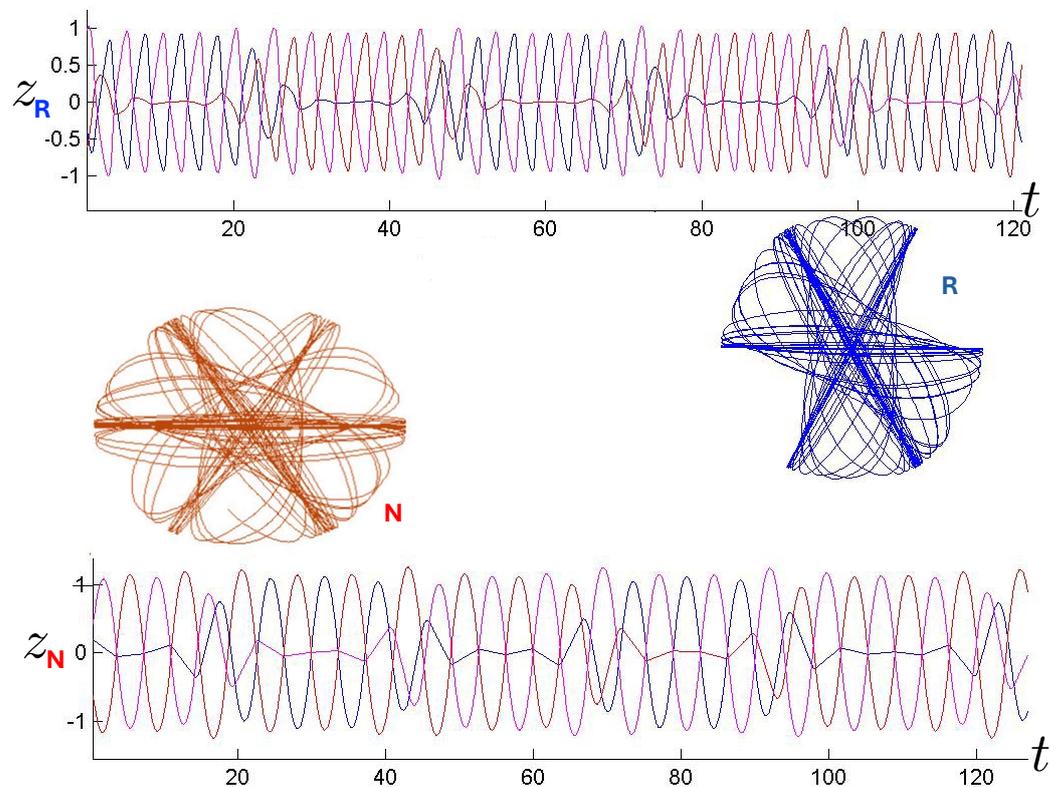

**Figure 28.** A comparison of chaotic orbits for the N (red) and R (blue) systems in the region of the phase space separating annulus and pretzel trajectories for 300 time steps, with their corresponding three-particle trajectories $z(t)$ (blue, red, magenta) truncated at 120 time steps. FE initial conditions were used, but with different initial values of $(\rho, \lambda, p_\rho)$ for each system. Most of the time, the middle ('m') particle remains nearly motionless between the other two particles, which oscillate about the centre of inertia with large amplitude. However, slight irregularities between the number of crossings for which one particle remains almost stationary result in the identity of the m-particle perpetually changing its identity, leading to chaos.

The transition from an annulus to a pretzel orbit through a chaotic region for the R system is shown in Figure 30. Proceeding from from left-to-right and top-to-bottom with decreasing initial angular momentum for the hex-particle, the system begins as an annulus, passes through a set of chaotic orbits, and then becomes a pretzel. The chaotic trajectories through the origin (or very close to it), a characteristic feature for this region of chaos in all three systems. In the R system, the transitional region shrinks as $\eta$ increases [30].

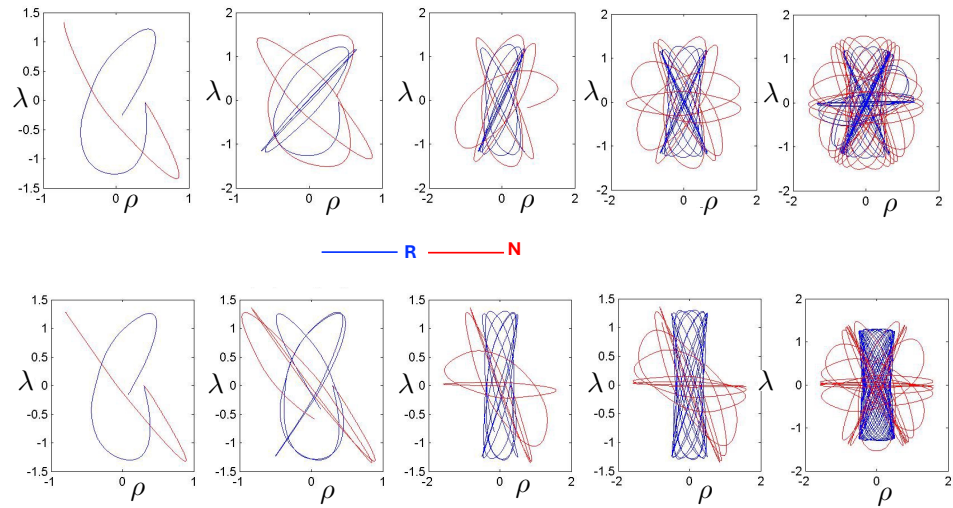

**Figure 29.** A comparison of the time-development of N (red) and R (blue) trajectories at $\eta = 0.5$ for FE initial conditions, as shown (in both rows), from left to right at $t = 5, 15, 30, 80$, and 110 units. For one set of identical FE conditions, the R trajectory is chaotic whereas its N counterpart forms a densely filled annulus (**top row**). For a different set of identical FE conditions (**bottom row**, with the same $\eta$), the N trajectory is chaotic whereas its R counterpart forms a densely filled cylinder in the pretzel class.

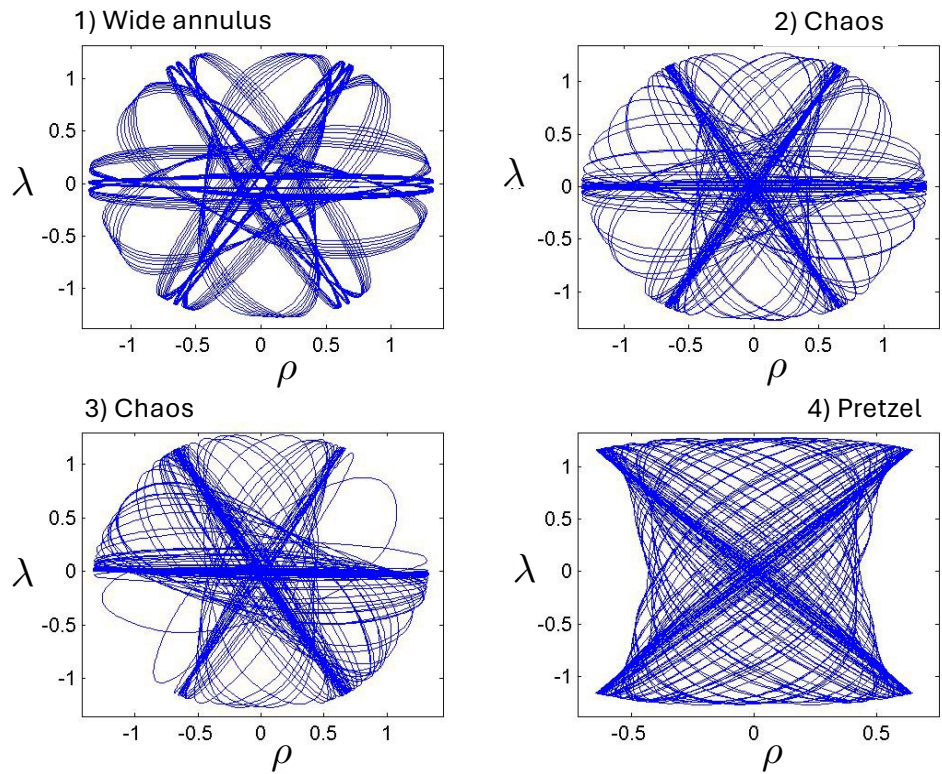

**Figure 30.** Transition in the R system from an annulus to a pretzel orbit through a chaotic region, with $\eta = 0.5$. The initial angular momentum in the $(\rho, \lambda)$ plane decreases from the upper left panel to the lower right one. Each plot is for FE initial conditions with 450 time steps. The chaotic trajectories pass very close to or through the origin.

There is a striking distinction between the pN system and the (N,R) systems for the chaotic class of motions. Unlike the latter two, the pN exhibits an additional area of chaos in the pretzel region. This feature will become evident in the next subsection.

### 6.4. Poincaré Plots

Poincaré sections for each of the N, pN, and R systems can be constructed by plotting the square of the angular momentum $p_\theta^2$, against the radial momentum $p_R$ (both scaled to be dimensionless as per (181)) of the hex-particle each time it crosses a bisector. Since all bisectors are equivalent in the equal mass case, all crossings can be plotted on the same surface of section. The plots then indicate regions of chaos, as well as periodicity and quasi-periodicity.

Since the Hamiltonian is time-independent, the total energy is a constant of the motion, and the phase space is a three-dimensional hypersurface in four dimensions for each system. The system is said to be integrable if an additional constant of the motion exists, in which case its trajectories are restricted to two-dimensional surfaces in the available phase space.

The types of motion that integrable systems can exhibit are either periodic or quasi-periodic. Periodic (1-dimensional) orbits have trajectories that always appear as lines or dots on the Poincaré section, since by definition, they comprise the intersection of two two-dimensional surfaces. By contrast, the extra degree of freedom for a non-integrable system permits orbits to visit all regions of phase space. In this case, the system typically displays strongly chaotic behavior and the associated trajectories appear as filled in areas on the Poincaré plot.

Small perturbations of an integrable system admit small regions of chaos, though most of its orbits remain confined to two-dimensional surfaces. The chaotic regions grow as the perturbation increases in magnitude and eventually become connected areas on the Poincaré section, a phenomenon known as the Kolmogorov, Arnold, and Moser (KAM) transition [119–122]. For sufficiently large perturbations, systems typically become almost fully ergodic [123], though islands of regularity may persist for quite some time prior to this and typically have an intricate fractal structure.

The Poincaré plots for the N, R, and pN systems appear, respectively, in Figures 31–33, using the same conventions as in [3], up to an overall normalization for each section. The energy conservation relation (176), which is

$$p_R^2 + p_\theta^2 \leq \frac{2}{3}\eta \tag{186}$$

determines the outer boundary of each plot, where the presence of $\eta$ reflects the different normalizations for each system. When the potential energy is zero, equality in (186) holds, corresponding to the hex-particle being at the origin. Any departure from the origin reduces the values of $(p_R, p_\theta)$ and so yields the phase-space limit.

Figure 31 for the N system reproduces the results for the wedge problem [3]. Here, $H - 3mc^2$ is normalized to unity and the RHS of (186) is 1. In the equal mass case, the three-body problem corresponds to motion of a body falling toward a wedge whose sides are each at angles $30^o$ relative to the vertical axis. The energy constraint after an *A*-collision has taken place yields another boundary

$$\left(p_R - 2\sqrt{3}|p_\theta|\right)^2 \leq 1 - p_\theta^2 \tag{187}$$

whose satisfaction yields all points in phase space that have undergone an *A*-collision (the *A*-region). Points violating (187) are those for which a *B*-collision has taken place (the *B*-region). Since the interaction is gravitational, collisions with the third particle cannot ultimately be avoided. Consequently, the *A*-region has no fixed points and any point in the *A*-region will inevitably venture into the *B*-region. However, fixed points can occur in a subset of the *B*-region: here, *B*-collisions are infinitely repeated, corresponding to the annulus orbits.

In the N system, the centre of the plot in Figure 31 is a fixed point surrounded by a subregion of near-integrable curves. The annulus orbits are all contained within the large triangle surrounding this region. The closed circles in this annulus region correspond to quasi-periodic orbits about the periodic annuli with higher period, such as in Figure 20. The boundary of the annulus region is a thin region of chaos, most prominent at the corners, as shown in the lower right inset. These chaotic regions are confined and not simply connected.

The region beyond this is the pretzel region, which has circles bounding the quasi-periodic near-integrable regions; these exhibit repeated self-similarity, as shown in the upper-right inset. The two large circles observed just below the annulus region correspond to the boomerang-shaped orbits ($\overline{AB^3}$) shown in Figure 27. The next set of circles will be $\overline{A^2B^3}$, and so on. Between these sets of circles, there are collections of crescents with sequences $\overline{AB^3A^2B^3}$, $\overline{AB^3AB^3A^2B^3}$, etc. Each circle is actually a continuum of possible circles, whose diameter depends on the initial conditions. At the centre of this family of circles is a dot corresponding to the periodic orbit in question.

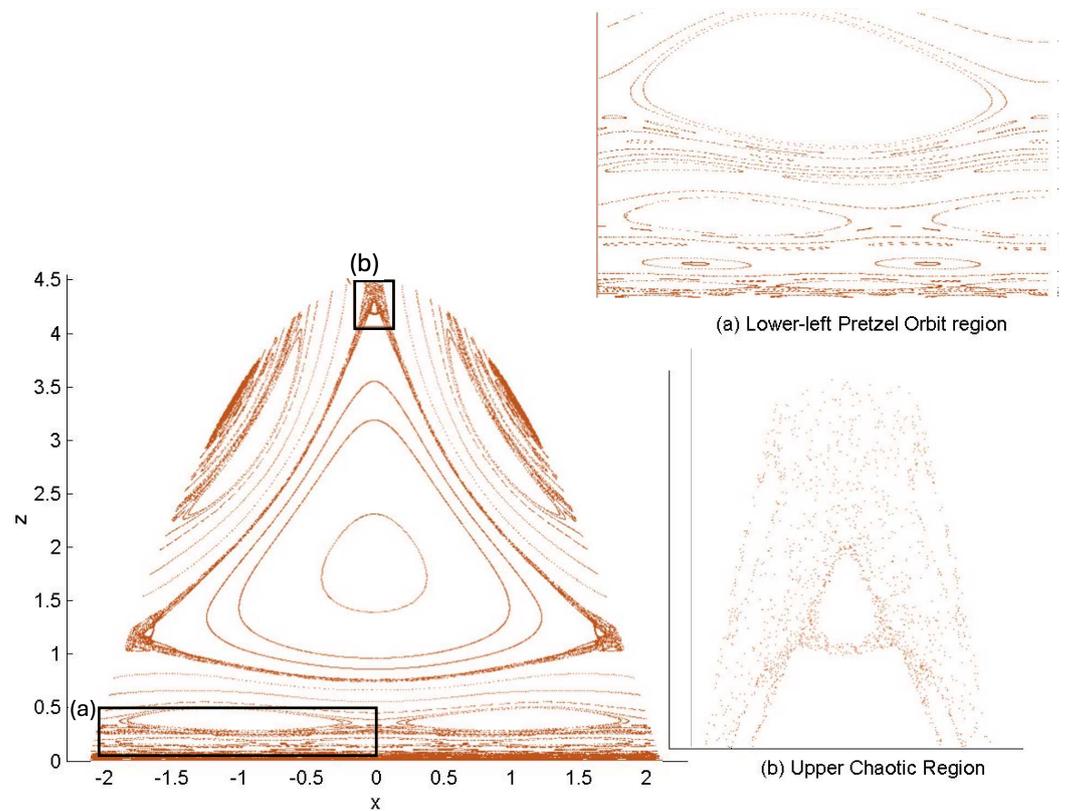

(a) Lower-left Pretzel Orbit region

(b) Upper Chaotic Region

**Figure 31.** The Poincaré plot of the N system. The squares denote the parts of the plot magnified in the insets.

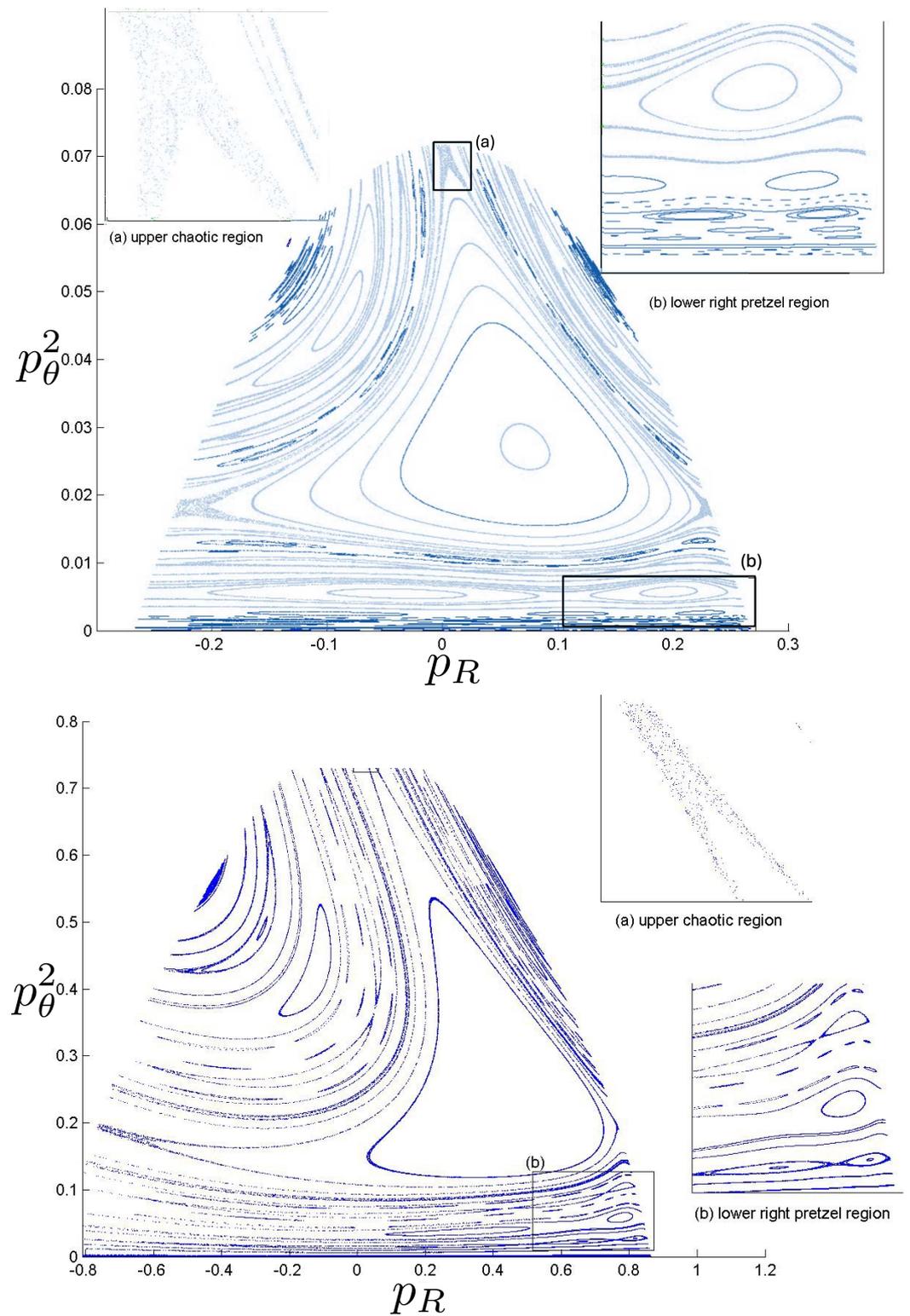

**Figure 32.** Poincaré plots of the R system. In the **upper plot**, $\eta = 0.1$, in the **lower plot**, $\eta = 0.5$. The upper insets provide close-ups of the chaotic region at the top of the diagram, which is similar to the N system, but distorted in shape. The lower insets are close-ups of the structure in the pretzel; it is likewise distorted relative to the N system, with the distortions growing as $\eta$ increases.

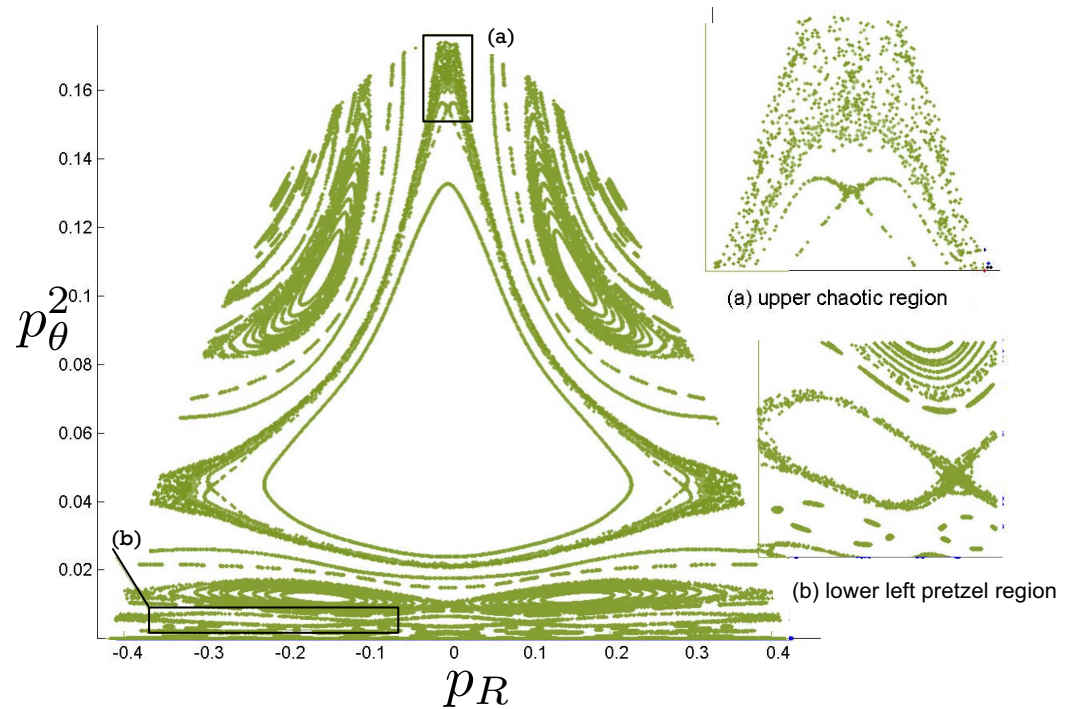

**Figure 33.** The Poincaré plot of the pN system at $\eta = 0.21$. Qualitatively similar to the N system in terms of symmetry, its chaotic regions are larger, and the pretzel region is on the threshold of KAM breakdown.

Somewhat remarkably, the highly nonlinear R system, shown for two different values of $\eta$ in Figure 32, retains all of the qualitative features of the N system, at least over the range of $\eta$ numerically accessible. The annulus, pretzel, and chaotic regions all retain their same basic structure, though asymmetrically deformed. This deformation increases as $\eta$ increases and occurs because the Hamiltonian given by Equation (152) is invariant under the discrete symmetry $(p_i, \epsilon) \to (-p_i, -\epsilon)$ instead of the $p_i \to -p_i$ symmetry in the N system. The discrete constant $\epsilon = \pm 1$ is a measure of the flow of time of the gravitational field relative to the particle momenta. In Figure 32, $\epsilon = +1$, the opposite choice would create the same distortion but towards the lower left. The situation is reminiscent of the two-body case of the previous section, where the gravitational coupling to the kinetic-energy causes a distortion of the trajectory from an otherwise symmetric pattern. Whether or not KAM breakdown occurs for higher $\eta$ values remains an interesting open question.

As compared to the N and R plots the pN system, shown in Figure 33, is notably different. It retains the $p_i \to -p_i$ symmetry of the N system but appears to undergo a KAM transition at $\eta \sim 0.3$, as shown in Figure 34. For small $\eta$, the distinction with the N system is mild, but for $\eta = 0.21$, the lines across the bottom of the figure slightly widen. Larger regions of chaos become evident around the edges of the groups of ellipses in the lower regions of the figure for $\eta = 0.26$. Further increasing $\eta \to 0.3$, the lower part of the Poincaré section becomes engulfed by a chaotic sea; only a few non-connected islands of regular motion remain. No such behavior is seen in R system for similar values of $\eta$. These distinctions are not artifacts due to differences in scalings between the systems. The underlying feature that enforces the structure on the phase space in the R system but not in the pN system remains to be understood.

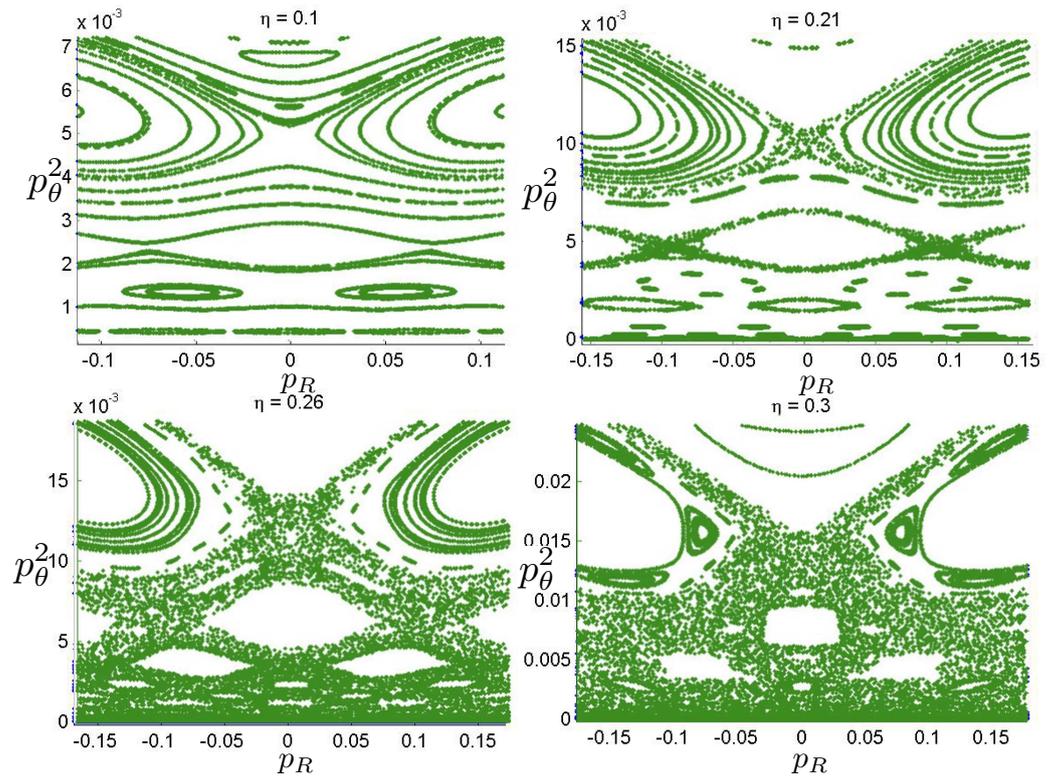

**Figure 34.** A series of successive close-ups of the lower section of the Poincaré plot of the N system. This illustrates the self-similar structure in the pretzel region that repeats at increasingly small scales. The limiting factor at very small scales is the number of trajectories that we included in the plot.

### 6.5. Unequal Masses

For unequal masses the hexagon becomes squashed, with two opposite corners moving inward, changing both the slopes of the straight edges and their relative lengths; relativistic corrections maintain this basic distortion, but with the straight edges becoming parabolic [116]. The exact R potential is given by (177), with its pN and N counterparts, respectively, given by (179) and (180).

The R potential is similar to the N potential except that the sides of the hexagon become concave, and have a much steeper slope as the radial variable in the $(\rho, \lambda)$ space increases. For $V_R = \hat{V}_R$ such that

$$\ln\left(\frac{\left(\hat{V}_R - m_j c^2\right)\left(\hat{V}_R - \left(M_{\text{tot}} - m_j\right)c^2\right)}{\left(M_{\text{tot}} - m_j\right)m_j c^4}\right) = \hat{V}_R\left[\frac{1}{\left(\hat{V}_R - m_j c^2\right)} + \frac{1}{\left(\hat{V}_R - \left(M_{\text{tot}} - m_j\right)c^2\right)}\right] \quad (188)$$

(for $j = 1, 2$ or 3), the slope of the R potential becomes infinite. In the equal mass case, this occurs at $\hat{V}_R \approx 6.71197 mc^2$, where $m = M_{\text{tot}}/3$. The maximal possible critical value of the potential occurs when one of the masses $m_j = M_{\text{tot}}/2$, for which $\hat{V}_R \approx 6.886682 m_j c^2$. For $m \longrightarrow 0, M_{\text{tot}}$, the potential $\hat{V}_R \longrightarrow M_{\text{tot}}c^2$—no energy is available for motion. A plot of $\hat{V}_R$ as a function of $m_j$ is shown in Figure 35. For values of $V_R > \hat{V}_R$, the size of the distorted hexagon decreases like $(\ln V_R)/V_R$.

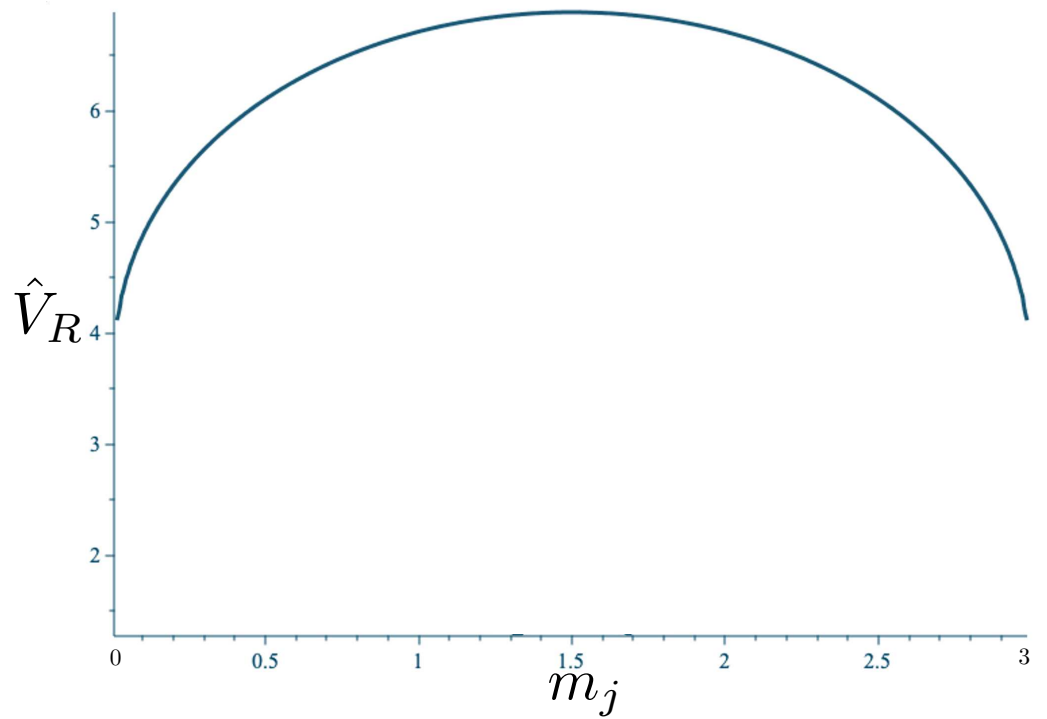

**Figure 35.** Critical values of the relativistic potential $\hat{V}_R$ plotted in units of $M_{tot}c^2$ (here set equal to 3) as a function of a given particle mass $m_j$. The maximum critical value occurs in the case when $m_j = M_{tot}/2$. As $m_j \to 0$ or $M_{tot}$, the minimal value $\hat{V}_R \to M_{tot}$ is attained.

The overall shape of the R potential is that of a distorted hexagonal carafe, whose distortion is analogous to that of the N potential, as shown in Figure 36. The relatively steeper growth of the R potential as a function of distance from the origin is manifest by the smaller scales for $\rho$ and $\lambda$ in the right-hand panel.

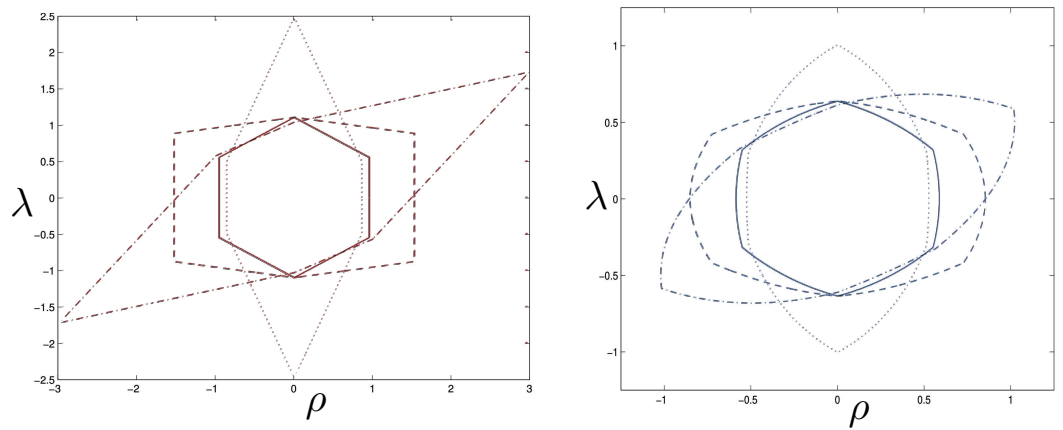

**Figure 36.** Cross-sections of the N potential (**left**) and R potential (**right**) at $V \approx 1.3 M_{tot}c^2$ for different particle mass ratios: solid—1:1:1; dashed—1:1:4; dotted—4:4:1; dash-dotted—1:4:8. All discontinuities lie on one of the three bisectors $\rho = 0$, $\rho = \sqrt{3}\lambda$, or $\rho = -\sqrt{3}\lambda$ regardless of the mass ratio in both systems. The smaller scales for $\rho$ and $\lambda$ are indicative of the steeper growth of the R potential as a function of distance from the origin.

### 6.5.1. Trajectories

Expressing relative masses as a ratio $m_1 : m_2 : m_3$, annulus, pretezel and chaotic trajectories are obtained and the symbol sequence always takes the form

$$\prod_{i,j,k} (A^{m_i}, B^{3n_j})^{l_k} \tag{189}$$

as in the equal mass case. Equation (189) implies that B motion always comes in multiples of three; no motions have been observed that depart from this situation [116].

The situation where two masses are equal ($m_1 : m_2 : m_3 = 1 : 1 : \alpha$) is instructive, and the plots of the relative motion of the particles for both the N and R systems is given in Figure 37 for decreasing values of $\alpha$. For equal mass ($\alpha = 1$), there is an annulus motion: no particle ever crosses another twice in a row. But as $m_3$ decreases ($\alpha$ decreases), its frequency of oscillation decreases while its amplitude increases with respect to the other two, which provide a gravitationally bound subsystem. The binding becomes tighter as $\alpha$ decreases, more so for the R system. Eventually, the binding becomes so tight that the more massive particles execute an additional $A$ motion before crossing the third particle, and the hex-particle transitions from annulus to pretzel motion. The greater the mass difference, the more difficult it is to set up initial conditions at a given energy so that particles 1 and 2 do not cross more than once during the long period oscillation of particle 3.

Large and small values of $\alpha$ are likewise instructive, and plots for $\alpha = 100$ (particle 3 very massive) and $\alpha = 0.01$ (particles 1 and 2 very massive) are, respectively, shown in Figures 38 and 39. As expected, for large $\alpha$, the heavy particle 3 barely moves as the other two oscillate about it, depicted in Figure 38. However, the passing of the other two particles causes small perturbations in the motion of the heavy body, as shown in the insets. The perturbation is smooth and regular in the N system, whereas in the R system, the velocity of the large mass increases much more suddenly, leading to a more erratic and jerky trajectory.

For small $\alpha$ (Figure 39), a stable gravitationally bound subsystem is formed by the two heavy particles, with the third oscillating about their centre of inertia. The oscillation amplitude is much larger and its frequency much smaller in the N system than in the R system, commensurate with the two-body system in Section 5. The effect of the light particle 3 is to cause the oscillatory motion of the centre of mass of the two more massive particles, which is clear from the insets in the N and R systems. This perturbation is almost imperceptible due to the two heavy particles being twice as massive as the single particle in the $\alpha = 100$ case (Figure 38); consequently, they are less susceptible to changes in motion from the smaller mass body.

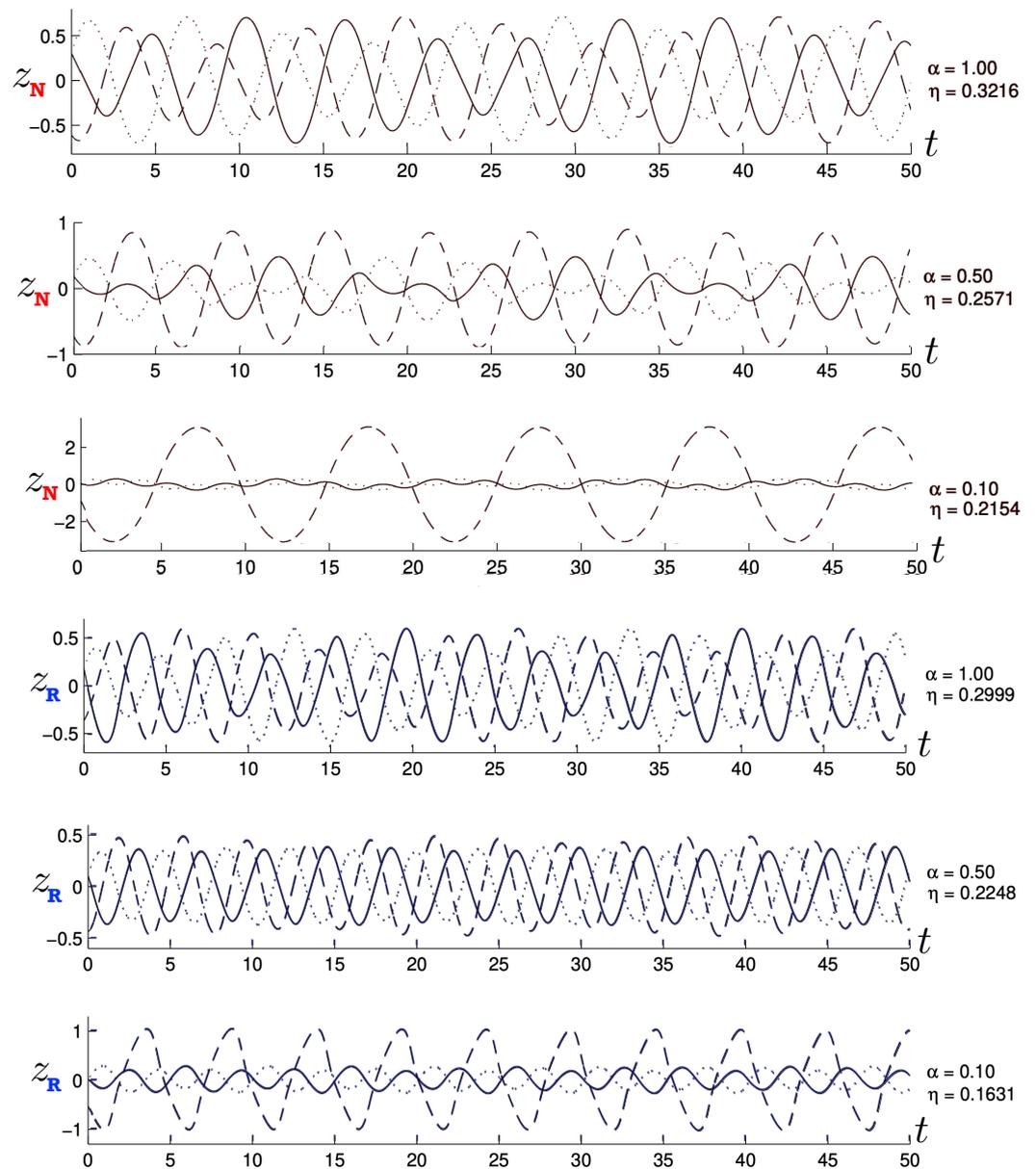

**Figure 37.** Plots of the relative position $z(t)$ of each particle with respect to the centre of mass in the N system (**top**) and R system (**bottom**). Particles 1 (solid), 2 (dotted), and 3 (dashed) have relative masses in the ratio $1 : 1 : \alpha$. The same initial values of $(\rho, \lambda, p_\rho, p_\lambda)$ are used in each plot, with (153) used to fix the total energy $\eta + 1$ for the R system and its non-relativistic limit used to fix $\eta$ in the N system. The top figure is an annulus trajectory ($\overline{B}$), while the next two are pretzels ($\overline{B^9 \overline{A}}$, $\overline{(B^3 A^2)^4 B^3 A^3}$) in the N system, whereas the bottom figure is a pretzel ($\overline{(B^6 A)^7 B^3}$) in the R system, and the two above that are both annuli in the R system. The smaller the value of $\alpha$, the more tightly bound the particles 1 and 2 are in each system, with relatively tighter binding in the R system, as can be seen by comparing the bottom figure with the third one from the top.

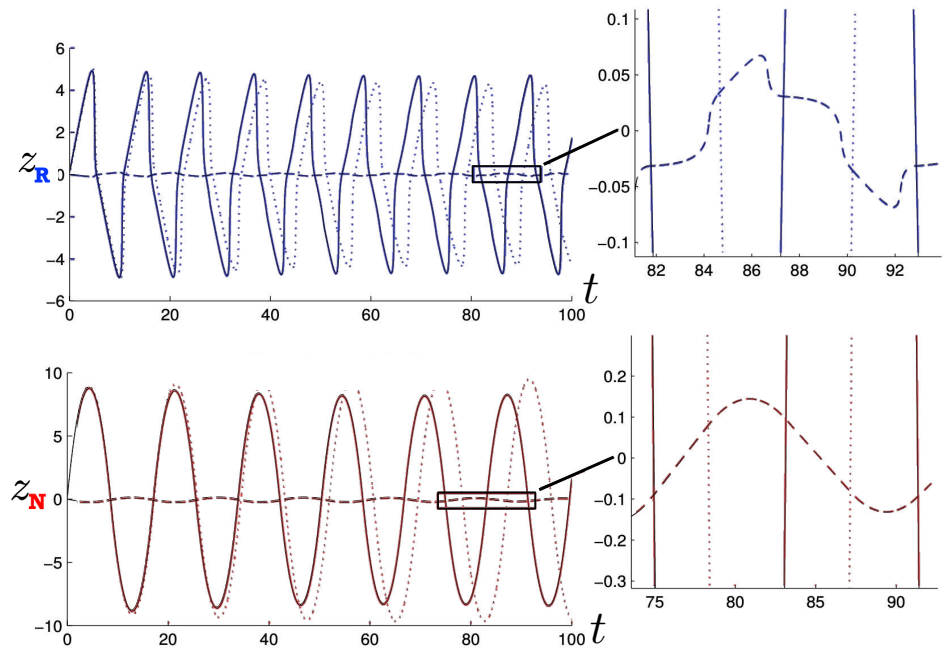

**Figure 38.** A comparison of the relative motion of the particles with respect to the centre of mass plotted as a function of time for the R system (**top**, $\eta = 0.2793$) and N system (**bottom**, $\eta = 0.1748$). Particles 1 (solid), 2 (dotted), and 3 (dashed) have relative masses in the ratio 1:1:100. Small perturbations in the motion of the large mass due to the crossing of the smaller masses are shown in the insets.

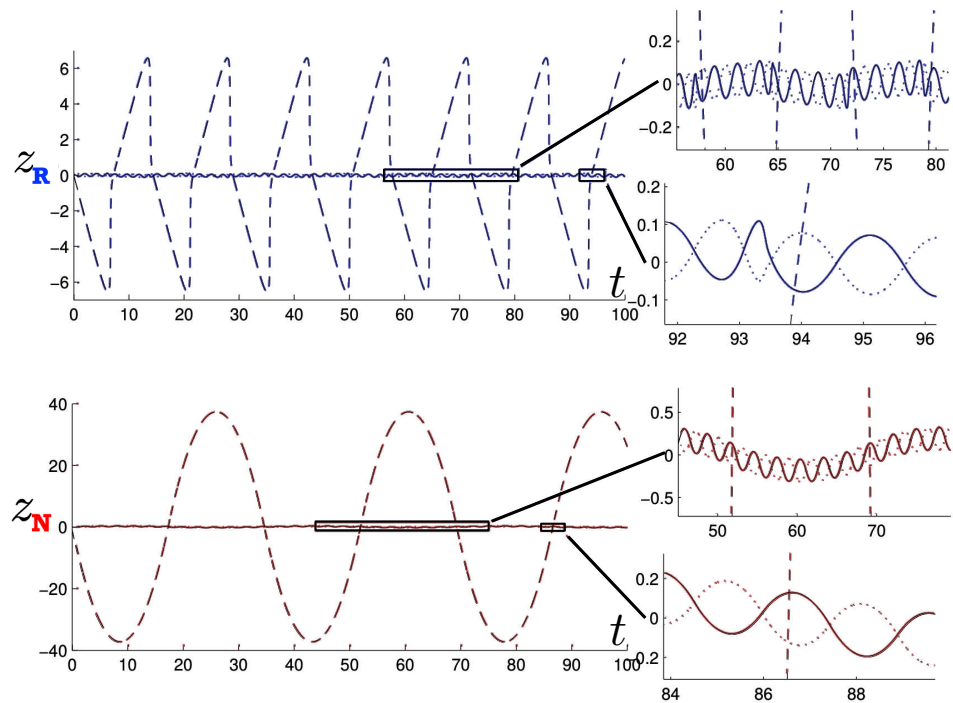

**Figure 39.** A comparison of the relative motion of the particles with respect to the centre of mass plotted as a function of time for the R system (**top**, $\eta = 0.2244$) and N system (**bottom**, $\eta = 0.2512$). Particles 1 (solid), 2 (dotted), and 3 (dashed) have relative masses in the ratio 1:1: = 0.01. The respective upper insets show the motion of the stable, two body sub-system made up of the two heavy particles, whereas the effect of encounters with the light particle are shown in the respective lower insets.

### 6.5.2. Poincaré Plots

As with the equal mass case, the annulus and pretzel trajectories are in similar regions of the Poincaré plot, separated by a region of chaos, but the shapes and sizes of the different regions change. The R system is topologically similar to the N system, but with the various regions distorted in a manner similar to the equal mass case.

Figure 40 compares the Poincaré sections for the N and R systems for the mass ratio 1:1:0.1 for $\eta = 0.3$. The triangular annulus region moves towards the top of the surface of section, and becomes smaller in both systems. This latter effect is a manifestation of the difficulty in attaining annulus motion noted above, when one particle is much less massive than the other two.

If one particle is much more massive than the other two, the annulus region becomes larger and extends towards the lower region of the plot, as shown in Figure 41 in both the R and N systems for the mass ratio 1:1:10. In this situation, the hex-particle needs less angular momentum to attain an annulus orbit in the $(\rho\text{-}\lambda)$ plane. Somewhat like a two-planet solar system in one-dimension, the two lighter bodies behave like members of two separate two-body systems, with the heavy particle taking the role of the second body for each, as shown in Figure 38. Furthermore, additional regions of chaos appear that are absent in the equal mass case.

An example of a Poincaré plot when all three masses are unequal is shown in Figure 42 for the N and R systems, with a mass ratio 1:5:10. Since none of the bodies have the same mass, the symmetry about the $p_R = 0$ axis in the N system is gone. Different regions are not as clearly segregated as in the $m_1 = m_2$ case, and instead extend over a larger region of the plot. A region of chaos separating outer pretzel regions from the inner annulus region is marked by A in the left diagram, and above and below B are new regions of chaos amongst pretzel trajectories.

The R system further distorts the N diagram to the lower right. The chaotic region separating annulus and pretzel trajectories is now two loops that were created by a single trajectory, both marked by a 1. The annulus region is confined to the area inside both of these loops (marked by a 2), where a single annulus trajectory will visit both regions.

The key feature of the unequal mass case, for both the N and R systems, is the presence of additional chaotic regions that are absent in the equal mass case in the corresponding constant energy hyper-surface. These additional chaotic regions appear within the pretzel regions of the corresponding equal mass plot, and are characterized by broadened lines in the pretzel region, evident in each of Figures 40–42. The origin of this additional chaos is not understood.

The unequal mass case is equivalent to the two-dimensional symmetric wedge billiard system in a uniform gravitational field [3], with the relative masses of the particles directly related to the wedge angle $\theta$ by

$$\tan \theta = \frac{\sqrt{1 + 2\alpha^{-2}}}{1 + 2\alpha^{-1}} \tag{190}$$

where $\alpha$ is as defined as above, and $\theta = \pi/6$ corresponds to $\alpha = 1$, the equal mass case. The angle of the wedge is related to the angle between the bisectors of the hexagonal well. The only distinction between the wedge system and the three-body system is the absence of collisions in the latter. If all masses are equal there is no distinction between a collision and a crossing of two bodies (apart from particle labelling) and so the Poincaré maps become identical in this case.

**Figure 40.** Poincaré plots with a mass ratio of 1:1:0.1 for the N (**left**) and R (**right**) systems, both with $\eta = 0.3$. The insets show the onset of chaos in the pretzel region for the R system.

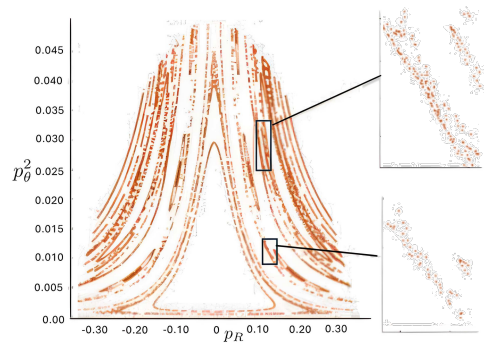

**Figure 41.** Poincaré plots with a mass ratio of 1:1:10 for the N (**left**) and R (**right**) systems, both with $\eta = 0.3$. Additional regions of chaos in the pretzel region appear that are not present in the corresponding regions of the equal mass Poincaré sections, shown in the insets.

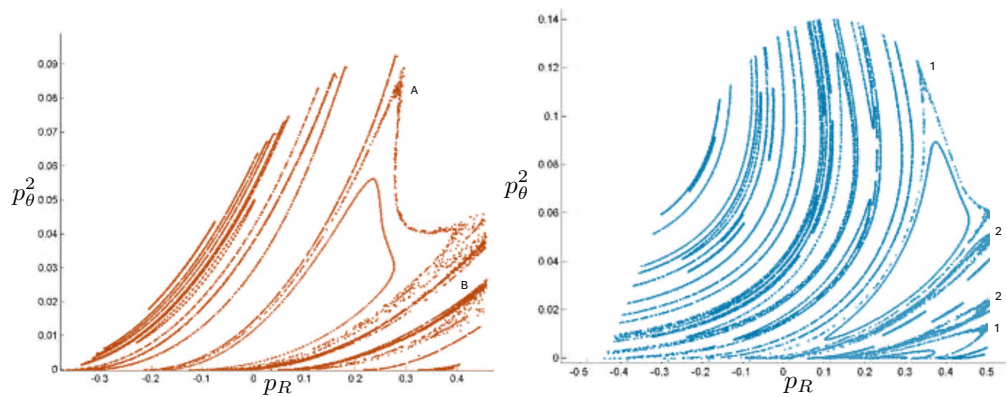

**Figure 42.** Poincaré plots with a mass ratio of 1:5:10 for the N (**left**) and R (**right**) systems, both for $\eta = 0.3$. The region of chaos separating annulus trajectories (inside) and predominantly pretzel trajectories (outside) is marked by A, whereas the densely filled area directly above and below B is a new region of chaos amongst the pretzel trajectories. In the R system (**right**), the densely filled regions (marked by a 1) were created by a single trajectory separating the annulus and pretzel orbits, while the chaotic regions (marked by a 2) were created by a trajectory within the pretzel region.

In both the N and R systems, there is an increase in the amount of chaos as the difference in the masses increases. Earlier studies explored a limited range of mass ratios; however, and it remains an open question as to whether or not one or both systems will undergo a transition to global chaos, or if integrable and near integrable regions exist for all mass ratios.

### 6.6. Charge and Cosmological Constant

Including a cosmological constant [117] and endowing the particles with charges [118] significantly increases the parameter space. The solutions to (63) and (64) still yield (65), but now

$$V = \begin{cases} 0 & \text{in region (1)} \\ -e_1 e_2 - e_1 e_3 & \text{in region (2)} \\ -e_1 e_3 - e_2 e_3 & \text{in region (3)} \\ 0 & \text{in region (4)} \end{cases} \tag{191}$$

where space is divided into four regions, chosen so that $z_1 < z_2 < z_3$ region (1) being to the left of particle 1 and region (4) to the right of particle 3. The determining equation becomes

$$\begin{aligned}
&\Big[ \Big( (\hat{M}_1 + \hat{K}_{1+}) \hat{K}_{3+}^2 + (\hat{M}_3 + \hat{K}_4) \hat{K}_{2+}^2 \Big) \tanh\left( \frac{\hat{K}_{3+} z_{32}}{4} \right) \tanh\left( \frac{\hat{K}_{2+} z_{21}}{4} \right) \\
&+ (\hat{M}_1 + \hat{K}_{1+}) (\hat{M}_3 + \hat{K}_4) \hat{M}_2 \tanh\left( \frac{\hat{K}_{3+} z_{32}}{4} \right) \tanh\left( \frac{\hat{K}_{2+} z_{21}}{4} \right) \\
&+ \Big( (\hat{M}_1 + \hat{M}_2 + \hat{K}_{1+}) (\hat{M}_3 + \hat{K}_4) + \hat{K}_{3+}^2 \Big) \hat{K}_{2+} \tanh\left( \frac{1}{4} \hat{K}_{3+} z_{32} \right) \\
&+ \Big( (\hat{M}_1 + \hat{K}_{1+}) (\hat{M}_2 + \hat{M}_3 + \hat{K}_4) + \hat{K}_{2+}^2 \Big) \hat{K}_{3+} \tanh\left( \frac{1}{4} \hat{K}_{2+} z_{21} \right) \\
&+ (\hat{M}_1 + \hat{M}_2 + \hat{M}_3 + \hat{K}_{1+} + \hat{K}_4) \hat{K}_{2+} \hat{K}_{3+} \Big] \\
&= 0
\end{aligned} \tag{192}$$

where $z_{ij} \equiv (z_i - z_j)$, $s_{ij} \equiv \text{sgn}(z_{ij})$ and

$$\hat{M}_i = \kappa \sqrt{p_i^2 + m_i^2} \tag{193}$$

$$\hat{K}_{1\pm} \equiv \sqrt{\kappa^2 \left[ X - \frac{\epsilon}{4} \left( \sum_{a=1}^3 p_a s_{a1} \pm p_1 \right) \right]^2 - \frac{\Lambda_e}{2}} \tag{194}$$

$$\hat{K}_{2\pm} \equiv \sqrt{\kappa^2 \left[ X - \frac{\epsilon}{4} \left( \sum_{a=1}^3 p_a s_{a2} \pm p_2 \right) \right]^2 - \frac{\kappa}{2}(e_1 e_2 + e_1 e_3) - \frac{\Lambda_e}{2}} \tag{195}$$

$$\hat{K}_{3\pm} \equiv \sqrt{\kappa^2 \left[ X - \frac{\epsilon}{4} \left( \sum_{a=1}^3 p_a s_{a3} \pm p_3 \right) \right]^2 - \frac{\kappa}{2}(e_1 e_3 + e_2 e_3) - \frac{\Lambda_e}{2}} \tag{196}$$

$$\hat{K}_4 = -2 \sqrt{\kappa^2 \left[ X + \frac{\epsilon}{4}(p_1 + p_2 + p_3) \right]^2 - \frac{\Lambda_e}{2}} \tag{197}$$

with

$$H = \frac{4}{\kappa} \sqrt{\kappa^2 X^2 - \frac{\Lambda_e}{2}} \geq \frac{4}{\kappa} \sqrt{-\frac{\Lambda_e}{2}} \tag{198}$$

which defines the constant of integration $X$. This relation is the same as (76) when the charges are zero. The inequality follows since $H$ and $X \geq 0$ must both be real, defining a negative critical value

$$\Lambda_{negcrit} \equiv -\frac{\kappa^2 H^2}{8} \leq \Lambda_e \tag{199}$$

The permutation of the particles yields the same determining equation with indices appropriately switched.

There are six inequivalent configurations: $(0\,0\,0)$, $(+\,0\,0)$, $(+\,+\,0)$, $(+\,-\,0)$, $(+\,+\,-)$, and $(+\,+\,+)$ — where $0$ denotes a neutral particle, since the potential is invariant for $e_i \to -e_i$ and the charge configurations interchange as the particles cross. The first of these has been considered in the previous subsections. The second of these, unless $\Lambda_e = 0$, is equivalent to three neutral particles with a cosmological constant. If one particle is neutral, the relative magnitudes of the charges are irrelevant since only products of their magnitudes matter. The effect of the electromagnetic energy is to yield a charge-dependent constant vacuum energy between the particles, with the size of the region and the magnitude of the vacuum energy changing as the particles move about. If one particle is neutral, then the magnitude of the vacuum energy between the particles does not change.

The parameter space has been explored in considerable detail [117,118]. As it is so vast, only a few key results shall be presented.

### 6.6.1. Neutral Configurations with $\Lambda_e \neq 0$

For a nonzero cosmological constant, the case of three neutral particles is equivalent to that of the (+ 0 0) configuration. In general, $\Lambda_e$ significantly modifies the chaotic properties of the relativistic three-body system in markedly different ways, depending on its sign. The following analysis sets all masses to be equal.

For $\Lambda_e < 0$, there is a rapid decrease in the size of the chaotic regions. These become even even smaller than in the N system in the Poincaré plot. Given the high degree of nonlinearity, this is quite surprising. This is manifest even at fairly small energies, as illustrated in Figure 43 for $H = 1.2 M_{\text{tot}} c^2$.

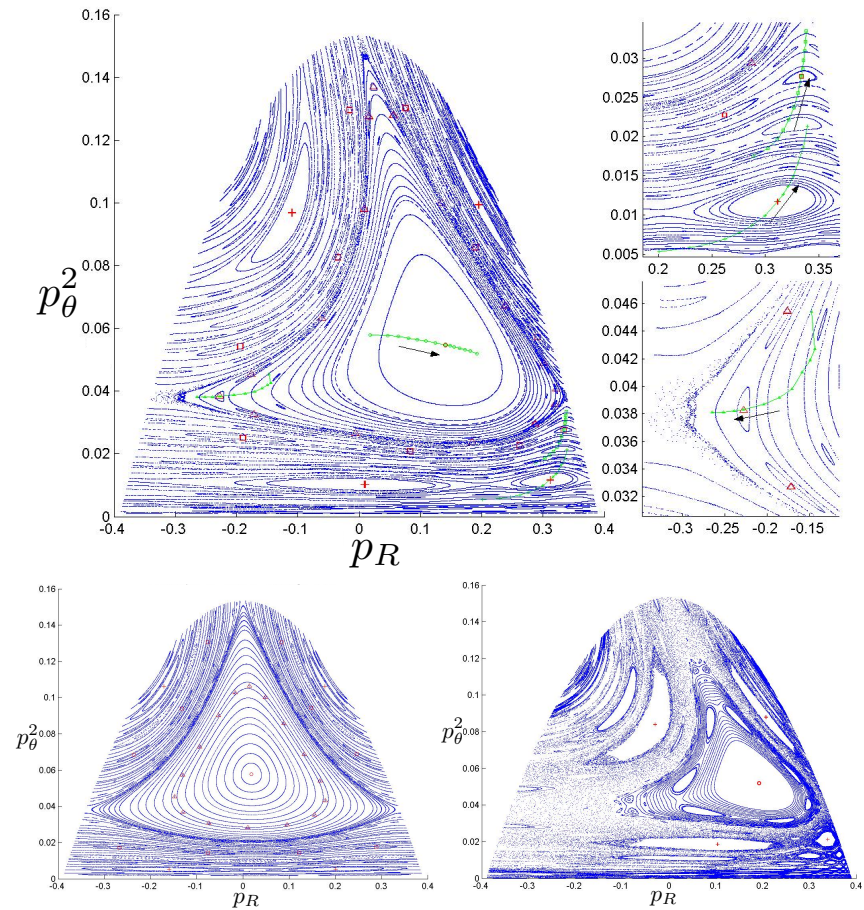

**Figure 43.** Poincaré plots for $H = 1.2 M_{\text{tot}}$ for three different values of $\frac{\Lambda_e}{\kappa^2 M_{\text{tot}}^2}$ : 0 (**top**), $-0.175$ (**lower left**), and 0.6 (**lower right**). The green curves track the orbits (a–d) shown in Figure 44, and indicate their trajectories in the plot as $\frac{\Lambda_e}{\kappa^2 M_{\text{tot}}^2}$ changes from $-0.175$ to 0.6; the arrow indicates the direction of increasing $\Lambda_e$. The insets show closeups of the trajectories near chaotic regions. The lower left plot shows that chaos diminishes when $\Lambda_e$ becomes more negative, whereas in the lower right diagram, orbits (b) and (c) have become chaotic and so do not appear; only the locations of orbits (a) and (d) are shown.

The upper diagram in Figure 43 corresponds to the $\Lambda_e = 0$ case, similar to the lower diagram in Figure 32 but with tracks (green curves) showing how different types of orbits move as $\Lambda_e$ changes from a negative value (lower left diagram) to a positive one (lower right diagram). These four different quasiperiodic (or stable) orbits are shown in Figure 44 and change as $\Lambda_e$ changes. Orbit (a) is located in the centre of the annulus region, whereas orbits (b), (c), and (d) produce a set of points on the Poincaré plot that follow the contours of the triangular shaped region. The stable and quasistable orbits remain so as $\Lambda_e$ becomes more negative, but $\Lambda_e > 0$ stable orbits can become chaotic. The transition point depends on the initial conditions of the orbit or its specific location in the phase space. These results support the intuitive higher-dimensional understanding of a negative cosmological constant as a parameter that provides stronger gravitational binding, leading to an increase in the integrability of the dynamics and thus an increase in the stability of trajectories.

Remarkably, as $\Lambda_e \to \Lambda_{negcrit}$, the chaotic regions nearly vanish. Since the area of the chaotic regions in the Poincaré section were found to be roughly proportional to $\left| \frac{\Lambda_e}{\Lambda_{negcrit}} \right|$, for the range of possible energies that could be numerically investigated, it has been conjectured that this holds for arbitrarily large values of $H$. Conversely, the area of the chaotic regions in the Poincaré section increases as $\Lambda_e$ becomes increasingly positive, as shown in the lower right panel of Figure 43. This occurs within the regions corresponding to the pretzel orbits and in the regions between annulus and pretzel orbits. This phenomenon has likewise been conjectured to occur at all energies [117].

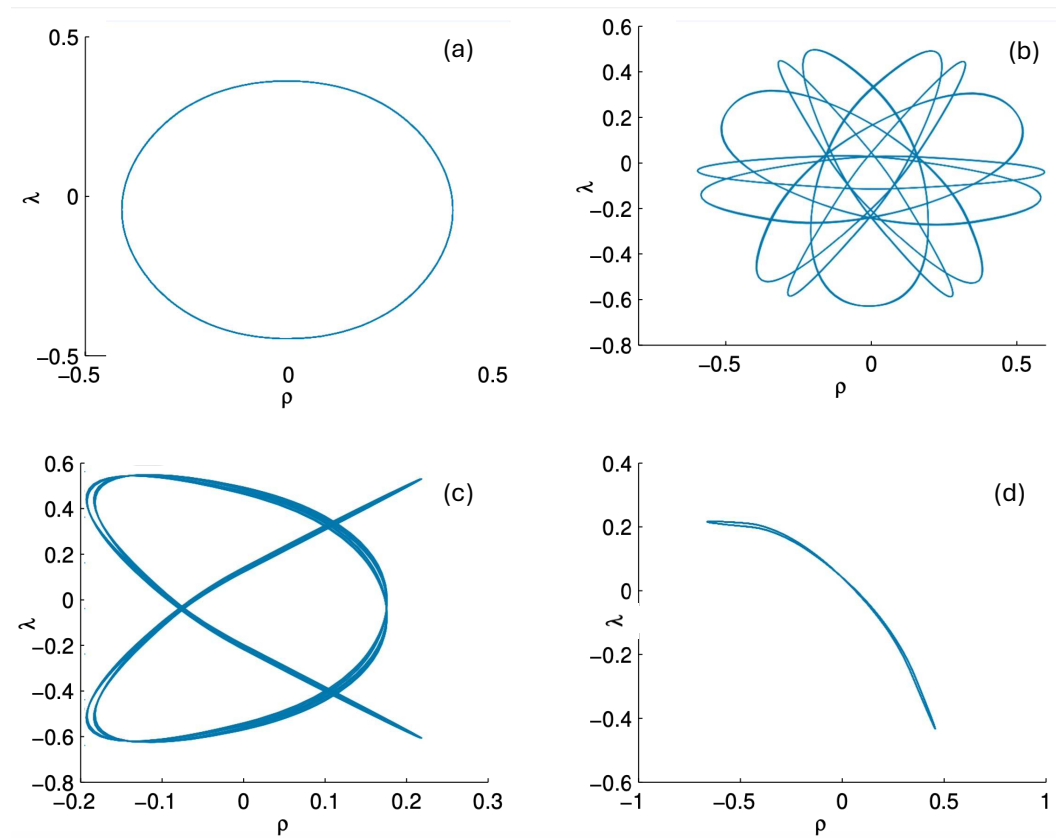

**Figure 44.** Four representative periodic and quasi-periodic orbits, labelled (**a**) a stable (nearly circular) orbit located in the centre of the annulus region, denoted by a '∘' symbol (**upper left**); (**b**) an annulus orbit located around the outside edge of the triangular annulus region, denoted by a '△' symbol (**upper right**); (**c**) a quasi-periodic pretzel orbit located halfway between the centre of the annulus region and the first large outer annulus regions, denoted by a '□' symbol (**lower left**); and (**d**) a banana-shaped $AB^3$ orbit located in the centre of that region, denoted by the '+' symbol (**lower right**).

### 6.6.2. Charged Configurations

The charged three-body case allows the study of additional novel phenomena such as localized vacuum energy and the breaking full hexagonal symmetry. Concerning the former, since the electromagnetic coupling between any pair of charges induces a vacuum energy between them, we can study how this localization of vacuum energy modifies the effects with $\Lambda_e \neq 0$. Concerning the latter, the shape of the hexagonal potential becomes elongated unless all particles have the same charge, as in the unequal mass case [116].

The diagrams of representative cases for the potential (relative to the total rest mass) are shown in Figures 45 and 46, taking (for simplicity) all charges to be equal in magnitude.

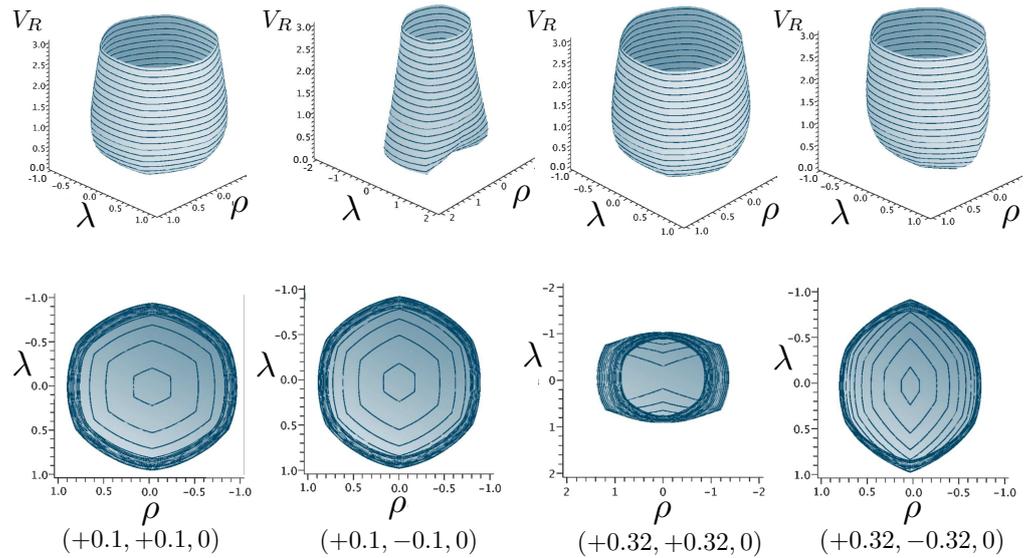

**Figure 45.** Plots of the potential for one neutral and two charged particles. Values of $\frac{e}{\sqrt{\kappa}M_{tot}}$ for each case are given at the bottom. Solid lines denote equipotentials.

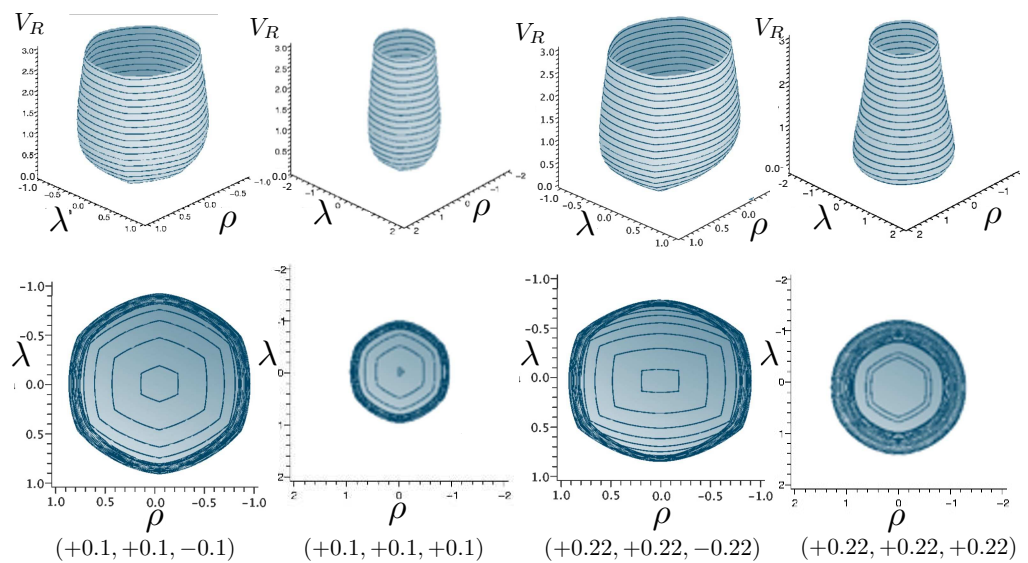

**Figure 46.** Plots of the potential for three charged particles. Values of $\frac{e}{\sqrt{\kappa}M_{tot}}$ for each case are given at the bottom. Solid lines denote equipotentials.

When all particles have identical charge, the potential has hexagonal symmetry, but for different charges, this symmetry becomes skewed. As shown in the first and third columns in Figure 45, the potential is stretched along the $\lambda = 0$ axis in the (+ + ne) case, corresponding to the decrease in the electric potential as particles 1 and 2 separate. This results in an increase in the magnitude of $\rho$. As the magnitude of the charge increases, the width of the potential at lower energies likewise increases and the sides of the hexagonal cross-section become more concave. The value of $V_{Rc}$ at which the cross section of $V$ is largest is also reduced. The same effects occur for the (+ 0 +) and (0 + +) configurations, but with the potential compressed along the $\rho + \sqrt{3}\lambda = 0$ and $\rho - \sqrt{3}\lambda = 0$ axis, respectively.

By contrast, in the (+ - 0) case, as shown in the second and fourth columns in Figure 45, the potential is compressed along the $\lambda = 0$ axis. This corresponds to a decrease in electric potential with increasing $\rho$.

When two of the particles have positive charges and the third negative (+ + −), as shown in the first and third columns in Figure 46, the hexagon becomes elongated in the $\rho$ direction as the magnitude of the charge increases. If all three particles have equal positive charge (+ + +), the hexagonal symmetry is preserved, as shown in the second and fourth columns in Figure 46. The width of the potential increases at lower energies and its sides become more convex as the magnitude of the charge increases.

As before, the annulus, pretzel, and chaotic trajectories are present depending on the initial conditions; samples are shown in Figure 47. We can gain more insights by considering Poincaré plots, as shown in Figure 48 (+ + 0), Figure 49 (+ − 0), and Figure 50 (+ + −).

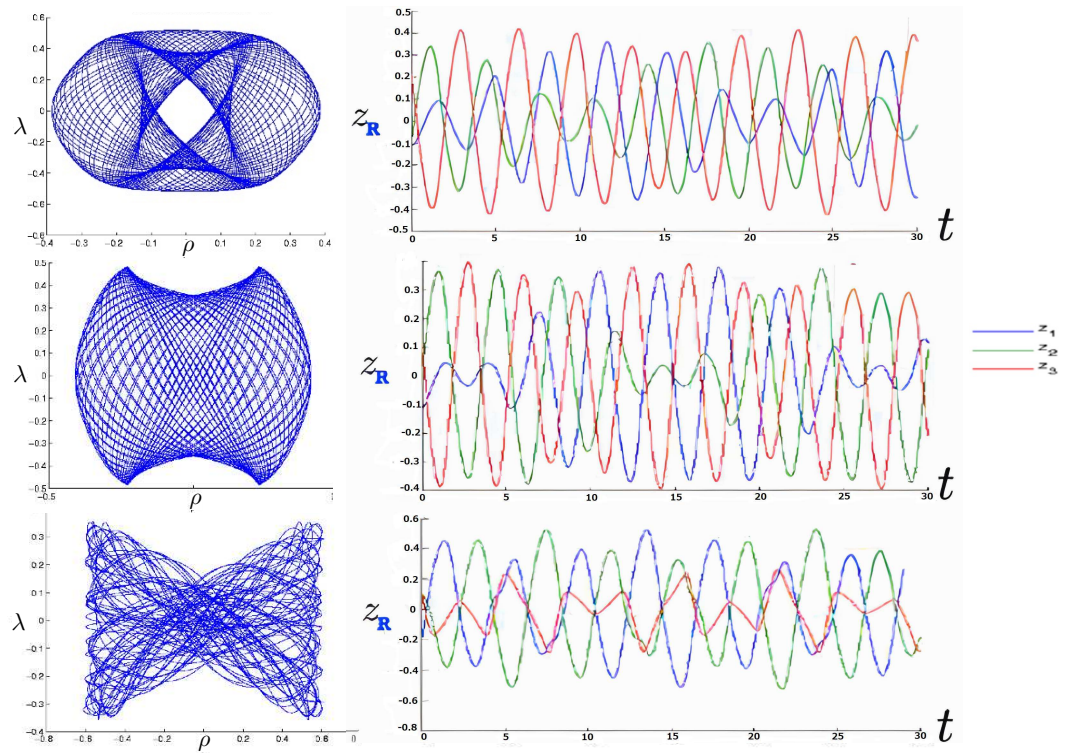

**Figure 47.** Sample trajectories in the charged case with FE conditions, for $H/M_{tot} = 1.2$, $\Lambda_e = 0$, and $\frac{e}{\sqrt{\kappa}M_{tot}} = (0.1, 0.1, -0.1)$ (**top** and **middle**) and $(0.2, 0.2, -0.2)$ (**bottom**). Each figure was run for 200 time steps; the corresponding three-particle trajectories at the right were truncated after 30 time steps.

If one particle is neutral, only the product of the charges is relevant and not their individual magnitudes; their relative sign determines the sign of $V$ from (191). The region between the two charged particles has constant vacuum energy throughout the motion, but the size of this region changes as the particles move. Consequently, using equal magnitude charges has no loss of generality.

For the (+ + 0) case in Figure 48, there are two different Poincaré plots: one corresponding to the crossing of the two identical charged particles and the other corresponding to the neutral particle crossing either of the identical particles. Comparing to the lower right panel of Figure 48, we see that chaotic behaviour emerges more rapidly with increasing repulsive charge than increasing energy, with the chaotic regions much more widespread in Figure 48. The chaotic behaviour is notably enhanced, filling the pretzel regions and pushing the annulus region in the opposite direction as the charge increases, leaving almost no circular periodic motions.

For three bodies of distinct charge, the relative magnitudes of the charges do matter, since the magnitude of the vacuum energy between them now changes as they interchange positions. The $(+ - 0)$ case depicted in Figure 49 is very different from that induced by a negative cosmological constant, shown in the lower left panel of Figure 43, where the amount of chaos is much less as compared to the non-relativistic case. In Figure 49, there is an increase in chaos throughout a band between the pretzel and annulus regions, even for comparatively small values of the total energy. As the energy increases (right panel) the expansion of the pretzel areas pushes out the annulus region. Some of the chaotic areas between the quasi annulus and pretzel regions transform into the substructures of the repeating circles.

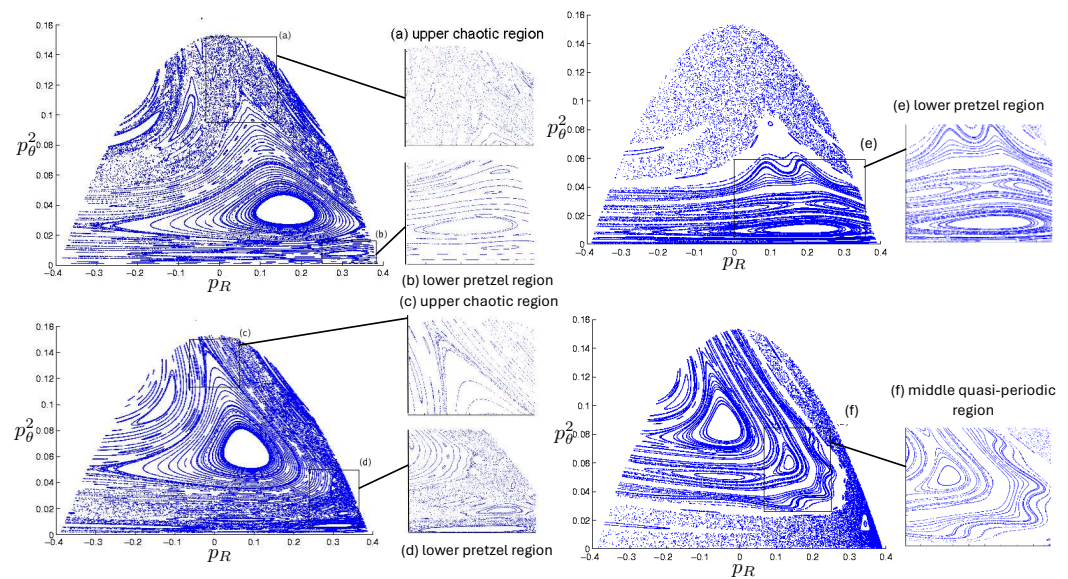

**Figure 48.** Poincaré plots of the system at $H/M_{tot} = 1.2$ corresponding to the crossings of (**upper**) the two positively charged particles and (**lower**) of the neutral particle with a positively charged particle. In the left panel, $\frac{e}{\sqrt{\kappa M_{tot}}} = (+0.1, +0.1, 0)$; in the right panel, $\frac{e}{\sqrt{\kappa M_{tot}}} = (+0.17, +0.17, 0)$. The left upper insets (**a,c**) show a close up of the upper chaotic regions; the left lower insets show the pretzel regions. The insets at the right show a close up of the structure in pretzel and quasi-periodic regions.

The $(+ + -)$ case in Figure 50 is yet again different. The recurring circles stay around the annulus region even for higher charges, and do not change into connected areas, though the annulus region is pushed to dissolve into the left pretzel region in the lower graphs and the lower pretzel in the upper graphs with a thin band of chaos between the two regions. The magnitude of the vacuum energy between the positive charges changes between 0 and $2e^2$, the former occurring whenever the negatively charged particle is between the other two. This leads to more widespread chaotic behaviour compared to the $(+ - ne)$ configuration, due to repeated changes in vacuum energy for a given total energy.

The behaviour of the charged case is quite rich and varied. A number of other scenarios have been studied at low energies [118], but the exploration of high-energy behaviour has yet to be carried out.

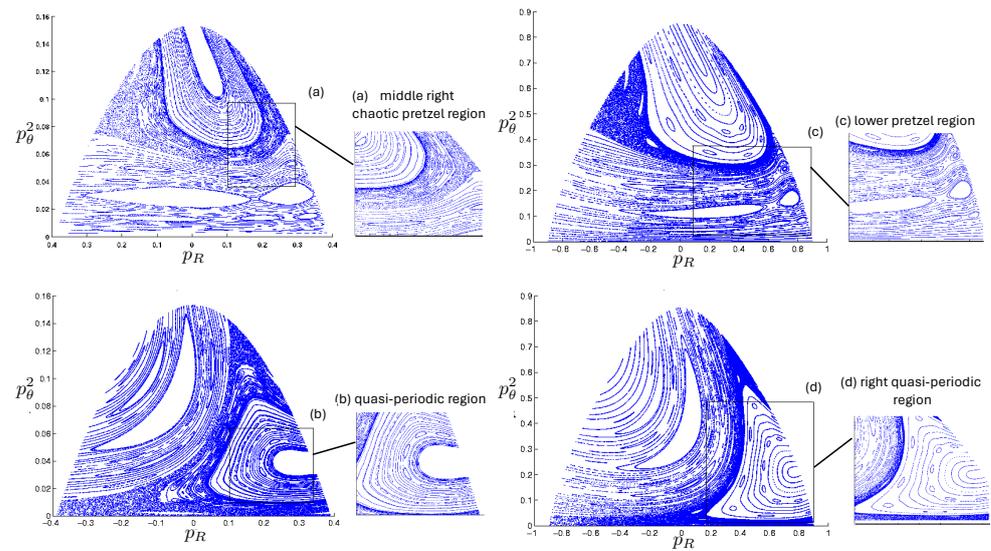

**Figure 49.** Poincaré plots of the system for charges $\frac{e}{\sqrt{\kappa}M_{tot}} = (+0.2, -0.2, 0)$; corresponding to the crossings of (**upper**) the two charged particles and (**lower**) the neutral particle with the positively charged particle. In the left panel $H/M_{tot} = 1.2$; in the right panel $H/M_{tot} = 1.8$. The upper insets (**a**,**c**) show a close up of pretzel regions; the lower insets (**b**,**d**) show quasi-periodic regions.

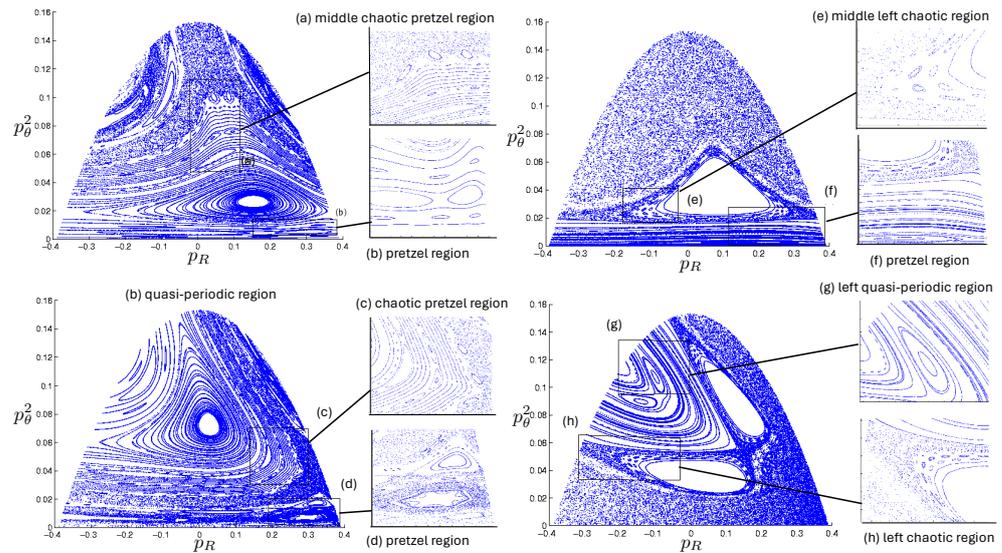

**Figure 50.** Poincaré plots of the system at $H/M_{tot} = 1.2$ corresponding to the crossings of (**upper**), the two positively charged particles and (**lower**) of the negative particle with a positively charged particle. In the left panel $\frac{e}{\sqrt{\kappa}M_{tot}} = (+0.1, +0.1, -0.1)$; in the right panel, $\frac{e}{\sqrt{\kappa}M_{tot}} = (+0.223, +0.223, -0.223)$. The left upper insets (**a**,**c**) show close ups of chaotic regions; the left lower insets (**b**,**d**) show pretzel regions. The upper and lower insets (**e**,**h**) at the right show close ups of chaotic regions; the lower and upper insets (**f**,**g**) of pretzel and quasi-periodic regions.

## 7. The Four-Body Problem

The $N$ particle OGS can be mapped to a single particle moving in $N − 1$ dimensions in a linear potential whose equipotential surfaces are that of an $N − 1$ simplex. Since the largest number of spatial dimensions that can be directly visualized is three, the $N = 4$ system—the four-body OGS—is of particular interest. Curiously, it has received almost no attention– only the non-relativistic system has been studied [124]. The relativistic 4-body system has yet to be investigated.

### 7.1. Four-Body Potential

The Hamiltonian for the non-relativistic four-body problem is given by (1) with $N = 4$. There are now six independent degrees of freedom: the three separations between the particles and their conjugate momenta. Writing

$$
\begin{aligned}
z_{12} &= \sqrt{2}\rho & z_{34} &= \sqrt{2}\alpha \\
z_{13} &= \tfrac{1}{\sqrt{2}}(\rho + \sqrt{3}\beta - \alpha) & z_{23} &= \tfrac{1}{\sqrt{2}}(-\rho + \sqrt{3}\beta - \alpha) \\
z_{24} &= \tfrac{1}{\sqrt{2}}(-\rho + \sqrt{3}\beta + \alpha) & z_{14} &= \tfrac{1}{\sqrt{2}}(\rho + \sqrt{3}\beta + \alpha)
\end{aligned}
\tag{200}
$$

where $z_{ij} = z_i - z_j$, the conjugate momenta are

$$
\begin{aligned}
p_1 &= \tfrac{1}{\sqrt{2}}\left(p_\rho + \tfrac{\sqrt{3}}{2}p_\beta\right) & p_2 &= \tfrac{1}{\sqrt{2}}\left(-p_\rho + \tfrac{\sqrt{3}}{2}p_\beta\right) \\
p_3 &= \tfrac{1}{\sqrt{2}}\left(p_\alpha - \tfrac{\sqrt{3}}{2}p_\beta\right) & p_4 &= \tfrac{1}{\sqrt{2}}\left(-p_\alpha - \tfrac{\sqrt{3}}{2}p_\beta\right)
\end{aligned}
\tag{201}
$$

with the conservation of momentum allowing us to set $p_1 + p_2 + p_3 + p_4 = 0$; the centre of mass can be fixed at the origin without loss of generality. When one of $z_{12}$, $z_{23}$, or $z_{13}$ vanish (two particles are placed directly on top of one another) this reduces to the three-body case studied in the previous section.

In the equal mass case, the Hamiltonian (1) becomes

$$
\begin{aligned}
H = \frac{1}{2m}(p_\rho^2 + p_\alpha^2 + \tfrac{3}{2}p_\beta^2) + \frac{8\pi Gm^2}{\sqrt{8}}\Big[&|\rho| + |\alpha| + \tfrac{1}{2}\big|\rho + \alpha + \sqrt{3}\beta\big| \\
&+ \tfrac{1}{2}\big|\rho - \alpha + \sqrt{3}\beta\big| + \tfrac{1}{2}\big|\rho + \alpha - \sqrt{3}\beta\big| + \tfrac{1}{2}\big|\rho - \alpha - \sqrt{3}\beta\big|\Big]
\end{aligned}
\tag{202}
$$

which is the Hamiltonian of a single particle (the box-particle) moving in three spatial dimensions in a linear potential whose shape is that of a three-simplex.

The potential

$$
V(\rho, \beta, \alpha) = H(p_\rho = 0, p_\beta = 0, p_\alpha = 0)
\tag{203}
$$

has equipotential surfaces which are that of a cube of pyramid-shaped sides, shown in Figure 51. As $V$ increases, the simplex likewise increases, as is clear from comparing the two diagrams. A cross-section of this surface through any of the edges of one of these pyramids yields a hexagon with sides of unequal length. For such cross-sections, the system reduces to that of a three-body case with unequal masses since two particles will occupy the same position.

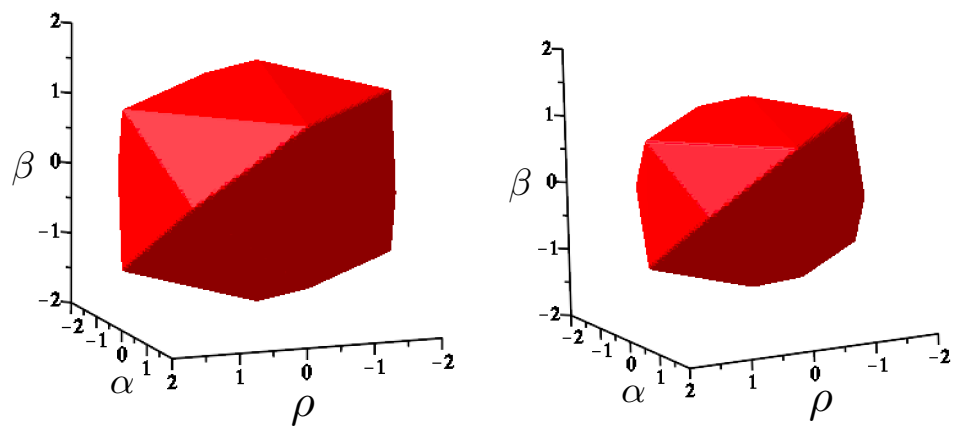

**Figure 51.** Two equipotential surfaces of the box-particle Newtonian potential in the equal mass case, with the right panel showing a smaller value of $V$ and the left one showing a larger value; axes are in units of $\kappa mc^2$.

### 7.2. Motion Classification

The four-body system has an interesting structure that can be described in terms of braid operators. This generalizes the A-type and B-type motions of the three-body case.

As in the two-body and three-body cases, each particle moves with a constant acceleration that is proportional to the difference between the total mass on its right and left sides prior to any collision. Assuming the particles move through each other, after a collision, the mass difference experienced by any given particle will, in general, change, and consequently, the acceleration of the particles also changes. From the viewpoint of the box particle, any crossing of a pair of particles corresponds to the box particle crossing a plane bisecting the three-simplex through its vertices and edges. There are a total of six such planes, obtained by setting any one of the six quantities in (200) to zero. The planes occur in pairs whose line of intersection is along each of the three principal axes.

Any sequence of crossings of $N$ bodies can be described using braid Group notation [125] using the set $\{\sigma_1, \sigma_2, \ldots, \sigma_{N-1}\}$, with $\sigma_j = \sigma_j^{-1}$, since the crossing direction is irrelevant. The positions of the particles (and not the particles themselves) are ordered as $(1, 2, 3, \ldots, N)$, where is the left-most particle is at position 1, next 2, and so on with the right-most particle being at position $N$. Note that it is the particular sequence of collisions that is important; any permutation of the operators would result in a loss of information about the motion in the system.

Applying this to the three-body system, the braid operators are $\{\sigma_1, \sigma_2\}$ and the motion can be classified into

$$
\begin{array}{ll}
\sigma_1\sigma_1 \,,\ \sigma_2\sigma_2 & \text{A motion} \\
\sigma_1\sigma_2 \,,\ \sigma_2\sigma_1 & \text{B motion}
\end{array}
\tag{204}
$$

or, in other words, the only interesting types of motion are when the same pair of particles crosses twice in a row (*A*-motion) or when one particle crosses each of its compatriots in succession (*B*-motion).

In the four-body case, there are only 3 possible crossings—$\{z_{12}, z_{23}, z_{34}\}$—at any given instant, and the braid operators $\{\sigma_1, \sigma_2, \sigma_3\}$, respectively, correspond to an interchange between the right-most, middle, and left-most pair of bodies. In this case, we have

$$
\begin{array}{ll}
\sigma_1\sigma_1 \,,\ \sigma_2\sigma_2 \,,\ \sigma_3\sigma_3 & \text{A motion} \\
\sigma_1\sigma_2 \,,\ \sigma_2\sigma_1 \,,\ \sigma_2\sigma_3 \,,\ \sigma_3\sigma_2 & \text{B motion} \\
\sigma_1\sigma_3 \,,\ \sigma_3\sigma_1 & \text{C motion}
\end{array}
\tag{205}
$$

which is depicted in Figure 52. The *A* and *B* motions represent the same physical situations as in the three-body case, but the *C* motion is new: two particles cross one another and then the other two cross one another.

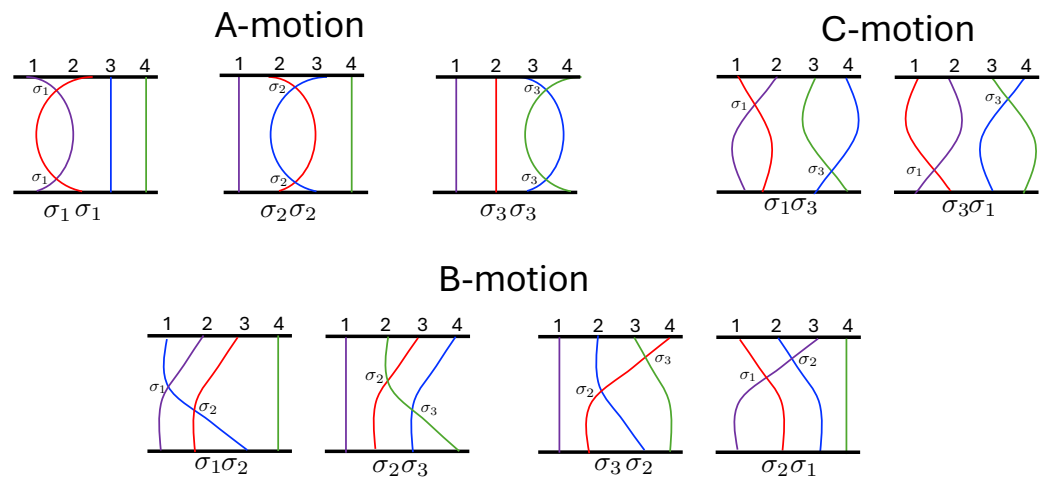

**Figure 52.** Four-body braid operations.

Multiple collisions can occur. In the three-body system, this happens when the hex particle crosses the origin, corresponding to all three bodies meeting at the same point at some instant of time. The analog of this in the four-body case occurs when the box particle crosses the line of intersection of any two bisecting planes of the three-simplexes. There are two kinds of these three-body collisions described by $\{\sigma_{1^3}, \sigma_{2^3}\}$. There are also two kinds of four-body collisions described by $\{\sigma_{1^4}, \sigma_{1^2 3^2}\}$. The former corresponds to all four bodies meeting at a single point, equivalent to the box particle crossing the origin. The latter occurs when one pair of particles crosses at one point and the other pair at a different point at the same time, corresponding to the box particle crossing one of the three lines connecting the opposite vertices of the pyramids in the simplex (see Figure 51).

### 7.3. Equal Mass Trajectories

It can be straightforwardly shown that, if one of the box-particle's position and momentum coordinates are initially zero, they will remain zero throughout the motion, and that all the phenomena seen in the three-body case in the previous section are recovered [124]. The more interesting situation is when the box particle exhibits motion in all spatial directions.

A comparison is shown in Figure 53, where $(\alpha, p_\alpha)$ remains fixed in the upper plots, but in the lower plots either $\alpha$ (lower left) or $p_\alpha$ (lower right) deviates from zero. The motion in the $\alpha$ direction simply perturbs the patterns in the upper figures, effectively giving a "thickness" to the original hex-particle patterns.

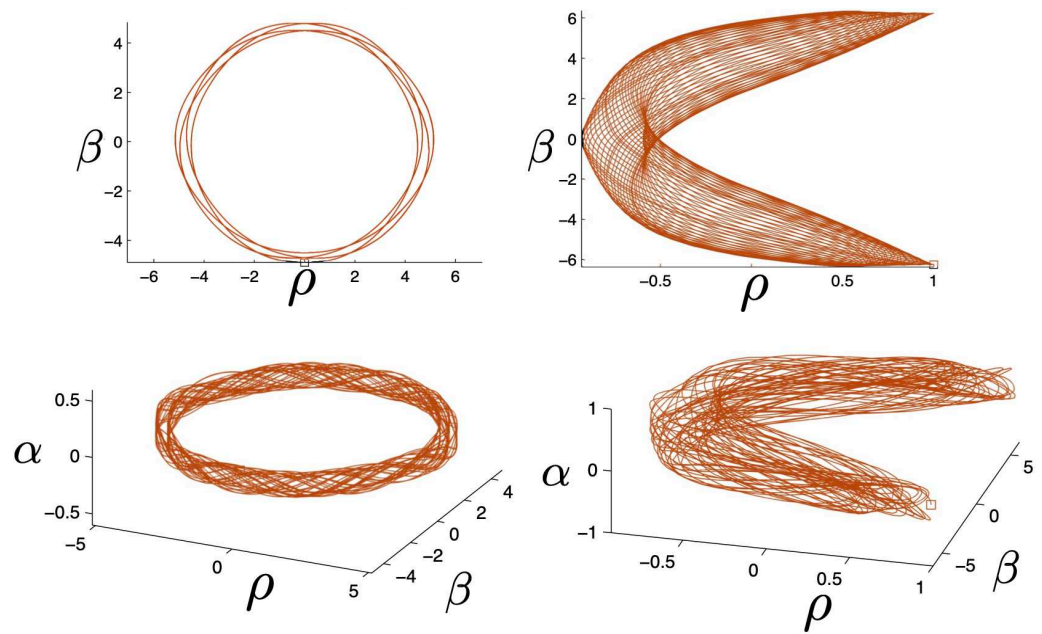

**Figure 53.** Annulus (**left**) and pretzel (**right**) orbits for the non-relativistic four-body system for 500 time steps and $\frac{H}{M_{tot}} = 2$. For the upper plots, the initial conditions are $\rho = 0$, $\alpha = 0$, $p_\rho = 0.5$, $p_\beta = 0$, $p_\alpha = 0$ (**left**) and $\rho = 1$, $\alpha = 0$, $p_\rho = 0$, $p_\beta = 0$, $p_\alpha = 0$ (**right**), with $\beta$ calculated so that (202) initially is satisfied. The initial conditions for the lower plots are $\rho = 0$, $\alpha = 0.1$, $p_\rho = 0.5$, $p_\beta = 0$, $p_\alpha = 0$ (**left**) and $\rho = 1$, $\alpha = 0$, $p_\rho = 0$, $p_\beta = 0$, $p_\alpha = 0.1$ (**right**). The respective Lyapunov exponents are $1.214 \times 10^{-2}$ (annulus, **lower left**) and $7.350 \times 10^{-3}$ (pretzel, **lower right**). The small square boxes in each diagram denote the initial conditions.

By carefully choosing the initial conditions, it is possible to obtain genuinely novel periodic orbits in three dimensions, as shown in Figure 54. This trajectory has a pretzel form when projected onto two of the planes $(\rho, \beta)$ (upper left) and $(\rho, \alpha)$ (similar to upper left panel), and has an annulus form when projected onto the third $(\beta, \alpha)$ plane (upper right). The full three-dimensional orbit is shown in the lower left panel (with no perspective so that the lines further away do not appear smaller). The motion here is $\overline{(CB^2CB^2CB^6)^2CB^6}$ and has no analogue in the three-body system. Motions of each particle are shown in the lower right panel, and have two bodies undergoing small amplitude oscillations (solid and dotted lines) with the other two undergoing a larger amplitude oscillation (dash and dot-dash lines).

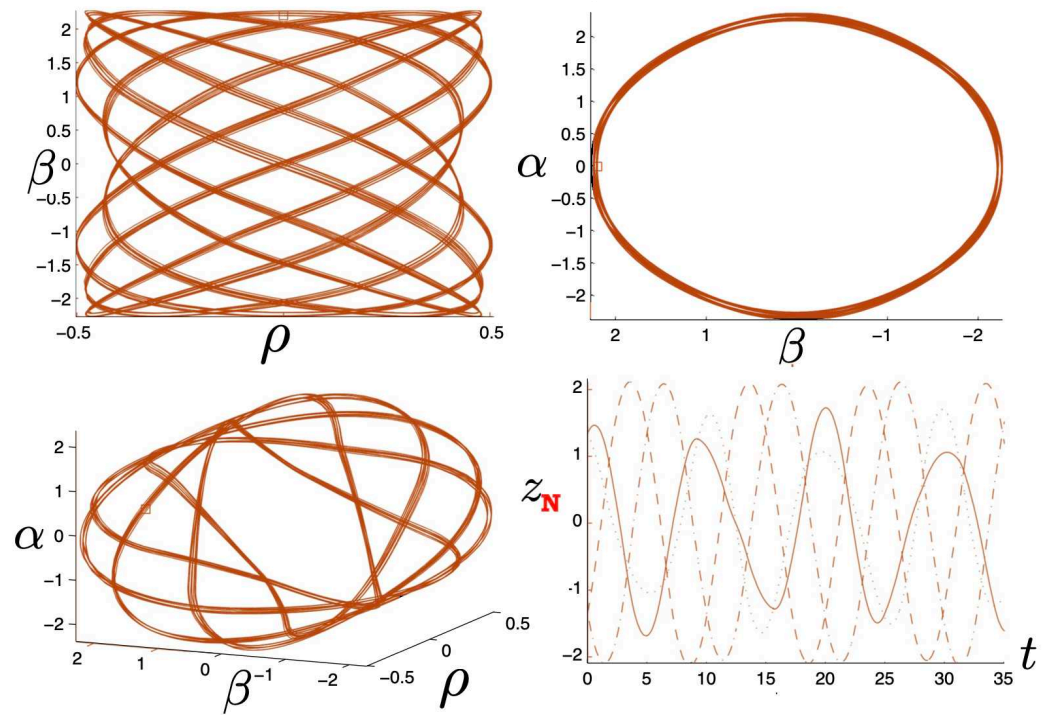

**Figure 54.** A three-dimensional periodic orbit (**lower left**) that has a pretzel form when projected into the $(\rho, \beta)$ (**upper left**) and $(\rho, \alpha)$ (similar to upper left panel) planes and annulus form when onto the $(\beta, \alpha)$ plane (**upper right**), using FE ($H/M_{tot} = 1$) conditions, where initially $\rho = \alpha = 0$, $\beta = 2.2$ and $p_\rho = 0.15$, $p_\alpha = 0.37509$, and $p_\beta = 0$. Particle trajectories are shown in the (**lower right**).

The upper panels of Figure 55 show a situation where the four particles begin at zero momentum and are equally spaced. As expected, they are all attracted together and cross at the same point, repeating this pattern indefinitely. The symbol sequence is undefined since all four particles always 'collide' at the same time step. The box particle oscillates along a line in $(\rho, \beta, \alpha)$ space (upper left), with all particles crossing at the origin simultaneously. The outer two particles (solid, dot-dash) undergo large amplitude oscillations and the inner two (dash, dotted) undergo small amplitude oscillations. However, the motion is unstable: a slight change in any of the initial conditions (via either a small perturbation in position or momentum) throws the system into chaos. This is shown in the bottom two panels of Figure 55, where the initial value of $\rho$ is slightly increased. All particle trajectories (lower right) continually vary their oscillation amplitudes. This is evident within 30 time steps, where we see the dashed-line grow in amplitude whilst the dot–dash one shrinks.

Entropy **2024**, 26, 612

71 of 84

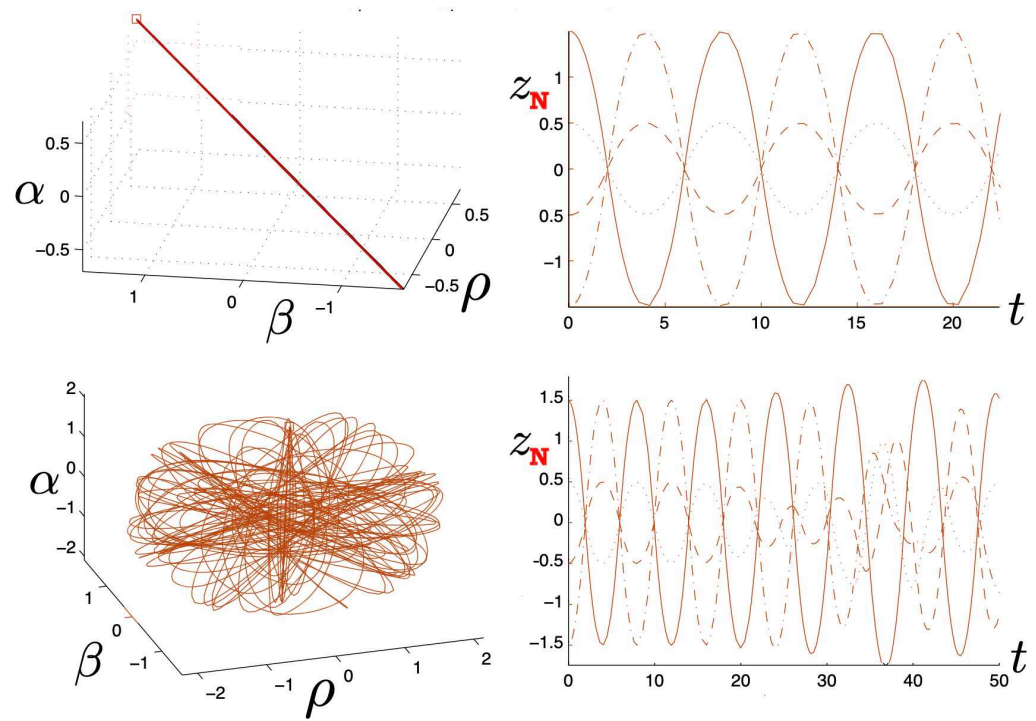

**Figure 55.** An example of how a regular motion (**upper left**) can become chaotic (**lower left**) from a small change in initial conditions, with corresponding particle trajectories shown at the right. In all figures, FE conditions are used with $H/M_{tot} = 0.625$, and initially $(p_\rho, p_\beta, p_\alpha) = (0, 0, 0)$, and $(\beta, \alpha) = (1.633, 0.70711)$. In the upper two figures $\rho = 0.70711$, initially, but in the lower two figures $\rho = 0.70712$ initially. The small box in each of the left figures marks the initial position of the box particle. The upper figures show regular motion, but the lower ones show a rapid onset of chaos.

## 7.4. Poincaré Plots

The most natural extension of a Poincaré plot to the 4-body case is the use of spherical coordinates, plotting the radial momentum $p_R$ against the squares of the two angular momenta $p_\theta^2$ and $p_\phi^2$. Writing

$$\sin\phi = \frac{\beta}{\sqrt{\rho^2 + \beta^2}}, \qquad \cos\phi = \frac{\rho}{\sqrt{\rho^2 + \beta^2}} \tag{206}$$

$$\sin\theta = \frac{\sqrt{\rho^2 + \beta^2}}{\sqrt{\rho^2 + \beta^2 + \alpha^2}}, \qquad \cos\theta = \frac{\alpha}{\sqrt{\rho^2 + \beta^2 + \alpha^2}}$$

$R$ is the distance from the origin to the point of crossing in $(\rho, \beta, \alpha)$ space and $(\theta, \phi)$ are the polar and azimuthal angles of this point. The unit vectors for these spherical coordinates are

$$\hat{R} = \begin{bmatrix} \cos\phi\sin\theta \\ \sin\phi\sin\theta \\ \cos\theta \end{bmatrix}, \quad \hat{\phi} = \begin{bmatrix} -\sin\phi \\ \cos\phi \\ 0 \end{bmatrix}, \quad \hat{\theta} = \begin{bmatrix} \cos\phi\cos\theta \\ \sin\phi\cos\theta \\ -\sin\theta \end{bmatrix} \tag{207}$$

The associated momenta are

$$p_R = \frac{p_\rho\rho + p_\beta\beta + p_\alpha\alpha}{\sqrt{\rho^2 + \beta^2 + \alpha^2}} = \hat{R}\cdot\vec{p}$$

$$p_\phi = \frac{-p_\rho\beta + p_\beta\rho}{\sqrt{\rho^2 + \beta^2}} = \hat{\phi}\cdot\vec{p} \tag{208}$$

$$p_\theta = \frac{p_\rho\rho\alpha + p_\beta\beta\alpha - p_\alpha(\rho^2 + \beta^2)}{\sqrt{(\rho^2 + \beta^2 + \alpha^2)(\rho^2 + \beta^2)}} = \hat{\theta}\cdot\vec{p}$$

and (208) can be used to compute $p_R$, $p_\phi^2$, and $p_\theta^2$ whenever two of the four particles cross one another.

Constructing the complete Poincaré plots is a challenge for two reasons. First, the standard approach of choosing a range of initial conditions that fill in the important regions is computationally much more formidable since, with five independent variables, the number of possible plots is very large. This can be dealt with by automating the generation of data over a specific range of initial conditions, but a significant reduction in the number of time steps must be employed for computational tractability. Visualizing the large number of discrete points in three-dimensional space is the second challenge. The plots in Figure 56 were constructed by separating out the space into millions of minute three-dimensional boxes, assigning a value corresponding to the number of Poincaré points that fall inside and a position corresponding to the location of the box. Although this limits the ability to zoom in to observe self-similar structures, highly saturated regions of chaos tend to show up well.

Figure 56 shows two special slices: $p_\phi = 0$ (the "bottom" slice) and $p_\theta^2 < 0.0005$ (the "side" slice). Out of 12 million points generated, 400,000 are in the bottom slice and 500,000 are in the side slice. In the latter case, any trajectory with $p_\theta = 0$ would remain in a cone rooted at the origin, and so the above bound on $p_\theta^2$ imposes the constraint of $p_\theta$ being "close" to zero.

The bottom slice in Figure 56 bears resemblance to the non-relativistic three-body case shown in Figure 31, exhibiting mixed regions of chaos and integrability. In contrast, the side slice does not display the same patterns and fractal-like properties as the bottom slice. Whether this is due to an insufficient number of time steps, a failure to cover a sufficient range of initial conditions, or an intrinsic lack of any patterns is not yet known.

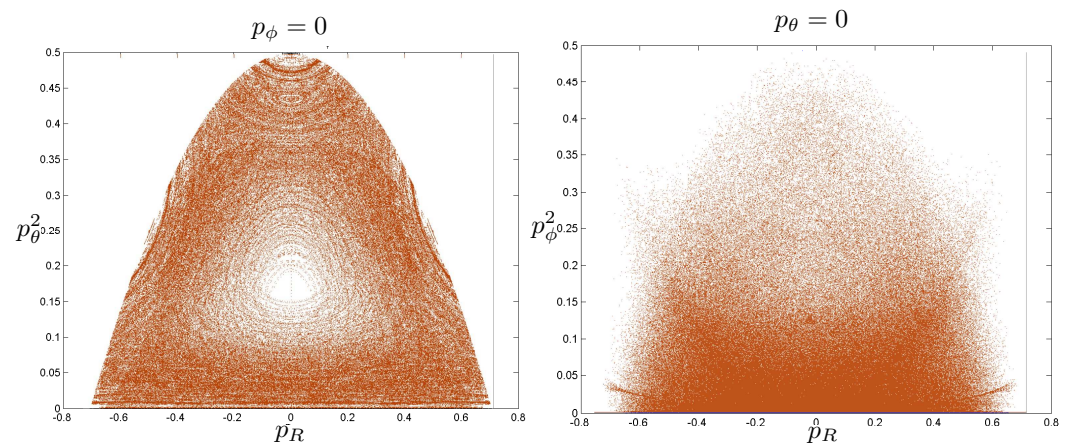

**Figure 56.** Slices of the complete Poincaré plot, with the bottom ($p_\phi = 0$) slice at **left** and the side ($p_\theta = 0$) slice at **right**. The bottom slice bears some resemblance to the 3-body non-relativistic case in Figure 31, but the side slice does not display similar fractal-like structures. Approximately 500,000 points were used to generate these figures.

The non-relativistic four-body problem has a number of other interesting aspects [124]. These include apparently chaotic motion in some projections with quasi-periodic motion in others, novel Poincaré plots for particular classes of orbits, and Lyapunov exponents that asymptote to constant values ranging between $10^{-2}$ for chaotic trajectories, $10^{-3}$–$10^{-4}$ for quasi-periodic trajectories and $10^{-5}$ (the limits of numerical precision) for periodic trajectories [124]. More complete studies remain to be carried out, not only for the non-relativistic system, but for its relativistic counterpart.

## 8. The *N*-Body Problem

Studies of the *N*-body OGS for $N > 4$ have generally been concerned with its statistical properties. Many unanswered questions remain despite extensive studies. Its ergodic and equipartition properties are still not well understood. Whether or not the OGS can attain a true equilibrium state from arbitrary initial conditions is also not clear. These issues remain outstanding in large part because the attractive interactions are cumulatively long range, unlike typical thermodynamic systems that have repulsive and short-range interactions between their constituents.

However, some statistical properties of the OGS are known. The single-particle distribution function in both the canonical and microcanonical ensemble has been derived [4]. These distribution functions reduce to the isothermal solution of the Vlasov equation in the large $N$ limit.

Even less is known about the $N$-body ROGS, with only one study of its statistical properties having been carried out to date [126]. In this section, a few of the basic results of this system will be summarized.

### 8.1. Motion Classification

Braid operators $\{\sigma_1, \sigma_2, \ldots, \sigma_N\}$ can be used to classify the motion in the $N$-body system. A sequence of $m$ pair crossings will be described by

$$\sigma_{f(1)}\sigma_{f(2)}\sigma_{f(3)}\cdots\sigma_{f(m)} \tag{209}$$

where $1 \leq f(x) \leq (N-1)$ for all $1 \leq x \leq m$ is a discrete integer function and $\sigma_{f(x)}$ means that the bodies currently in the positions $f(x)$ and $f(x) + 1$ cross. Crossing directions are irrelevant, and for a given trajectory, any given sequence of $m$ braid operators forms a unique ordered list of crossings.

It is possible to define a metric that describes the relative "distance" between any pair of crossings via

$$g(x) \equiv |\Delta f(x)| = |f(x+1) - f(x)| \tag{210}$$

which implies that $0 \leq g(x) \leq (N-2)$ for all $1 \leq x \leq (m-1)$. The motion can be then classified as follows:

$$\begin{array}{cc} g(x) & \text{Motion Class} \\ 0 & A \\ 1 & B \\ 2 & C \\ 3 & D \\ \cdots & \end{array} \tag{211}$$

denoting each type by increasing the letters of the alphabet. $A$-motion corresponds to any two crossings in nearest proximity—two particles cross each other twice in succession. $B$-motion corresponds to any two crossings in the next-nearest proximity–two particles cross each other, and then one of them crosses its other nearest neighbour. $C$-motion corresponds to any two crossings in the next-to-next-nearest proximity: two particles cross each other and then a neighbouring pair cross each other. This continues until the left-most pair of particles cross one another followed by the crossing of the right-most pair (or vice versa), which is the extreme case. In the four-body case, for example, $\sigma_1\sigma_2\sigma_1\sigma_3\sigma_2$, yields from (205), the symbol sequence $BBCB$. Computing successive values of $g(x)$ ( 1, 1, 2, 1) gives same result.

A collision of $m$ particles simultaneously corresponds to a single-particle crossing through an $(N-m)$-dimensional surface in the interior of the $(N-1)$ simplex. Such collisions can be further classified by extending the braid group notation with the set $\{\sigma_{1^m}, \sigma_{2^m}, \ldots, \sigma_{(N+1-m)^m}\}$. The subscript denotes which set of particles is involved, beginning with the left-most; the superscripts denote the number of particles in the collision, with the superscript "2" dropped when pairwise collisions occur. For example, $\sigma_{8^5}$ denotes a five-particle collision that involves particles 8–12. All collisions yield crossings with the exception of the initial conditions causing $m$ particles to occupy the same point throughout the motion (as in the upper left panel of Figure 55 in the four-body case). In this situation, the system reduces to that of an (unequal mass) $(N-m)$-body problem.

After a multiple particle collision, it is always possible to predict the new order of particles given their preceding order. Defining rightward velocity as positive, for two adjacent particles a small time just before the collision, the left one must have a larger velocity than right one or else the latter would be moving away and not toward the left one and no collision would occur. Applying this reasoning to every adjacent pair implies that, on moving from left to right, the velocity of each particle decreases in the sequence, with the left-most particle having the largest velocity. Immediately after the collision, the original order will be reversed, since the previously left-most particle will be travelling rightward faster than all other particles in the collision, and emerge afterwards as the right-most particle, and so on for all particles in the collision. If any of the $n$ particles does not satisfy the increasing velocity condition, there will be fewer than $n$ particles in the multiple collision.

### 8.2. Post-Newtonian Canonical Ensemble

The canonical one-particle distribution function can be shown to be

$$f_c^R(p,z) = \frac{1}{\mathcal{Z} N!} \int \int d\mathbf{p} d\mathbf{z} \delta(\overline{p}) \delta(\overline{z}) \exp(-\beta H) N^{-1} \sum_a \delta(z - z_a) \delta(p - p_a) \tag{212}$$

by making use of momentum conservation and translation invariance [126], where $H$ is the Hamiltonian of the system. This quantity is straightforward to compute for the N system [4], since its Hamiltonian (1) is at most quadratic in the canonical variables. However, the Hamiltonian (76) for the R system is a highly nonlinear function of these variables, and the computation of (212) is not obvious. For practical reasons, the post-Newtonian Hamiltonian $H_{pN}$ in (28) has been used to gain insights into the statistical properties of the ROGS [126].

The quantity

$$\mathcal{Z} = \frac{1}{N!} \int \int d\mathbf{p} d\mathbf{z} \delta(\overline{p}) \delta(\overline{z}) \exp(-\beta H_{pN}) \tag{213}$$

is the partition function. A somewhat tedious calculation yields the result [126]

$$\mathcal{Z} = \frac{\exp\left[-\beta Mc^2 - \frac{3(N-1)}{2}\ln(\beta mc^2) - \left\{\frac{(5N+3)(N-1)+8N\sum_{k=1}^{N-1}\sum_{l=k+1}^{N-1}\frac{(l-k)}{l(N-k)}}{8N\beta mc^2}\right\}\right]}{\sqrt{N}\left(\sqrt{2\pi}G/c^3\right)^{(N-1)}[(N-1)!]^2} \tag{214}$$

to the lowest relativistic order, where $M = \sum_{a=1}^{N} m_a$. The average energy is then

$$
\begin{aligned}
\langle E \rangle &= -\frac{\partial}{\partial \beta} \ln \mathcal{Z} \\
&= Mc^2 + \frac{3}{2\beta}(N-1) - \frac{(5N+3)(N-1)+8N\sum_{k=1}^{N-1}\sum_{l=k+1}^{N-1}\frac{(l-k)}{l(N-k)}}{8\beta^2 Mc^2}
\end{aligned} \tag{215}
$$

to the relevant order in $c^{-2}$. For fixed $M = Nm$, the relativistic correction grows quadratically with $N$ and is negative. Consequently, the average energy of the ROGS is lower than its non-relativistic counterpart at the same temperature.

When the thermal energy $kT = \beta^{-1}$ is sufficiently small relative to the rest of the energy $Mc^2$ of the system, the average energy $\zeta \equiv \frac{(E - Mc^2)}{Mc^2}$ is not significantly different from its non-relativistic value of $\frac{3(N-1)}{2\beta Mc^2}$. As $\beta$ decreases, the value of $\langle \zeta \rangle$ increases more slowly than its non-relativistic counterpart, and reaches a maximum at

$$\beta = \beta_{\max} \equiv \frac{(5N+3)(N-1)+8N\sum_{k=1}^{N-1}\sum_{l=k+1}^{N-1}\frac{(l-k)}{l(N-k)}}{6(N-1)Mc^2} \xrightarrow[N>>1]{} \left(\frac{7}{2} - \frac{2\pi^2}{9}\right)\frac{N}{Mc^2} \tag{216}$$

which is half the value of its non-relativistic counterpart. This maximum value is plotted in Figure 57 as a function of $N$, and asymptotes to the constant value $\langle \zeta \rangle = 0.573940872$ as $N \to \infty$. For $\beta > \beta_{\max}$, the average energy $\zeta$ decreases with increasing $\beta$, vanishing at $\beta = \frac{1}{2}\beta_{\max}$. The post-Newtonian expansion (28) breaks down well before this value of $\beta$.

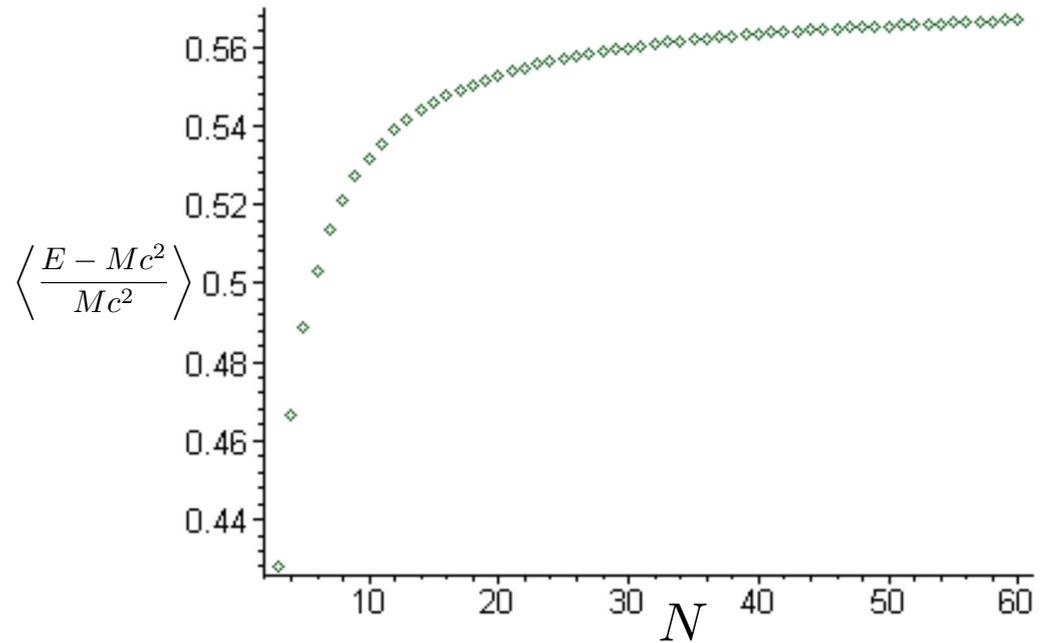

**Figure 57.** The maximum value of the average relativistic energy as a function of $N$ to leading order in $1/c^2$.

The canonical momentum distribution function can be obtained by integrating the single particle distribution function. The result is

$$
\begin{aligned}
\vartheta_{cn}(p) &= \int_{-\infty}^{\infty} dz\, f_{cn}(p,z) \\
&= \sqrt{\frac{(N\beta)}{2\pi m(N-1)}} \exp\left[-\frac{N\beta p^2}{2m(N-1)}\right] \\
&\quad \times \left(1 + \frac{1}{\beta mc^2}\left(\frac{N\beta^2 p^4\left(N^2-3N+3\right)}{8m^2(N-1)^3} - \frac{\beta p^2\left(4N^2-7N+6\right)}{4m(N-1)^2} + \frac{5N(N-1)+3}{8N(N-1)}\right)\right)
\end{aligned}
\tag{217}
$$

and corrects the standard non-relativistic Gaussian expression by a polynomial in $p^2$.

The preceding expression can be rewritten in terms of dimensionless variables

$$
\eta \equiv \frac{p}{mV} \qquad V^2 \equiv \frac{4(E-Mc^2)}{3M} = \frac{4\zeta c^2}{3}
\tag{218}
$$

and is plotted in Figure 58.

The central momentum density grows with increasing $\zeta$, but falls off more rapidly than its non-relativistic counterpart does. However, for $\eta > 2$, the momentum density grows exponentially relative to its non-relativistic counterpart, overtaking it for sufficiently large $\eta$. This is clearly seen in the right-hand panels of Figure 58. The differences become less pronounced as $N$ increases, although the basic features remain the same for all $N$.

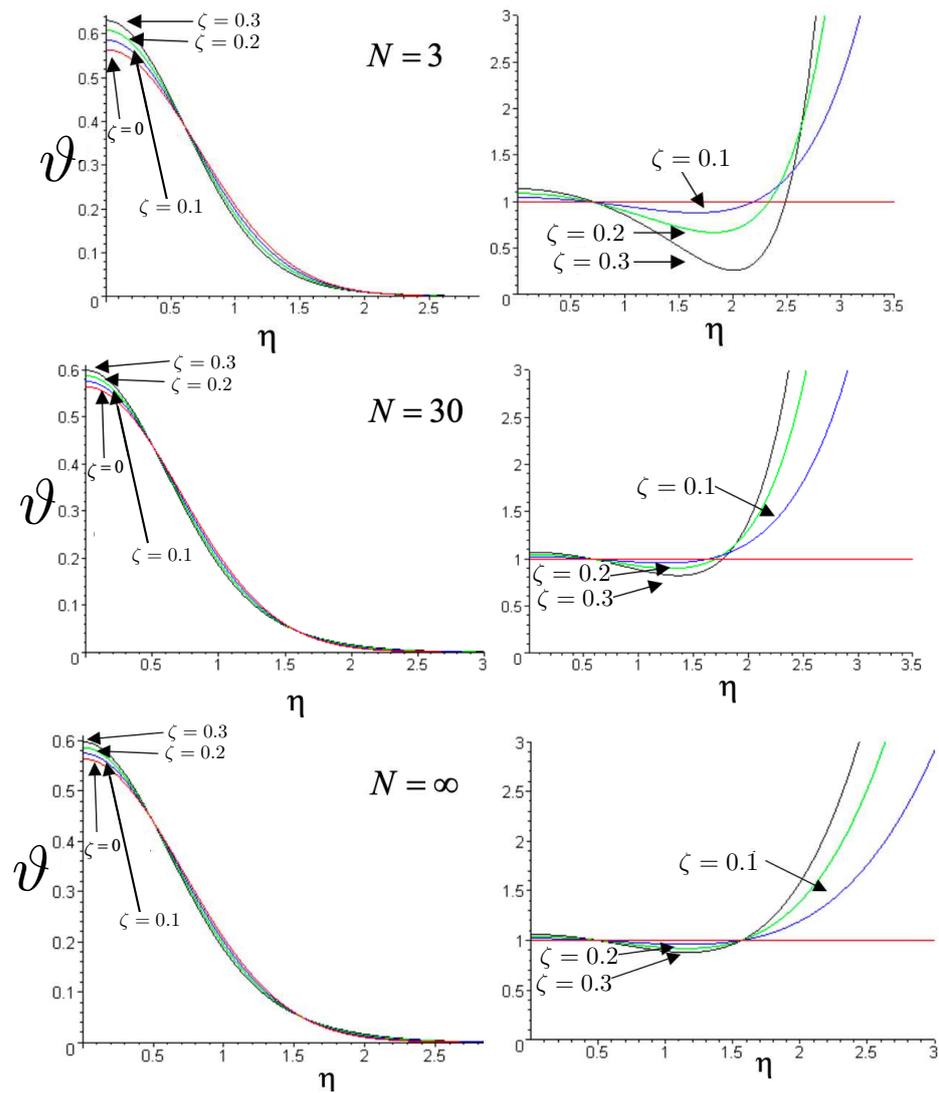

**Figure 58.** Plots of the momentum distribution (217) as a function of $\eta = \frac{p}{mV}$ (**left** column) and its value relative to the non-relativistic case (**right** column) for three different values of $N$ and various values of $\zeta$.

### 8.3. Other Statistical Features

The computation of other quantities, such as the canonical density distribution, the full-single particle canonical distribution, and the microcanonical distribution functions involve a considerable amount of tedious algebra. Some general features emerge from this analysis [126]. One is that relativistic effects cool the system: at a given energy, the ROGS temperature is smaller than the OGS temperature. Another is that the ROGS density and distribution functions become more sharply peaked than their OGS counterparts as $\zeta$ increases for any given $N$. For the sufficiently large values of the position parameter $z$, $\rho_{OGS} > \rho_{ROGS}$ for both the canonical and microcanonical distribution functions. The momentum densities exhibit a different behaviour, with $\vartheta_{OGS} > \vartheta_{ROGS}$ for small values to intermediate values $\eta$, but for large enough $\eta$, the inequality is reversed and $\vartheta_{OGS} < \vartheta_{ROGS}$.

Further exploration of the $N$-body ROGS will be challenging. A natural first step would be a consideration of the charged and cosmological systems at the post-Newtonian level. The unequal mass case is likely similarly tractable, though it will be considerably more difficult. However, a full understanding will almost certainly require a numerical approach, particularly to go beyond leading order corrections in $1/c^2$.

### 9. The Circular $N$-Body Problem

Thus far, the discussion has been concerned with lineal topology. For non-relativistic gravity, this is the only option. The OGS equations (4) for a general potential $V$ are

$$V'' = 4\pi Gm\, \delta(x - z(t)) \tag{219}$$

$$\dot{p} = -V'(z) \tag{220}$$

$$\dot{z} = \frac{p}{m} \tag{221}$$

for a single particle. The first equation implies $V = 2\pi Gm|x|$, which has a vanishing derivative at $x = 0$ and yields $p = z = 0$ as a consistent single particle solution. However, if the topology is circular, then both $V(L) = V(-L)$ and $V'(L) = V'(-L)$ for some $L$, where $2L$ is the circumference of the circle. These matching conditions have no solution unless another point source of negative mass is introduced. For $N$ bodies modelled as compact smeared sources, the problem remains: the potential grows linearly with the increasing distance from the source(s) and the matching conditions cannot be satisfied for physically reasonable (i.e., positive mass) sources.

However, the ROGS, being in a dynamical space–time, does not suffer from this problem, since the space–time can expand or contract in response to the presence of sources. It is possible to solve the canonical equations of motion and obtain both single-particle [127] and exact $N$-body static equilibrium solutions [128,129]. This latter solution corresponds to a space–time that expands/contracts in response to $N$ equal mass bodies at equidistant proper separations from one another. These are the first $N$-body dynamic equilibrium solutions in any relativistic theory of gravity.

The action is still given by (10), but the extrinsic curvature $K$ in (17) is now taken to be a time coordinate $\tau(t)$. This allows the elimination of $\pi$ from all canonical field equations. The reduced canonical action (22) instead becomes

$$I_R = \int d^2x \left\{ \sum_a p_a \dot{z}_a \delta(x - z_a) + \Pi \frac{\partial}{\partial t}(\Psi + \ln \gamma) - \mathcal{H} \right\}, \tag{222}$$

where now

$$H = \int dx\, \mathcal{H} = \frac{2\dot{\tau}}{\kappa} \int dx \sqrt{\gamma} \tag{223}$$

which is the circumference functional of the circle when $\dot{\tau}$ is constant. The $N$-body system is now time-dependent. The spatial metric can be chosen so that $\gamma = \gamma(t)$, as in general relativity on $(2 + 1)$-dimensional spatially compact manifolds [130]. The Hamiltonian will be time-independent if a time parametrization is chosen so that $\dot{\tau}\sqrt{\gamma}$ is constant.

Equilibrium solutions correspond to a situation in which all particles are motionless at various points around the circle. Consequently, they are characterized by $\dot{z}_a = 0 = p_a$, and so the canonical field Equations (49)–(54) become [128,129]

$$2\Pi' - \Pi\Psi' = 0; \tag{224}$$

$$\Psi'' - \frac{1}{4}(\Psi')^2 - (\kappa\Pi)^2 + \gamma\left(\tau^2 - \Lambda_e\right) + \kappa \sum_a \sqrt{\gamma}\, m\, \delta(x - z_a(x^0)) = 0 \tag{225}$$

$$N_0'' - \dot{\tau}\gamma - N_0\left\{(\tau^2 - \Lambda_e)\gamma + \kappa/2 \sum_a \sqrt{\gamma}\, m\, \delta(x - z_a(x^0))\right\} = 0 \tag{226}$$

$$N_1' - \dot{\gamma}/2 + \gamma\tau N_0 = 0 \tag{227}$$

$$\dot{\Pi} + \partial_1\left(-\frac{1}{\gamma}N_1\Pi + \frac{1}{2\kappa\sqrt{\gamma}}N_0\Psi' + \frac{1}{\kappa\sqrt{\gamma}}N_0'\right) = 0 \tag{228}$$

$$\dot{\Psi} + 2N_0\left(\kappa\frac{\Pi}{\sqrt{\gamma}} + \tau\right) - N_1(\frac{1}{\gamma}\Psi') = 0 \tag{229}$$

with

$$\frac{\partial N_0}{\partial x}\Big|_{x=z_a} = 0 \qquad N_1(z_a) = 0 \tag{230}$$

which follows from the geodesic equations.

Solutions to (224)–(229) must be appropriately matched at the locations $z_a$ of each particle and the identification point $|x| = L$. The solutions for $\Psi$ and $N_0$ must be continuous but are not differentiable at the particle locations; for example, $\lim_{\epsilon \to 0}[\Psi'(z_a + \epsilon) - \Psi'(z_a - \epsilon)] = \kappa\sqrt{\gamma}m$. Setting $r = |z_a - z_{a+1}| = \frac{2L}{N}$ corresponding to $N$ bodies of equal mass $m$ at equal time-varying proper separations, the solution is

$$\sqrt{\gamma} = \frac{N}{L\sqrt{\tau^2 - \Lambda_e}}\operatorname{arctanh}\left(\frac{\xi\sqrt{\frac{\kappa^2 M^2}{16N^2} + (\xi^2 - 1)(\tau^2 - \Lambda_e)} - \frac{\kappa M}{4N}}{\frac{\kappa^2 M^2}{16N^2} + \xi^2(\tau^2 - \Lambda_e)}\sqrt{\tau^2 - \Lambda_e}\right) \tag{231}$$

for the metric function $\gamma(t)$, where $\xi$ is an integration constant, $\hat{c}^2 \equiv \gamma(\tau^2 - \Lambda_e) = c_+^2 > 0$, and $M = mN$ is the total mass of the system.

The remaining functions are

$$\kappa\Pi = \frac{\pm c_+\sqrt{\beta^2 - 1}}{\frac{\sinh(\frac{c_+ L}{N})}{\sinh(c_+ L)}\sum_{a=1}^{n}\cosh(c_+(|x - z_a| - L)) - \beta} \tag{232}$$

$$\Psi = -2\ln\left(\beta - \frac{\sinh(\frac{c_+ L}{N})}{\sinh(c_+ L)}\sum_{a=1}^{n}\cosh(c_+(|x - z_a| - L))\right) - 2\ln(\frac{\kappa\Pi_0(t)}{c_+\sqrt{\beta^2 - 1}}) \tag{233}$$

$$N_0 = \frac{\dot{\tau}\gamma}{c_+^2\beta}\left(\frac{\sinh(\frac{c_+ L}{N})}{\sinh(c_+ L)}\sum_{a=1}^{n}\cosh(c_+(|x - z_a| - L))) - \beta\right) \tag{234}$$

$$N_1 = \gamma\frac{c_+}{c_+}x - \frac{\gamma^2\tau\dot{\tau}}{c_+^3\beta}\frac{\sinh(\frac{c_+ L}{N})}{\sinh(c_+ L)}\sum_{a=1}^{n}\epsilon(x - z_a)[\sinh(c_+(|x - z_a| - L)) + \sinh(c_+ L)] \tag{235}$$

where

$$\beta = \frac{4N}{\kappa M}\sqrt{\frac{\kappa^2 M^2}{16N^2} + (\xi^2 - 1)(\tau^2 - \Lambda_e)} \tag{236}$$

and $\epsilon(x) = \frac{|x|}{x}$ is a step function with $\epsilon(0) = 0$.

The system is cyclically symmetric, with $N_1(L) = N_1(-L)$, and so we can choose the origin to be halfway between any two particles in the $N = $ even case, or on a particle in the $N = $ odd case. The spatial periodicity of the solution can be better seen using the relation

$$\sum_{a=1}^{n}\cosh(c_+(|x - z_a| - L)) = \cosh(c_+ f(x))\frac{\sinh(c_+ L)}{\sinh(\frac{c_+ L}{N})} \tag{237}$$

where $f(x)$ is the saw-tooth function that peaks with a value of $L/N$ (in other words, $f(z_a) = L/N$ at the particle locations and vanishes half-way between the particles (i.e., $f(\frac{z_a - z_{a+1}}{2}) = 0$). A simple shift in the origin of the cosh function and a subsequent manipulation of the sum yields the equivalence.

If $\hat{c}^2 = c_-^2 < 0$, another class of solutions exists with

$$\sqrt{\gamma} = \frac{N}{L\sqrt{\Lambda_e - \tau^2}}\left[\arctan\left(\frac{\xi\sqrt{\frac{\kappa^2 M^2}{16N^2} + (1 - \xi^2)(\Lambda_e - \tau^2)} - \frac{\kappa M}{4N}}{\frac{\kappa^2 M^2}{16N^2} - \xi^2(\Lambda_e - \tau^2)}\sqrt{\Lambda_e - \tau^2}\right) + k\pi\right] \tag{238}$$

and $c_+ \to ic_-$ in (232)–(235). The integration constant $|\xi| < \sqrt{1 + \frac{(\kappa M)^2}{16\Lambda_e N^2}}$, due to the periodicity of $N_1(x, t)$, and $k$ is an integer. No solutions exist for $c^2 = 0$.

The solution (231)–(235) (and its $c^2 < 0$ counterpart) corresponds to an expanding/contracting space–time of a circle with $N$ bodies at equal time-varying proper separations from one another around the circle. The solution for the spatial metric is equivalent to that of single-particle solution [127] upon rescaling $L \to L/N$ and $M \to M/N$. However, this rescaling equivalence does not hold for the remaining functions. For $\Lambda_e = 0$, the space–time expands but perpetually decelerates due to the presence of the point masses. If $\Lambda_e < 0$, the proper circumference of the circle expands from zero to some maximal size and then recontracts. The most interesting behaviour occurs if $\Lambda_e > 0$. In this case, the cosmological expansion opposes the decelerating effects due to the point masses and the space–time can expand from zero size to some finite value, evolve from some minimal/maximal circumference to a maximal/minimal size, or undergo perpetual oscillation.

A generalization of this solution to one in which there are an even number of bodies with a charge alternating in sign but equal in magnitude has also been obtained [129]. This solution, and its neutral counterpart in (231)–(235), almost certainly describes an unstable equilibrium, since the masses are all equal and the particles are evenly separated. Perturbations from equilibrium would be interesting to investigate, as they would form model inhomogeneous self-gravitating cosmological systems. This remains an interesting avenue for further study.

## 10. Conclusions

Relativistic one-dimensional self-gravitating systems provide an interesting (and in the view of this author) undervalued theoretical laboratory for studying a number of physical effects of which rather little is known. These include exact two-body motion, static $N$-body equilibrium, relativistic thermodynamics and statistical mechanics, relativistic chaos, and the interplay between gravity, electromagnetism, and cosmological evolution. I shall close this review with a brief overview of seven different research avenues that warrant further study.

1. Relativistic Chaos at High Energy
   All studies of relativistic three-body chaos have been at energies below the cross-sectional maximum of the potential $V_R$ in (177). The regions of chaos in the R system are distortions of their N system counterparts, but do not increase. It would be of great interest to know what the chaotic behaviour is for energies larger than the cross-sectional maximum, where very strong relativistic effects are present. Will the chaotic regions in the Poincaré plots grow or shrink? Such studies would provide further insights into the effects of strong gravity on chaotic systems.
2. Four-Body Chaos
   The largest value of $N$ in the $N$-body problem for which equipotential surfaces can be visualized is $N = 4$. As noted in Section 7, only the N system has been investigated for its motions and chaotic behaviour. The R system has yet to be investigated along these lines. It is conceivable that qualitatively new features will be observed in this case.
3. Fully Relativistic Statistical Mechanics
   The scope for exploration here is very broad. Only the statistical properties of the neutral pN system have been studied. The effects of charge and cosmological expansion are not known, and a full study of the statistical properties of the R system remain to be carried out. This latter problem will be quite technically challenging, since the distribution functions cannot be analytically integrated. Some novel blend of analytic and numerical methods will need to be employed.
4. Circular $N$-Body Dynamics
   A novel feature of the $N$-body ROGS is that it admits two distinct spatial topologies: linear and circular. In the latter case, there are no dynamical solutions for $N \geq 2$. These will likely need to be found numerically. It would be particularly interesting to investigate three-body chaos in this setting to see what effects circular topology has on chaotic phenomena.
5. The two-dimensional $N$-body problem

Since there is no gravitational radiation in two spatial dimensions, the *N*-body problem in this setting is of considerable interest, all the more so since general relativity will provide the foundation for the field equations. This problem has been considered from a topological perspective [131], from which an implicit solution for the metric and the motion of *N* particles was obtained [132]. The solution becomes explicit for $N = 2$. However, the relationship between this approach and the canonical approach has only been explored to a limited extent [133,134]. A thorough analysis should be carried out, particularly since particle collisions can form black holes [135] and quite possibly lead to other interesting space–time effects.

6. Extensions to dilaton gravity

    The $R = T$ theory (10) has provided the context for exploring the relativistic *N*-body problem since it is the $D \to 2$ limit of general relativity [28]. However, a broad class of two-dimensional theories of gravity exist [72] and are of physical interest for a variety of reasons. Exploring the *N*-body problem in this broader context could lead to new physical insights into chaos, relativity, and quantum gravity.

7. The Quantum *N*-body Problem

    The Hamiltonian (76) is the exact energy functional of all degrees of freedom in the relativistic charged two-body system. Consequently, its quantization will be tantamount to the full quantization of gravity coupled to charged matter in one spatial dimension. The N system can be fully quantized, with energy eigenfunctions given in terms of airy functions, and the associated eigenvalues in terms of their zeroes. Perturbative solutions to the quantum pN system were obtained [136], but a full analysis of the quantum R system has yet to be carried out. This problem is of considerable interest, since there is experimental evidence that the energy states of neutrons are given by the eigenstates of the N system [137], confirming the test-mass limit of the quantum N-system. A better understanding of the quantum R-system could conceivably lead to the experimental tests of relativistic quantum gravity.

**Funding:** This work was supported in part by the Natural Sciences and Engineering Research Council of Canada.

**Data Availability Statement:** No new data were created or analyzed in this study. Data sharing is not applicable.

**Acknowledgments:** I am grateful to the students and collaborators that have contributed to the work presented in this review: Sven Bachmann, Fiona Burnell, Philip Chak, Pierre Farrugia, Peter Gustainis, Ryan Kerner, Michael Koop, Andrew Lauritzen, Justin Malecki, Geoff Potvin, Daniel Robbins, Simon Ross, Marco Raiteri, Mina Rohanazadegan, Tony Scott, Mike Trott, Alex Yale, and Matt Young. And I am especially grateful to Tadayuki Ohta (now deceased) for introducing me to the *N*-body problem and the canonical methods that allowed us to explore this interesting aspect of gravitational physics.

**Conflicts of Interest:** The author declares no conflicts of interest